\documentclass[aps,superscriptaddress,showpacs,showkeys,preprintnumbers,amsmath,amssymb]{revtex4}
\usepackage{txfonts}
\usepackage{dcolumn}
\usepackage{mathrsfs}
\usepackage{bm}
\usepackage{amsmath,amssymb,epsfig,float,graphics}
\usepackage{epstopdf}
\usepackage{amsfonts}
\usepackage{subfigure}
\begin{document}

\title{Entanglement dynamics in $\kappa$-deformed spacetime}

\author{Xiaobao Liu \footnote{Corresponding author, Email:xiaobaoliu@hotmail.com}}
\affiliation{Department of physics and electrical engineering, Liupanshui Normal University, Liupanshui 553004, Guizhou, P. R. China}

\author{Zehua Tian \footnote{Corresponding author, Email:tianzh@ustc.edu.cn}}
\affiliation{School of Physics, Hangzhou Normal University, Hangzhou, Zhejiang 311121, P. R. China}
\affiliation{CAS Key Laboratory of Microscale Magnetic Resonance and School of Physical Sciences,
University of Science and Technology of China, Hefei 230026, China}

\author{Jiliang Jing\footnote{Corresponding author, Email: jljing@hunn.edu.cn}}
\affiliation{Department of
Physics, Key Laboratory of Low Dimensional Quantum Structures and
Quantum Control of Ministry of Education, and Synergetic Innovation
Center for Quantum Effects and Applications, Hunan Normal
University, Changsha, Hunan 410081, P. R. China}

\begin{abstract}
We treat two identical and mutually independent two-level atoms that are coupled to a quantum field as an open quantum system. The master equation that governs their evolution is derived by tracing over the degree of freedom of the field. With this, we compare the entanglement dynamics of the two atoms moving with different trajectories in $\kappa$-deformed and Minkowski spacetimes. Notably, when the environment-induced interatomic interaction does not exist, the entanglement dynamics of two static atoms in $\kappa$-deformed spacetime are reduced to that in Minkowski spacetime in the case that the spacetime deformation parameter $\kappa$ is sufficiently large as theoretically predicted. However, if the atoms undergo relativistic motion, regardless of whether inertial or non-inertial, their entanglement dynamics in $\kappa$-deformed spacetime behave differently from that in Minkowski spacetime even when $\kappa$ is large. We investigate various types of entanglement behavior, such as decay and generation, and discuss how different relativistic motions, such as uniform motion in a straight line and circular motion, amplify the differences in the entanglement dynamics between the $\kappa$-deformed and Minkowski spacetime cases. In addition, when the environment-induced interatomic interaction is considered, we find that it may also enhance the differences in the entanglement dynamics between these two spacetimes. Thus, in principle, one can tell whether she/he is in $\kappa$-deformed or Minkowski spacetime by checking the entanglement behavior between two atoms in certain circumstances.
\end{abstract}

\maketitle

\section{Introduction}
Quantum entanglement, as a crucial physical resource in technologies based on quantum effects, is the essential feature underlying quantum information, cryptography, and quantum computation \cite{Nielsen2000,Bouwmeester2000,Raimond2001}.
Recently, quantum entanglement has been investigated together with the theories of relativity and quantum fields.
For example, entanglement dynamics~\cite{Benatti2004,Benatti2005,Cliche2011,Hu2011,Martinez2012,TianAOP2013,Martinez2014,TianAOP2014,Yu2015,Tian2017,TianAOP2017,Kukita2017CQG,She2019,He2020,Zhou2021,Yan2022,Soares2022} and entanglement harvesting~\cite{Salton2015,Pozas2015,Kukita2017,Henderson2018,Ng2018,Cong2019,Cong2020,Tjoa2020,Zhang2021,Suryaatmadja2022,Barman2022} have been investigated in a variety of different relativistic scenes and settings.
The aforementioned studies aimed to explore the effects of relativistic motion and gravity on quantum entanglement, and in turn, entanglement could be used to probe the structure of spacetime and analyze gravitational matter interactions. On the one hand, the unavoidable coupling between a quantum system and the external environment usually leads to decoherence and dissipation of the quantum system and may also cause disentanglement or even entanglement sudden death~\cite{Yu2004,Dodd2004,Eberly2007}, and thus, it has been considered one of the major obstacles to the realization of quantum information technologies. On the other hand, a common bath may induce indirect interactions between otherwise independent atoms immersed as a consequence of environment correlations. Thus, in this case, entanglement could be generated for atoms, although initially in separable states~\cite{Braun2002,Kim2002,Ficek2008,Tanas2010}, and even the destroyed entanglement may also be revived~\cite{Ficek2006}. Furthermore, the entanglement of two atoms with vanishing separation persists even in the asymptotic equilibrium regime~\cite{Benatti2004, Benatti2005}. In Ref.~\cite{Chen2022}, the authors showed that the environment-induced interatomic interaction would enable entanglement generation.  In particular, the entanglement character of two particle detectors in two different but quite similar spacetimes, e.g., the de Sitter spacetime and the thermal Minkowski spacetime~\cite{Steeg2009,Hu2013,Nambu2011,Nambu2013}, has been recently discussed. Notably, the relevant differences in the entangling power of spacetime could be used to distinguish these different universes in principle~\cite{Tian2017,Steeg2009,Hu2013,Nambu2011,Nambu2013,Tian2014,Tian2016,LiuPRD2018}. The aforementioned investigations remind us that the study of entanglement dynamics within a framework of relativity can elucidate the nature of spacetime, which motivates us to explore related fields.

The Minkowski spacetime has a continuous structure whose spacetime coordinates are commutative, and the scalar field theory is well constructed in this commutative spacetime. However, on the microscopic level, one of the significant topics of quantum gravity theories concerns the modification of the notion of spacetime and the quantization of spacetime coordinates. This requirement may result in modification of the notions of symmetry of the spacetime. The symmetric algebra of a certain quantum gravity model is known to be the $\kappa$-Poincar\'{e} algebra. The corresponding ``Lie algebra"-type noncommutative spacetime is named the $\kappa$-Minkowski spacetime~\cite{Lukierski1995,Majid1994}. In this respect, the exploration of this noncommutative spacetime can help us deepen our understanding of the structure and properties of spacetime in the microscopic areas. Substantially related works on the $\kappa$ spacetime, as well as the construction and relevant investigations of field theory models of this spacetime, exist~\cite{Kosinski2000,Kowalski2002,Amelino2002,Dimitrijevi2003,Kim2007,Freidel2007,Daszkiewicz2008,Kovacevic2012,Meljanac2012} (and references cited therein). Usually, the quantum field theory in $\kappa$-deformed spacetime is complicated as a result of the non-commutation of spacetime coordinates. However, recently, in Ref.~\cite{Harikumar2012}, an interesting model starts with the $\kappa$-deformed Klein-Gordon theory, which is invariant under a $\kappa$-Poincar\'{e} algebra and written in commutative spacetime. Inspired by this approach, we, in Ref.~\cite{Liu2018}, investigated the quantum Fisher information in $\kappa$-deformed spacetime and determined that the relativistic motion can effectively improve the quantum precision in the estimation of spacetime deformation. For a review of recent results regarding resonance interaction energy in $\kappa$-Minkowski spacetime, we refer the reader to the work presented in Ref.~\cite{Xu2023}. In this regard, we note that the possible and expected $\kappa$ deformation of spacetime is weak, such that all of the physical theories are usually consistent with the Minkowski spacetime. Hence, it is worthy asking whether it is possible to distinguish the $\kappa$-deformed spacetime from the Minkowski spacetime?

In this study, we investigate the entanglement dynamics of a two-atom system coupled to a scalar field in $\kappa$-deformed spacetime and compare the results with that in Minkowski spacetime. First, the master equation that governs the system evolution is derived, and the standard field theory techniques developed for the commutative spacetime are used to explore this $\kappa$-deformed scalar theory~\cite{Harikumar2012,Liu2018}. Then, we consider the evolution behavior of entanglement between two atoms with different trajectories in $\kappa$-deformed spacetime and Minkowski background. Our results indicate that the relativistic motion of the atoms can help us distinguish the entanglement dynamics between $\kappa$-deformed and Minkowski spacetimes. In addition, when choosing special initial states that may introduce the environment-induced interatomic interaction, the atomic entanglement dynamics could also behave quite differently in these two universes. Thus, the differences in the entanglement dynamics of two atoms can be used to distinguish $\kappa$-deformed and Minkowski spacetimes in principle.

This paper is organized as follows: In Section \ref{section2}, the basic formulas of the master equation for the two-atom system interacting with the scalar field are reviewed. We also review the $\kappa$-deformed scalar theory written in commutative spacetime and its propagator. In Section \ref{section3}, we consider the entanglement dynamics of the two-atom system in $\kappa$-deformed and Minkowski spacetimes without the environment-induced interatomic interaction. In Section \ref{section4}, the influence of the environment-induced interatomic interaction on the behavior of the entanglement dynamics is explored in detail. Finally, in Section \ref{section5}, we present our conclusions.

Throughout the paper, we employ natural units $c=\hbar=1$. Relevant constants are restored when needed for the sake of clarity.

\section{Master equation and Scalar field propagator} \label{section2}
We simply review the master equation of two atoms that interact with a scalar field in the vacuum fluctuation and introduce the corresponding scalar field propagator in $\kappa$-deformed spacetime with the commutative theory.

\subsection{Master equation of two-atom system}
Let us consider two identical and mutually independent atoms weakly coupled to a bath of fluctuating scalar field in its vacuum state. We also applied this model in the relativistic scenario recently~\cite{TianJHEP2013,Jia2015,LiuAOP2016,LiuQIP2016,Yang2018,Liu2020,Liu2021,Tian2023,Jing2022,Jing20222,Jing20231}.

The total Hamiltonian $H$ for the complete system, i.e., the two atoms together with the external scalar field, is expressed as follows:
\begin{eqnarray}\label{Hamiltonian0}
H=H_S+H_{E}+H_I.
\end{eqnarray}
Here, the free two-atom Hamiltonian $H_S$ is written as follows:
\begin{eqnarray}\label{atoms Hamiltonian}
H_S=\frac{1}{2}\omega_0\sigma_3^{(1)}+\frac{1}{2}\omega_0\sigma_3^{(2)},
\end{eqnarray}
where $\sigma_i^{(1)}=\sigma_i\otimes \mathbf{I}$, $\sigma_i^{(2)}=\mathbf{I}\otimes \sigma_i$, $\sigma_i$ with $i\in \{1,2,3\}$ are the Pauli matrices and $\mathbf{I}$ is the $2\times2$ unit matrix, and $\omega_0$ is the energy level spacing of an individual atom. Note that $H_{E}$ is the free Hamiltonian of the scalar field, whose explicit expression is not required here, and $H_I$ is the interaction Hamiltonian between atoms and the scalar field. We assume that the coupling between the two atoms and the scalar field takes the following form, specifically in analogy to the electric dipole interaction~\cite{Audretsch1994}:
\begin{eqnarray}
H_I=\mu [\sigma_2^{(1)}\phi(\mathbf{x}_1(\tau))+\sigma_2^{(2)}\phi(\mathbf{x}_2(\tau))].
\end{eqnarray}
Here, $\mu$ is the coupling constant that we assume to be small, and $\phi(\mathbf{x}_{\alpha}(\tau))$ with $\alpha\in \{1,2\}$ corresponds to the scalar field operator with $\tau$ being the atomic proper time.

In the frame of atoms, the evolution time of the total system is governed by the von Neumann equation:
\begin{eqnarray}\label{whole}
\frac{\partial\rho_{\mathrm{tot}}(\tau)}{\partial\tau}=-i[H,\rho_{\mathrm{tot}}(\tau)].
\end{eqnarray}
We assume the initial density matrix of the atoms-field system as follows:
\begin{eqnarray}
\rho_{\mathrm{tot}}=\rho(0)\otimes|0\rangle\langle0|,
\end{eqnarray}
where $\rho(0)$ is the initial reduced density matrix of the two-atom system, and $|0\rangle$ is the vacuum state of the scalar field. We are particularly interested in the evolution time of the two-atom system; thus, we trace over the field degrees of freedom, i.e., $\rho(\tau)=\mathrm{Tr}_E[\rho_{t\mathrm{ot}}(\tau)]$. Under the Born-Markov approximation~\cite{Breuer2002}, the reduced dynamics of the two-atom system can be expressed based on the Kossakowski-Lindblad form~\cite{Lindblad1,Lindblad2,Benatti1} in the limit of weak coupling, as follows:
\begin{eqnarray}\label{Lindblad equation}
\frac{\partial\rho(\tau)}{\partial\tau}&=&-i[H_{\mathrm{eff}},\rho_A(\tau)]+\mathcal{D}[\rho(\tau)],
\end{eqnarray}
where the effective Hamiltonian $H_{\mathrm{eff}}$ is
\begin{eqnarray}\label{Effective H}
H_{\mathrm{eff}}=H_S-\frac{i}{2}\sum_{\alpha,\beta=1}^2\sum_{i,j=1}^3 H_{ij}^{(\alpha\beta)}\sigma_i^{(\alpha)}\sigma_j^{(\beta)},
\end{eqnarray}
and the dissipator $\mathcal{D}[\rho(\tau)]$ is
\begin{eqnarray}\label{Effective D}
\mathcal{D}[\rho(\tau)]&=&\frac{1}{2}\sum_{\alpha,\beta=1}^2\sum_{i,j=1}^3 C_{ij}^{(\alpha\beta)}\nonumber\\
&&\;\;\;\times[2\sigma_j^{(\beta)}\rho\sigma_i^{(\alpha)}-\sigma_i^{(\alpha)}\sigma_j^{(\beta)}\rho-\rho\sigma_i^{(\alpha)}\sigma_j^{(\beta)}].
\end{eqnarray}
In the master equation (\ref{Lindblad equation}), the dissipator $\mathcal{D}[\rho(\tau)]$ describes the environment-induced decoherence and dissipation, which means that the evolution of the quantum system is nonunitary.
The free Hamiltonian of the two-atom system, which incarnates in the effective Hamiltonian $H_{\mathrm{eff}}$, is also modified. The coefficients of the matrix $C_{ij}^{(\alpha\beta)}$ in Eq.~(\ref{Effective D}) are expressed as follows:
\begin{eqnarray}\label{coefficient matrix}
C^{(\alpha\beta)}_{ij}=
A^{(\alpha\beta)}\delta_{ij}-iB^{(\alpha\beta)}\epsilon_{ijk}\delta_{3k}-A^{(\alpha\beta)}\delta_{3i}\delta_{3j},
\end{eqnarray}
where
\begin{eqnarray}\label{AB}
&&A^{(\alpha\beta)}=\frac{\mu^2}{4}[\mathcal{G}^{(\alpha\beta)}(\omega)+\mathcal{G}^{(\alpha\beta)}(-\omega)],\nonumber\\
&&B^{(\alpha\beta)}=\frac{\mu^2}{4}[\mathcal{G}^{(\alpha\beta)}(\omega)-\mathcal{G}^{(\alpha\beta)}(-\omega)].
\end{eqnarray}
Based on the previous expressions, we have defined
\begin{eqnarray}\label{Fourier transform}
\mathcal{G}^{(\alpha\beta)}(\lambda)=\int_{-\infty}^{+\infty}d \Delta\tau e^{i\lambda\Delta\tau} G^{(\alpha\beta)}(\Delta\tau),
\end{eqnarray}
which is the Fourier transform of the following field correlation functions:
\begin{eqnarray}\label{correlation functions0}
G^{(\alpha\beta)}(\Delta\tau)=\langle \phi(\mathbf{x}_\alpha(\tau)) \phi(\mathbf{x}_\beta(\tau')) \rangle.
\end{eqnarray}
Similarly, $H_{ij}^{(\alpha\beta)}$ in the previous expressions can be derived by replacing the Fourier transform $\mathcal{G}^{(\alpha\beta)}(\lambda)$ with the Hilbert transform $\mathcal{K}^{(\alpha\beta)}(\lambda)$, as follows:
\begin{eqnarray}\label{Hilbert transform}
\mathcal{K}^{(\alpha\beta)}(\lambda)=\frac{P}{\pi i}\int_{-\infty}^{+\infty} d\omega\frac{\mathcal{G}^{(\alpha\beta)}(\lambda)}{\omega-\lambda},
\end{eqnarray}
with $P$ denoting the principal value. In Refs.~\cite{Benatti2004,Benatti2005}, the effective Hamiltonian $H_{\mathrm{eff}}=\tilde{H}_S+H_{\mathrm{eff}}^{(12)}$ includes two terms. The first term is the renormalization of the transition frequencies, i.e., the Lamb shift of each atom, and is derived by replacing $\omega$ in the Hamiltonian $H_S$ (\ref{atoms Hamiltonian}) of the atom with a renormalized energy level spacing, as follows:
\begin{eqnarray}
\tilde{\omega}=\omega+\frac{i\mu^2}{2}[\mathcal{K}^{(11)}(-\omega)-\mathcal{K}^{(11)}(\omega)].
\end{eqnarray}
Note that this term can be regarded as a rescaling of the gap in the energy level; thus, we shall not consider it any further. The second term is an environment-induced coupling between the atoms, which can be expressed as follows:
\begin{eqnarray}
H_{\mathrm{eff}}^{(12)}=-\sum_{i,j=1}^3 \Omega_{ij}^{(12)}\sigma_i\otimes\sigma_j,
\end{eqnarray}
where
\begin{eqnarray}\label{D}
\Omega_{ij}^{(12)}=D\delta_{ij}-D\delta_{3i}\delta_{3j},
\end{eqnarray}
with
\begin{eqnarray}\label{D1}
D=\frac{i\mu^2}{4}[\mathcal{K}^{(12)}(-\omega)+\mathcal{K}^{(12)}(\omega)].
\end{eqnarray}
As a result, we can rewrite the master equation (\ref{Lindblad equation}) as follows:
\begin{eqnarray}\label{master equation}
&&\frac{\partial\rho(\tau)}{\partial\tau}
=-i\tilde{\omega}\sum_{\alpha=1}^2[\sigma_{3}^{(\alpha)},\rho_(\tau)]+
i\sum_{i,j=1}^3 \Omega_{ij}^{(12)}[\sigma_i\otimes\sigma_j,\rho(\tau)]\nonumber\\
&&\;\;\;+\frac{1}{2}\sum_{\alpha,\beta=1}^2\sum_{i,j=1}^3C^{(\alpha\beta)}_{ij}
[2\sigma_j^{(\beta)}\rho\sigma_i^{(\alpha)}-\sigma_i^{(\alpha)}\sigma_j^{(\beta)}\rho-
\rho\sigma_i^{(\alpha)}\sigma_j^{(\beta)}].\nonumber\\
\end{eqnarray}

To analyze the dynamics of the two-atom system, we need to solve the master equation~(\ref{Lindblad equation}) in terms of the coupled basis, i.e., in the set: $\{|G\rangle=|00\rangle,|A\rangle=\frac{1}{\sqrt{2}}(|10\rangle-|01\rangle),|S\rangle=\frac{1}{\sqrt{2}}(|10\rangle+|01\rangle),|E\rangle=|11\rangle\}$. Moreover, based on Eqs. (\ref{coefficient matrix})-(\ref{D}), we can write the coefficient of the dissipator in the master equation (\ref{master equation}) as follows:
\begin{eqnarray}\label{coefficient matrix1}
&&C^{(11)}_{ij}=A_1\delta_{ij}-iB_1\epsilon_{ijk}\delta_{3k}-A_1\delta_{3i}\delta_{3j},\nonumber\\
&&C^{(22)}_{ij}=A_2\delta_{ij}-iB_2\epsilon_{ijk}\delta_{3k}-A_2\delta_{3i}\delta_{3j},\nonumber\\
&&C^{(12)}_{ij}=A_3\delta_{ij}-iB_3\epsilon_{ijk}\delta_{3k}-A_3\delta_{3i}\delta_{3j},\nonumber\\
&&C^{(21)}_{ij}=A_4\delta_{ij}-iB_4\epsilon_{ijk}\delta_{3k}-A_4\delta_{3i}\delta_{3j},\nonumber\\
&&\Omega^{(12)}_{ij}=D\delta_{ij}-D\delta_{3i}\delta_{3j}.
\end{eqnarray}
Then, the master equation~(\ref{master equation}) in terms of the coupled basis can be rewritten as follows \cite{Ficek2002}:
\begin{eqnarray}
\dot{\rho}_{GG}&=&-2(A_1+A_2-B_1-B_2)\rho_{GG}+(A_1+A_2-A_3-A_4+B_1+B_2-B_3-B_4)\rho_{AA}\nonumber\\
&&+(A_1+A_2+A_3+A_4+B_1+B_2+B_3+B_4)\rho_{SS}+(A_1-A_2-A_3+A_4+B_1\nonumber\\
&&-B_2-B_3+B_4)\rho_{AS}+(A_1-A_2+A_3-A_4+B_1-B_2+B_3-B_4)\rho_{SA},\nonumber\\
\dot{\rho}_{EE}&=&-2(A_1+A_2+B_1+B_2)\rho_{EE} + (A_1+A_2-A_3-A_4-B_1-B_2+B_3+B_4)\rho_{AA}\nonumber\\
&&+(A_1+A_2+A_3+A_4-B_1-B_2-B_3-B_4)\rho_{SS}+(-A_1+A_2+A_3-A_4+B_1\nonumber\\
&&-B_2-B_3+B_4)\rho_{AS}+(-A_1+A_2-A_3+A_4+B_1-B_2+B_3-B_4)\rho_{SA},\nonumber\\
\dot{\rho}_{AA}&=&-2(A_1+A_2-A_3-A_4)\rho_{AA} +(A_1+A_2-A_3-A_4-B_1-B_2+B_3+B_4) \rho_{GG},\nonumber\\
&&+(A_1+A_2-A_3-A_4+B_1+B_2-B_3-B_4)\rho_{EE}+(-B_1+B_2+B_3-B_4)\rho_{AS}\nonumber\\
&&+(-B_1+B_2+B_3-B_4)\rho_{SA},\nonumber\\
\dot{\rho}_{SS}
&=&-2(A_1+A_2+A_3+A_4)\rho_{SS} +(A_1+A_2+A_3+A_4-B_1-B_2-B_3-B_4) \rho_{GG},\nonumber\\
&&+(A_1+A_2+A_3+A_4+B_1+B_2+B_3+B_4)\rho_{EE}+(-B_1+B_2+B_3-B_4)\rho_{AS}\nonumber\\
&&+(-B_1+B_2+B_3-B_4)\rho_{SA},\nonumber\\
\dot{\rho}_{AS}&=&(A_1-A_2-A_3+A_4-B_1+B_2+B_3-B_4)\rho_{GG}+(-A_1+A_2+A_3-A_4-B_1\nonumber\\
&&+B_2+B_3-B_4)\rho_{EE}-2(A_1+A_2+2iD)\rho_{AS},\nonumber\\
\dot{\rho}_{SA}&=&(A_1-A_2+A_3-A_4-B_1+B_2-B_3+B_4)\rho_{GG}+(-A_1+A_2-A_3+A_4-B_1\nonumber\\
&&+B_2-B_3+B_4)\rho_{EE}-2(A_1+A_2-2iD)\rho_{SA},\nonumber\\
\dot{\rho}_{GE}&=&-2(A_1+A_2)\rho_{GE}, \;\;\;\;\;\;\;\;\; \;\;\;\;\;\;\;\dot{\rho}_{EG}=-2(A_1+A_2)\rho_{EG},\label{state1}
\end{eqnarray}
where $\rho_{IJ}=\langle I|\rho|J\rangle$, $I,J\in \{G,A,S,E\}$, and $\dot{\rho}_{IJ}$ is the derivative with respect to the atomic proper time $\tau$.

Note that the parameter $D$ contains the environment-induced interatomic interaction; thus, if $D=0$, then Eq.~(\ref{state1}) recovers to a scenario where the environment-induced interatomic interaction does not exist for the two atoms coupled to the scalar field, as reported in Ref.~\cite{Ficek2002}. Notably, if the initial matrix takes the X form, i.e., states with nonzero elements only along the diagonal and antidiagonal of the density matrix, in the decoupled basis $\{|00\rangle,|01\rangle,|10\rangle,|11\rangle\}$, then the X structure will be maintained during evolution.
To learn the entanglement dynamics of the two-atom system, we use concurrence to characterize quantum entanglement, which was introduced by Wootters~\cite{Wootters1998}. For the X states, the concurrence is analytically expressed as follows~\cite{Ficek2004}:
\begin{eqnarray}\label{concurrence0}
C[\rho(\tau)]=\textrm{max}\{0,K_1(\tau),K_2(\tau)\},
\end{eqnarray}
where
\begin{eqnarray}\label{concurrence1}
&&K_1(\tau)=\sqrt{[\rho_{AA}(\tau)-\rho_{SS}(\tau)]^2-[\rho_{AS}(\tau)-\rho_{SA}(\tau)]^2}\nonumber\\
&&\;\;\;\;\;\;\;\;\;\;\;\;\;\;\;-2\sqrt{\rho_{GG}(\tau)\rho_{EE}(\tau)},
\end{eqnarray}
\begin{eqnarray}\label{concurrence2}
&&K_2(\tau)=-\sqrt{[\rho_{AA}(\tau)+\rho_{SS}(\tau)]^2-[\rho_{AS}(\tau)+\rho_{SA}(\tau)]^2}\nonumber\\
&&\;\;\;\;\;\;\;\;\;\;\;\;\;\;\;+2|\rho_{GE}(\tau)|.\
\end{eqnarray}

\subsection{Scalar field propagator in $\kappa$-deformed and Minkowski spacetimes}
We are interested in the entanglement dynamics in $\kappa$-deformed and Minkowski spacetimes. Here, we simply review the $\kappa$-deformed Klein-Gordon theory, especially the field correlation function in $\kappa$-deformed spacetime, which plays an important role in the following calculation.

First, let us provide the basic ingredients for the field correlation function of the scalar field in $\kappa$-deformed spacetime. For more details, please refer to Refs.~\cite{Harikumar2012,Liu2018}, where the $\kappa$-deformed Klein-Gordon theory has been investigated in the commutative spacetime itself. This processing enables us to explicitly define the trajectory of the motion of atoms in the commutative spacetime. Specifically, in $\kappa$-deformed spacetime, the time and space coordinates are not commutative but obey the following Lie algebra-type commutation relations
\begin{eqnarray}\label{commutation relation}
[\hat{x}_{i},\hat{x}_{j}]=0,\;\;\;[\hat{x}_{0},\hat{x}_{i}]= \frac{i}{\kappa}\hat{x}_{i},
\end{eqnarray}
with $i,j\in\{1,2,3\}$ and the positive parameter $\kappa$ representing the deformation of the spacetime. In Refs.~\cite{Lukierski1995,Majid1994}, the authors indicated that the symmetry of $\kappa$-deformed spacetime is well known to be the $\kappa$-Poincar\'{e} algebra, in which the defined relations of this algebra involve the deformation parameter $\kappa$, and when $\kappa\rightarrow\infty$, it is reduced to the Poincar\'{e} algebra. To construct the $\kappa$-Poincar\'{e} algebra, we seek the realizations of the noncommutative coordinate $\hat{x}_\mu$ in terms of the ordinary commutative coordinate $x_{\mu}$ and the corresponding derivative $\partial_{\mu}$: $\partial_{\mu}=\frac{\partial}{\partial x_\mu}$. These realizations define a unique mapping between the functions in noncommutative space and the functions in commutative space. In such references, a general ansatz for noncommutative coordinates satisfying the algebraic expression \eqref{commutation relation} is written as follows:
\begin{eqnarray}\label{solution0}
\hat{x}_{i}=x_{i}\varphi(A),\;\;\;\;
\hat{x}_{0}=x_{0}\psi(A)+\frac{i}{\kappa} x_i \partial_i \gamma(A),
\end{eqnarray}
where $\varphi$, $\psi$, and $\gamma$ are functions of $A=-\frac{i}{\kappa} \partial_0$.
By inserting the general ansatz \eqref{solution0} into \eqref{commutation relation}, we derive the following expression:
\begin{eqnarray}\label{solution1}
\gamma=1+\frac{\varphi'}{\varphi}\psi,
\end{eqnarray}
where $\varphi'$ is the derivative of $\varphi$ with respect to $A$. Note that, here, $\varphi$, $\psi$, and $\gamma$ are positive functions with the following boundary conditions:
\begin{eqnarray}
\varphi(0)=1,\;\;\;\; \psi(0)=1,
\end{eqnarray}
and $\gamma(0)=1+\varphi'(0)$ has to be finite. Notably, in the previous equations, $\varphi$ characterizes arbitrary realizations of the noncommutative coordinates in terms of the commutative coordinates and their derivatives.

Furthermore, let $M_{\mu\nu}$ denote the generators obeying the ordinary undeformed ${\textit so}(n-1,n)$ algebraic expression:
\begin{eqnarray}\label{generators}
&&[M_{\mu\nu},M_{\lambda\rho}]=\eta_{\nu\lambda}M_{\mu\rho}-\eta_{\mu\lambda}M_{\nu\rho}
-\eta_{\nu\rho}M_{\mu\lambda}+\eta_{\mu\rho}M_{\nu\lambda},\nonumber\\
&&M_{\mu\nu}=-M_{\nu\mu},\;\;\;\;\eta_{\mu\nu}=\textrm{diag}(-1,1,1,1).
\end{eqnarray}
The commutator $[M_{\mu\nu},\hat{x}_\lambda]$ between the generator $M_{\mu\nu}$ and the noncommutative coordinate $\hat{x}_\lambda$ needs to be antisymmetric with respect to the indices $\mu$ and $\nu$ and linear functions of $\hat{x}_\lambda$ and $M_{\mu\nu}$. Note that, as $\kappa\rightarrow\infty$, we have a smooth commutative limit. Therefore, in this regard, two classes of possible realizations, expressed as $\psi=1$ and $\psi=1+2A$, emerge. We will focus on $\psi=1$, and the explicit corresponding form of $M_{\mu\nu}$ is expressed as follows:
\begin{eqnarray}
M_{i0}&=&x_i\partial_0 \varphi\frac{e^{2A}-1}{2A}-x_0\partial_i\frac{1}{\varphi}
+\frac{i}{\kappa}x_i\Delta\frac{1}{2\varphi}-
\frac{i}{\kappa}x_k\partial_k\partial_i\frac{\gamma}{\varphi},\nonumber\\
M_{ij}&=&x_i\partial_j-x_j\partial_i,
\end{eqnarray}
where $\Delta=\partial_k\partial_k$. In Refs.~\cite{Meljanac2006,Kresic-Juric2007,Meljanac2008epjc,Govindarajan2009,Meljanac2008jpa,Govindarajan2008,Meljanac2009}, the Dirac derivative  $D_\mu$ and invariant Laplace operator $\Box$ have been introduced. The generalized Klein-Gordon equation invariant under the $\kappa$-Poincar\'{e} algebra can be obtained using the following relations:
\begin{eqnarray}
&&[M_{\mu\nu},D_\lambda]=\eta_{\nu\lambda}D_\mu-\eta_{\mu\lambda}D_{\nu},\;\;\;[D_\mu,D_\nu]=0,\nonumber\\
&&\;\;\;\;\;\;[M_{\mu\nu},\Box]=0,\;\;\;\;\;[\Box,\hat{x}_\mu]=2D_\mu,
\end{eqnarray}
with
\begin{eqnarray}\label{D and M}
&&D_{i}=\partial_i \frac{e^{-A}}{\varphi},\;\;\;D_{0}=\partial_0 \frac{\sinh A}{A}+\frac{i}{\kappa}\Delta\bigg(\frac{e^{-A}}{2\varphi^2}\bigg),\nonumber\\
&&\Box =\Delta\bigg(\frac{e^{-A}}{\varphi^2}\bigg)+2\partial_0^2 \frac{1-\cosh A}{A^2}.
\end{eqnarray}
Note that $D_\mu$ and $M_{\mu\nu}$ generate the $\kappa$-Poincar\'{e} algebra, whose relations are the same as that of the usual Poincar\'{e} algebra. However, the explicit forms of these generators are modified, and those modifications are dependent on the deformation parameter.

Based on Eq.~(\ref{D and M}), one can find that the Casimir element of the algebraic expression $D_{\mu}D_{\mu}$ can be expressed in terms of the $\Box$ operator as follows:
\begin{eqnarray}\label{Casimir}
D_{\mu}D_{\mu}=\Box \bigg(1-\frac{1}{4\kappa^2}\Box \bigg).
\end{eqnarray}
When $\kappa\rightarrow\infty$, we derive $D_{\mu}D_{\mu}\rightarrow\partial_\mu\partial_\mu$, which is reduced to the usual relativistic dispersion relation. We refer the reader to the works reported in Refs.~\cite{Harikumar2019,Harikumar2017}, which generalize the core-envelope model of a superdense star to a noncommutative spacetime. The positivity condition for tangential pressure and the expression for central density are given with a bound on the deformation parameter $\kappa>10^{16}\;m^{-1}$. Moreover, the authors analyzed the hydrogen atom model in Ref.~\cite{Harikumar2011} and compared the shift in the spectra of atoms with the experimental data. Then a most optimistic bound on the spacetime deformation parameter $\kappa$, i.e., $\kappa>10^{19} \;m^{-1}$, is derived from the $\kappa$-deformed Dirac equation.
As reported in Refs.~\cite{Harikumar2012,Liu2018,Xu2023,Harikumar2011}, we generalize the notions from commutative space, and the generalized Klein-Gordon equation is written using the Casimir element, which is invariant under the $\kappa$-Poincar\'{e} algebra, as follows:
\begin{eqnarray}
\bigg[\Box \bigg(1-\frac{1}{4\kappa^2}\Box \bigg)-m^2\bigg]\phi(\mathbf{x})=0\;.\label{KGequation}
\end{eqnarray}
As a result of the realizations of the noncommutative coordinates in terms of the commutative coordinates and the corresponding derivatives, both the generators and Casimir element of the $\kappa$-Poincar\'{e} algebra can be expressed in terms of the commutative coordinates and the corresponding derivatives. The scalar field and operators appearing in the $\kappa$-deformed Klein-Gordon equation~(\ref{KGequation}) are well defined in the commutative spacetime. Therefore, we can use the standard tools of field theory defined in the commutative spacetime to analyze the $\kappa$-deformed Klein-Gordon theory. The deformed dispersion relation based on Eq.~(\ref{KGequation}) is expressed as follows:
\begin{eqnarray}\label{deformed dispersion relation}
&&4\kappa^2\sinh^2\bigg(\frac{p_0}{2\kappa}\bigg)-p_i^2\frac{e^{-\frac{p_0}{\kappa}}}{\varphi^2(\frac{p_0}{\kappa})}
-m^2\nonumber\\
&&\;\;\;\;\;\;\;\;\;\;\;\;\;\;\;\;\;\;\;\;
=\frac{1}{4\kappa^2}\bigg[4\kappa^2\sinh^2\bigg(\frac{p_0}{2\kappa}\bigg)-p_i^2\frac{e^{-\frac{p_0}{\kappa}}}{\varphi^2
(\frac{p_0}{\kappa})}\bigg]^2,
\end{eqnarray}
where $p_0=i\partial_0$ and $p_i=-i\partial_i$. We observe from~(\ref{deformed dispersion relation}) that the characteristics of the field, i.e., nonlocal and noncausal, emerge. The Hamiltonian for this field is difficult to express in a compact form. Therefore, to obtain the two-point correlation function in $\kappa$-deformed spacetime for simplicity, we chose $\varphi(\frac{p_0}{\kappa})=e^{-\frac{p_0}{2\kappa}}$, as reported in Refs.~\cite{Harikumar2012,Liu2018}. Moreover, from here onwards, we only retain terms up to the second order in $1/\kappa$ because $\kappa$ is expected to be large in theory. Through the specific calculation, we determine that the two-point correlation function can be expressed as follows:
\begin{eqnarray}\label{deformed correlation function}
&&G^+(\mathbf{x},\mathbf{x}')=\frac{1}{4\pi^2}\frac{1}{(\mathbf{x}-\mathbf{x}')^2-(t-t')^2}\nonumber\\
&&\;\;\;\;\;\;\;\;\;\;\;\;\;\;\;\;\;\;\;-\frac{1}{16\pi^2\kappa^2}
\frac{(\mathbf{x}-\mathbf{x}')^2+3(t-t')^2}{[(\mathbf{x}-\mathbf{x}')^2-(t-t')^2]^3}
\nonumber\\
&&\;\;\;\;\;\;\;\;\;\;\;\;\;\;\;\;\;\;\;-\frac{1}{4\pi^2\kappa^2}
\frac{[(\mathbf{x}-\mathbf{x}')^2+(t-t')^2](t-t')^2}{[(\mathbf{x}-\mathbf{x}')^2-(t-t')^2]^4}.
\end{eqnarray}
Note that, for $\kappa\rightarrow\infty$, the two-point correlation function in Eq.~(\ref{deformed correlation function}) recovers to the Minkowski spacetime case~\cite{Birrell1982}, as expected, which is written as follows:
\begin{eqnarray}\label{correlation function}
G^+(\mathbf{x},\mathbf{x}')=\frac{1}{4\pi^2}\frac{1}{(\mathbf{x}-\mathbf{x}')^2-(t-t')^2}.
\end{eqnarray}
In the subsequent section, these two correlation functions will be used to explore the two-atom system entanglement dynamics.

\section{Entanglement dynamics of two atoms without the environment-induced interatomic interaction}\label{section3}
Now, let us consider the entanglement dynamics of a two-atom system interacting with an external environment. Three different initial state cases will be considered: (1) the separable state $|E\rangle$, (2) the symmetric entangled state $|S\rangle$, and (3) the antisymmetric entangled state $|A\rangle$. Note that all of these chosen initial states will not result in the environment-induced interatomic interaction. We focus on how the relativistic motion affects the entanglement dynamics. Specifically, we will analyze the entanglement dynamics of static atoms, inertial atoms moving with a constant velocity, and circularly accelerated atoms, which are coupled to a massless scalar field in $\kappa$-deformed and Minkowski spacetimes. In particular, we will comparatively investigate the phenomena of entanglement generation and degradation in these two universes.

\subsection{Entanglement dynamics of two static atoms}
We first consider the entanglement dynamics of two static atoms, which are separated by a distance $L$ along the following trajectories:
\begin{eqnarray}\label{trajectories0}
&&t_1(\tau)=\tau,\;\;\;x_1(\tau)=0,\;\;\;y_1(\tau)=0,\;\;\;z_1(\tau)=0,\nonumber\\
&&t_2(\tau)=\tau,\;\;\;x_2(\tau)=0,\;\;\;y_2(\tau)=0,\;\;\;z_2(\tau)=L.
\end{eqnarray}
Substituting the aforementioned trajectories into the two-point correlation function in $\kappa$-deformed spacetime~(\ref{deformed correlation function}), we obtain the following expression:
\begin{eqnarray}\label{rest deformed0}
G^{11}(x,x')&=&G^{22}(x,x')
=-\frac{1}{4\pi^2}\frac{1}{\triangle\tau^2}-\frac{1}{16\pi^2\kappa^2}
\frac{1}{\triangle\tau^4},\nonumber\\
G^{12}(x,x')&=&G^{21}(x,x')
=-\frac{1}{4\pi^2}\bigg[\frac{1}{\triangle\tau^2-L^2}
-\frac{1}{4\kappa^2}\frac{3\triangle\tau^2+L^2}{(\triangle\tau^2-L^2)^3}\nonumber\\
&&\;\;\;\;\;\;\;\;\;\;\;\;\;\;\;\;\;\;\;\;\;\;\;\;\;\;\;\;\;\;\;\;\;\;
+\frac{1}{\kappa^2}
\frac{\triangle\tau^4+L^2\triangle\tau^2}{(\triangle\tau^2-L^2)^4}\bigg].
\end{eqnarray}
By invoking the residue theorem method, the corresponding Fourier transforms in (\ref{Fourier transform}) to the field correlation functions (\ref{rest deformed0}) can be rewritten as follows:
\begin{eqnarray}\label{rest deformed Fourier0}
\mathcal{G}^{11}(\lambda)&=&\mathcal{G}^{22}(\lambda)
=\frac{\lambda}{2\pi}\bigg[1-\frac{7\lambda^2}{96\kappa^2}
\bigg]\theta(\lambda),\nonumber\\
\mathcal{G}^{12}(\lambda)&=&\mathcal{G}^{21}(\lambda)
=\frac{\lambda}{2\pi}\bigg[\frac{\sin \lambda L}{\lambda L}-\frac{\lambda^2 \cos \lambda L}{24\kappa^2}\bigg]\theta(\lambda),
\end{eqnarray}
where $\theta(x)$ is the step function. By substituting the Fourier transform (\ref{rest deformed Fourier0}) into Eq.~(\ref{AB}), we derive the following expression:
\begin{eqnarray}\label{coefficient krest}
&&A_1=A_2=B_1=B_2=\frac{\Gamma_0}{4}\bigg[1-\frac{\omega^2}{24\kappa^2}\bigg],\nonumber\\
&&A_3=A_4=B_3=B_4=\frac{\Gamma_0}{4}\bigg[\frac{\sin \omega L}{\omega L}-\frac{\omega^2 \cos \omega L }{24\kappa^2} \bigg],
\end{eqnarray}
with $\Gamma_0=\frac{\mu^2\omega}{2\pi}$ being the spontaneous emission rate of each individual atom.
Note that, for $\kappa\rightarrow\infty$, the function (\ref{rest deformed Fourier0}) in $\kappa$-deformed spacetime recover to that of two atoms at rest in Minkowski spacetime as expected. Thus, we can rewrite the relevant coefficients of Eq.~(\ref{coefficient matrix1}) for this case as follows:
\begin{eqnarray}\label{coefficient mrest}
&&A_1=A_2=B_1=B_2=\frac{\Gamma_0}{4},\nonumber\\
&&A_3=A_4=B_3=B_4=\frac{\Gamma_0}{4}\frac{\sin \omega L}{\omega L}.
\end{eqnarray}

By preparing the initial state, e.g., $|E\rangle$, $|S\rangle$, or $|A\rangle$, and inserting the relevant coefficients into Eq.~(\ref{state1}), we can solve the master equation correspondingly. Then, we find that the corresponding entanglement in Eq.~(\ref{concurrence0}) can be written as follows:
\begin{eqnarray}\label{concurrence0-1}
C[\rho(\tau)]=\textrm{max}\{0,K_1(\tau)\},
\end{eqnarray}
where
\begin{eqnarray}\label{concurrence0-2}
K_1(\tau)=\sqrt{[\rho_{AA}(\tau)-\rho_{SS}(\tau)]^2}
-2\sqrt{\rho_{GG}(\tau)\rho_{EE}(\tau)},
\end{eqnarray}
from which we determine that concurrence is independent of the environment-induced interatomic interaction. The explicit entanglement dynamics of various situations are investigated in the subsequent sections.

\subsubsection{Two static atoms initially prepared in a separable state $|E\rangle$}
We start with the entanglement dynamics of static atoms initially prepared in a separable state $|E\rangle$. From Eqs.~(\ref{concurrence0-1}) and (\ref{concurrence0-2}), we determine that entanglement can be generated only when the factor $\sqrt{[\rho_{AA}(\tau)-\rho_{SS}(\tau)]^2}$ outweighs the threshold factor $2\sqrt{\rho_{GG}(\tau)\rho_{EE}(\tau)}$. We note that such an entanglement generation phenomenon inevitably occurs when the system undergoes only spontaneous emission evolution for a finite time. This phenomenon is called the delayed sudden birth of entanglement~\cite{Ficek2008}.

\begin{figure}[H]
\centering
{\includegraphics[height=1.65in,width=2.65in]{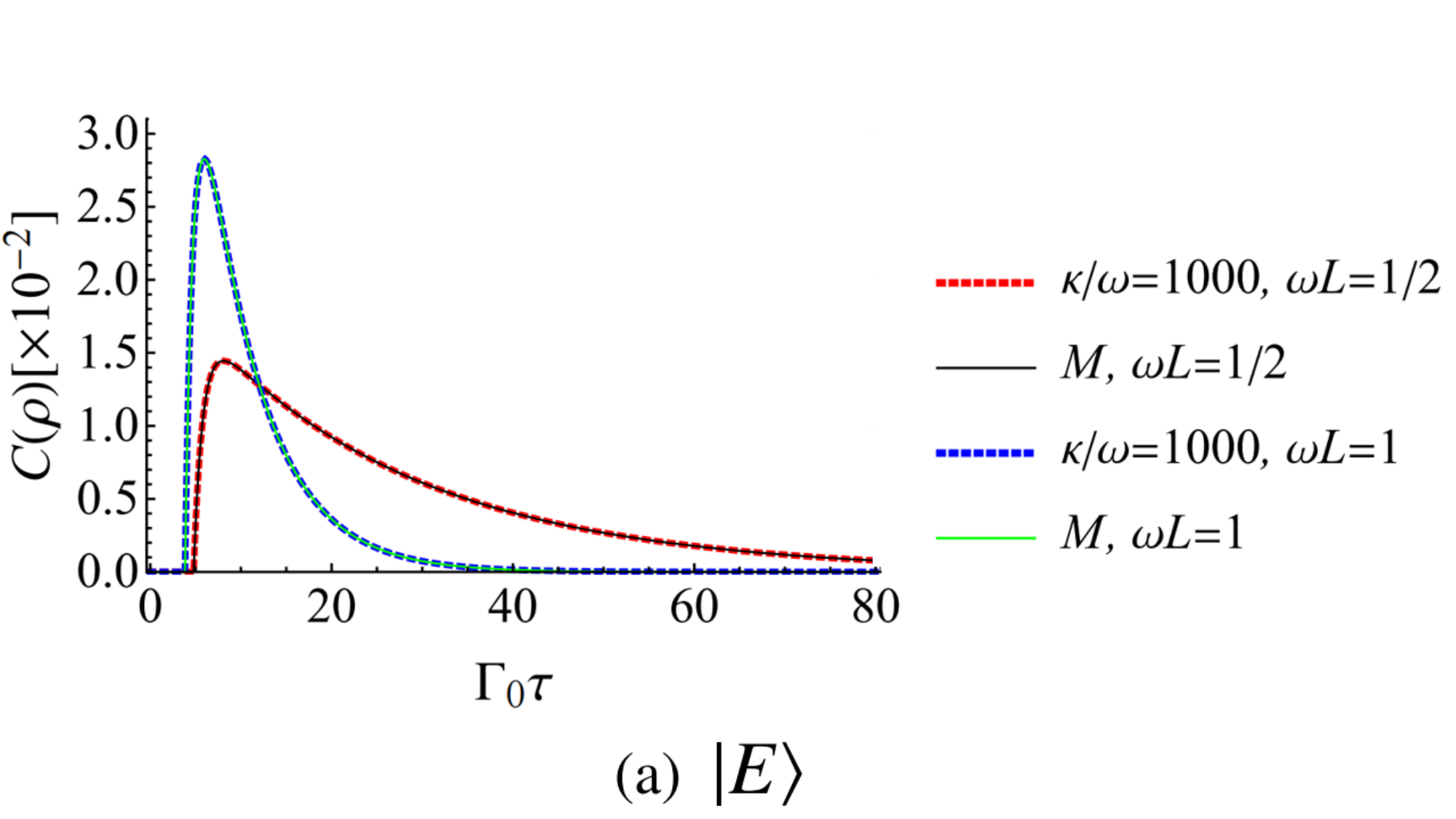}}
{\includegraphics[height=1.65in,width=2.65in]{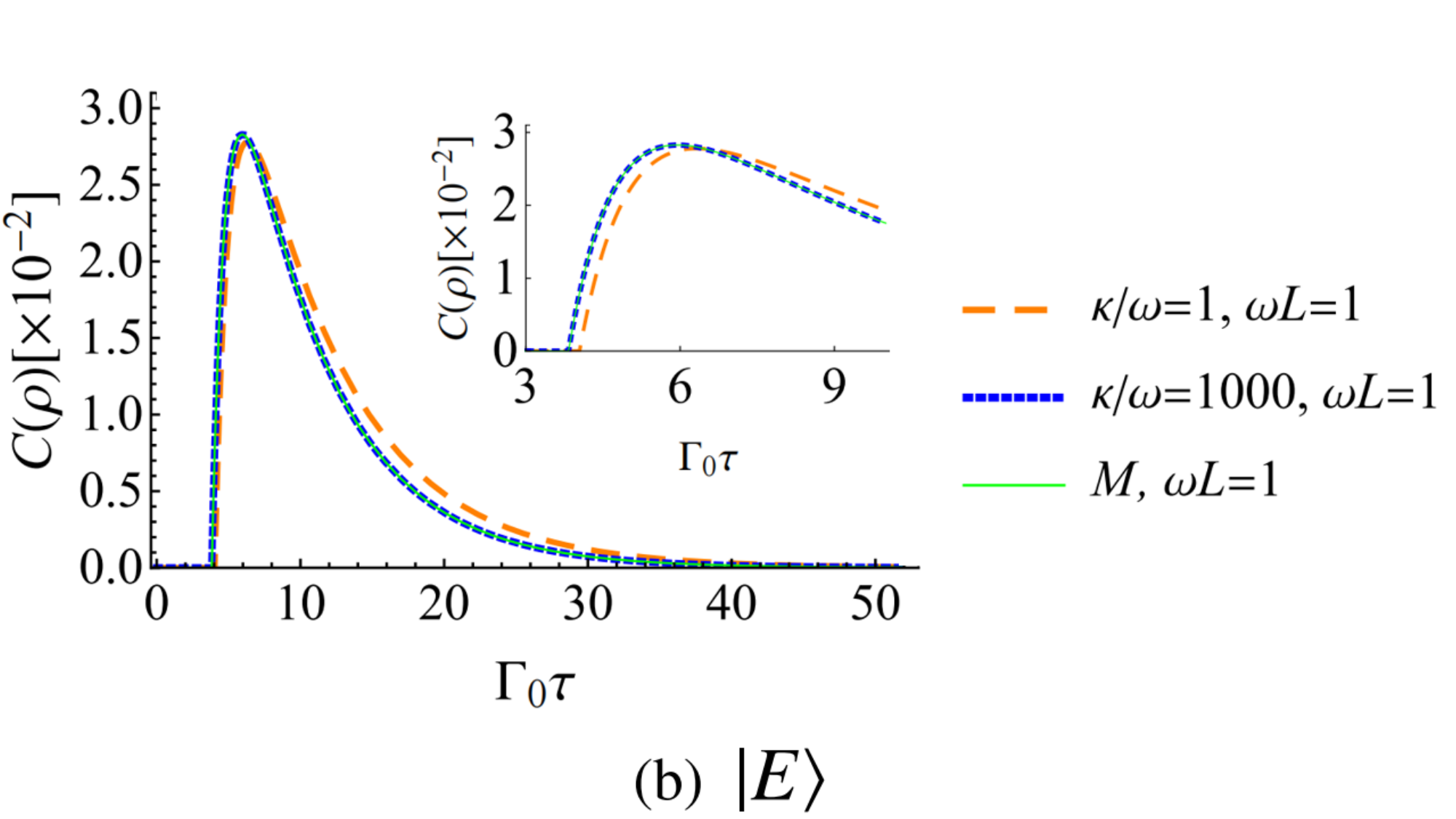}}
{\includegraphics[height=1.65in,width=2.65in]{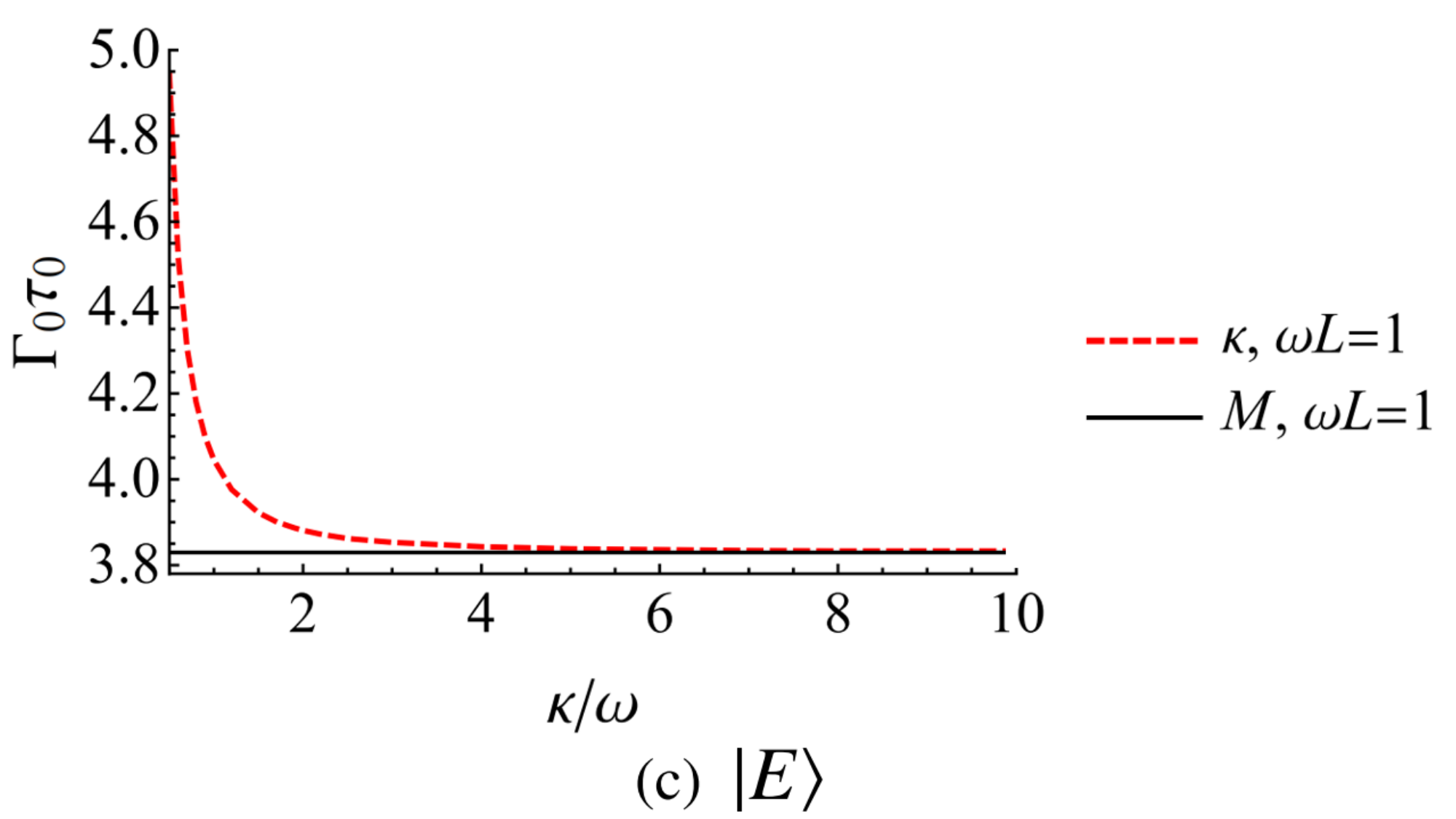}}
\caption{Time evolution of concurrence of different values of $\omega L=1/2,\;1$ (a) and $\kappa/\omega=1,\;1000$ (b) and the waiting time to generate entanglement $\Gamma_0\tau_0$ as function of the deformation parameter $\kappa/\omega$ (c) for two static atoms initially prepared in $|E\rangle$. Here, \emph{$\kappa$} denotes the $\kappa$-deformed spacetime and \emph{M} denotes the Minkowski spacetime.
}\label{R3}
\end{figure}
\begin{figure}[H]
\centering
{\includegraphics[height=1.65in,width=2.65in]{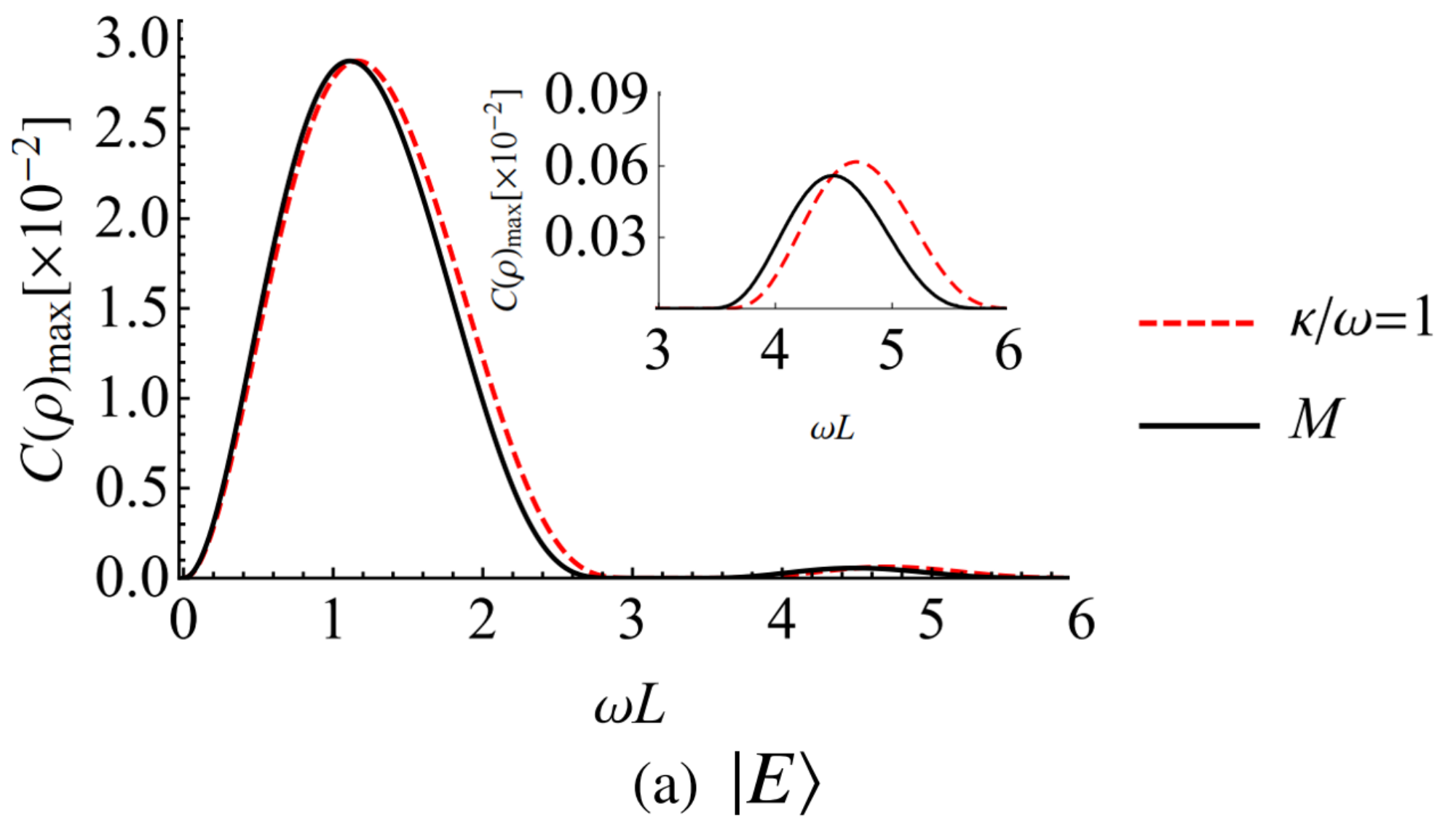}}
{\includegraphics[height=1.65in,width=2.65in]{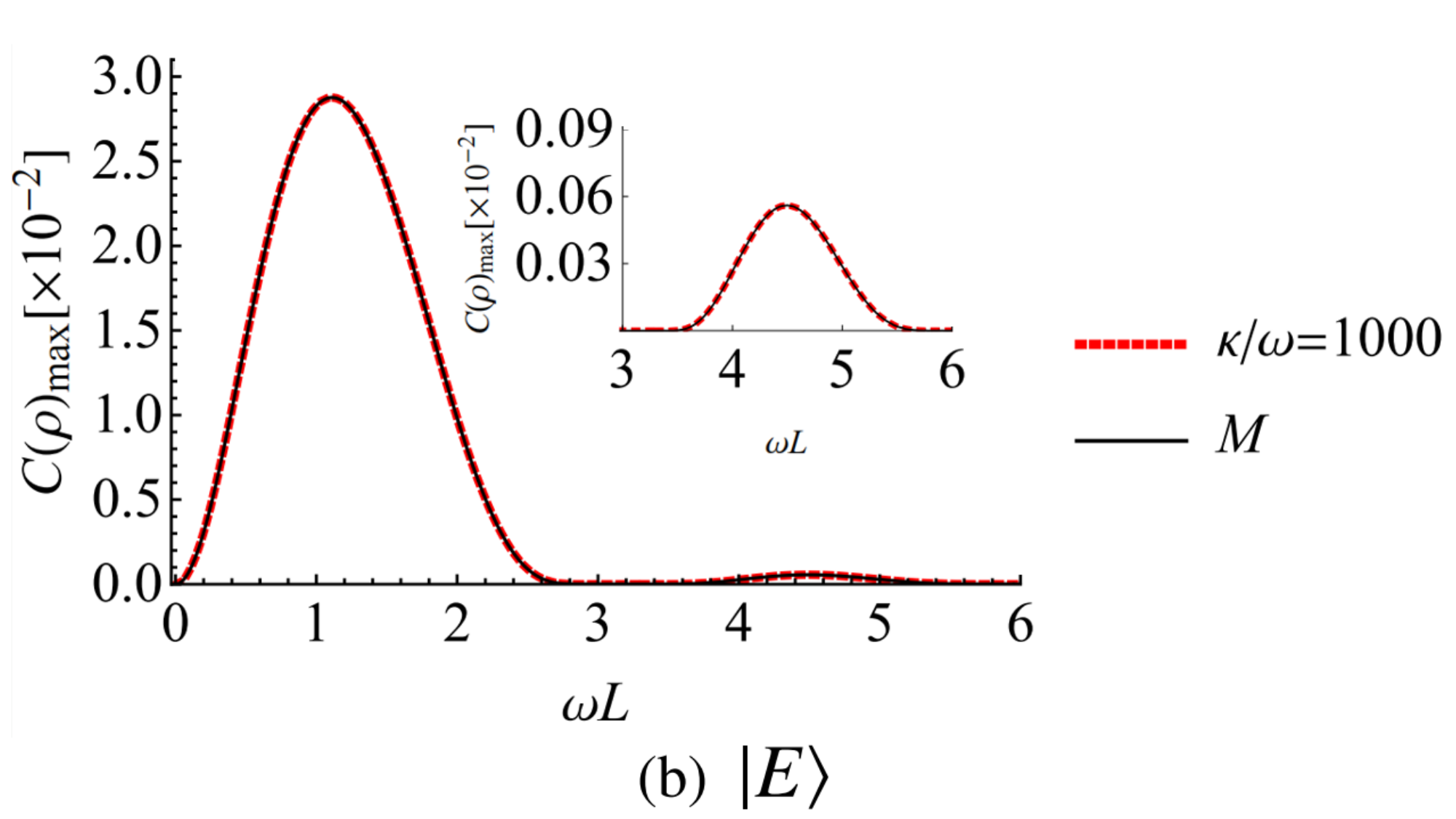}}
\caption{Maximum of concurrence during the evolution of two static atoms via the interatomic separation initially prepared in $|E\rangle$ with fixed $\kappa/\omega=1$ (a) and $\kappa/\omega=1000$ (b).
}\label{R4}
\end{figure}

Let us consider the case where the interatomic separation vanishes $(L\rightarrow0)$. Then, $A_i=B_i=\frac{\Gamma_0}{4}[1-\frac{\omega^2}{24\kappa^2}]$ in Eq.~(\ref{coefficient krest}) and $A_i=B_i=\frac{\Gamma_0}{4}$ in Eq.~(\ref{coefficient mrest}) with $i\in\{1,2,3,4\}$. Therefore, $\rho_{AA}(\tau)$ remains zero during evolution in both $\kappa$-deformed and Minkowski spacetimes. In such a case, the threshold always outweighs the population $\rho_{SS}(\tau)$, and no quantum entanglement can be generated in these two universes. For the interatomic separation comparable to the transition wavelength $(L\sim\omega^{-1})$, we solve Eq.~(\ref{state1}) numerically. We show the corresponding results in Fig.~\ref{R3}. Here, we fixed $\omega L=1/2$ and $\omega L=1$ because the ranges of the maximum of concurrence during evolution are mainly located in $\omega L\in[0,3]$ and $\omega L\in[4,6]$, whereas in other ranges, concurrence is either too small or zero, as shown in Fig.~\ref{R4}. We find that, in contrast to the vanishing interatomic separation case, the delayed sudden birth of entanglement occurs for both $\kappa$-deformed and Minkowski spacetime cases. Specifically, one can note that the time when entanglement generation begins is related to the interatomic separation. Moreover, the deformation of spacetime may delay entanglement generation [Figs.~\ref{R3} (b) and (c)]. Furthermore, the amplitude of the generated entanglement is influenced by the interatomic separation and spacetime deformation, which we will subsequently discuss in detail. When the value of the spacetime deformation parameter is large,  regardless of how long the atomic separation is, the dotted line substantially coincides with the solid line, i.e., the entanglement generation for two static atoms during evolution in $\kappa$-deformed spacetime is nearly identical to that in Minkowski spacetime.

As shown in Fig.~\ref{R4}, we analyze the effects of atomic separation on the maximum entanglement generated during evolution. Notably, the maximum entanglement is a periodic function of interatomic separation, and the amplitude decays with the increase in interatomic separation in both $\kappa$-deformed and Minkowski spacetimes. Remarkably, when the modification of spacetime is relatively strong, e.g., $\kappa/\omega=1$, we find that the entanglement behavior of two atoms via the interatomic distance in $\kappa$-deformed spacetime is different from that in Minkowski spacetime [Fig.~\ref{R4} (a)]. However, when the modification of spacetime is relatively weak, e.g., $\kappa/\omega=1000$, we find that the entanglement in the $\kappa$-deformed and the Minkowski spacetime cases behaves nearly the same [Fig.~\ref{R4} (b)]. In this regard, we infer that, for the larger spacetime deformation parameter (LDP) case, all of the laws of physics in $\kappa$-deformed spacetime substantially recover to that in flat spacetime. Thus, in this case, distinguishing these two spacetimes is difficult.

\subsubsection{Two static atoms initially prepared in entangled states $|A\rangle$ and $|S\rangle$}
Here, we investigate the entanglement degradation of two static atoms initially prepared in the antisymmetric entangled state $|A\rangle$ and the symmetric entangled state $|S\rangle$, both of which are maximally entangled.

\begin{figure}[H]
\centering
{\includegraphics[height=1.65in,width=2.65in]{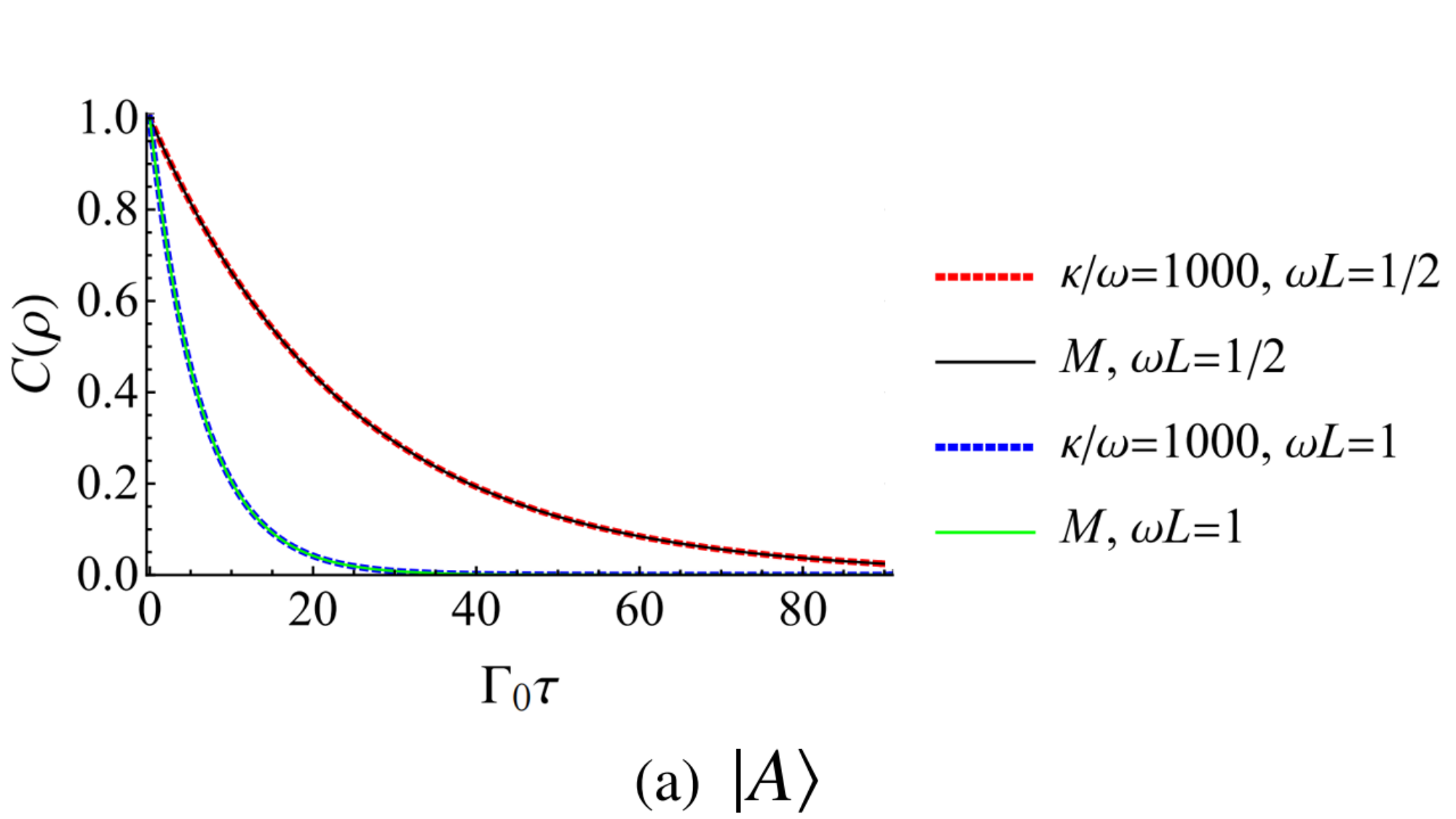}}
{\includegraphics[height=1.65in,width=2.65in]{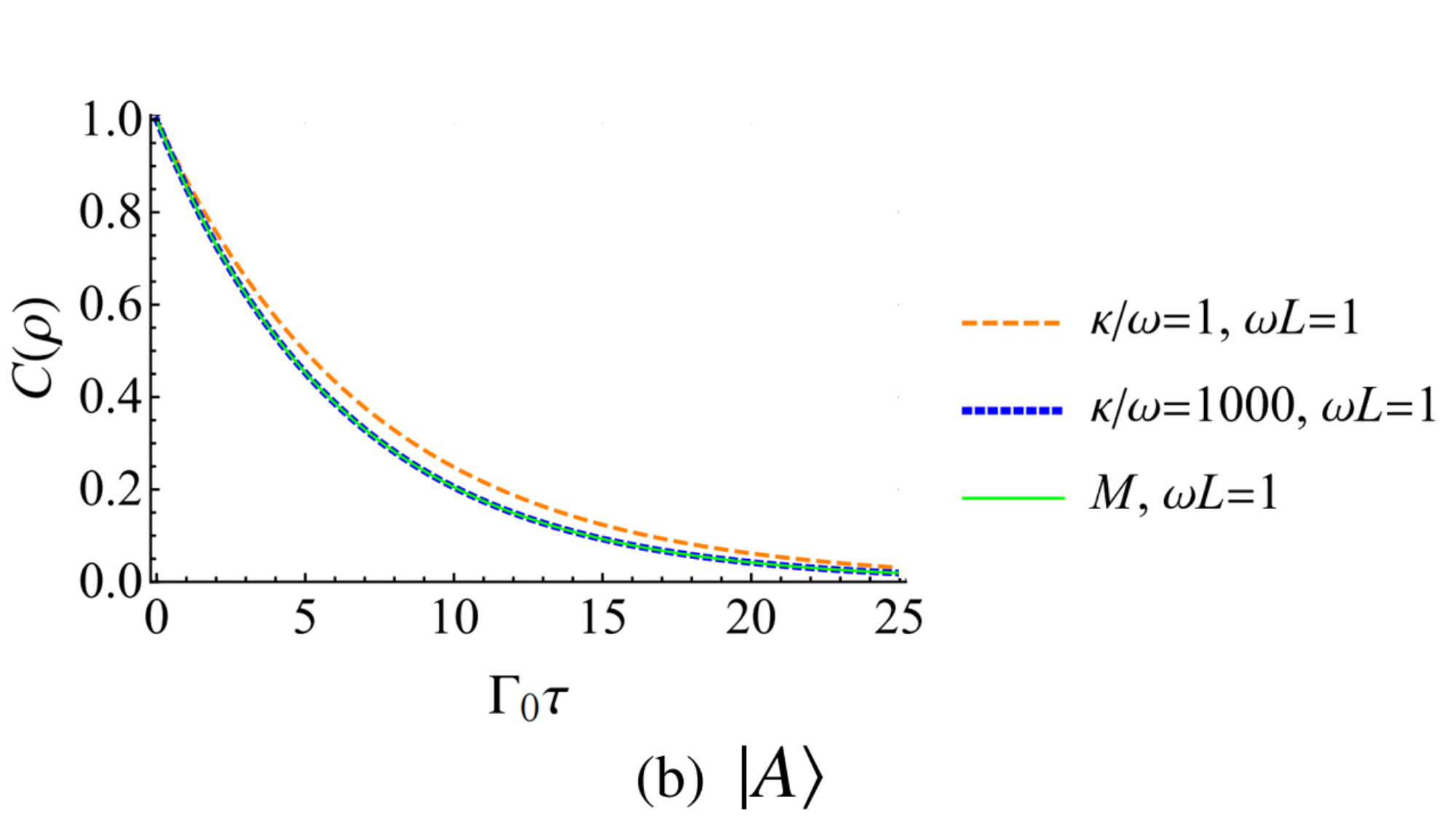}}
{\includegraphics[height=1.65in,width=2.65in]{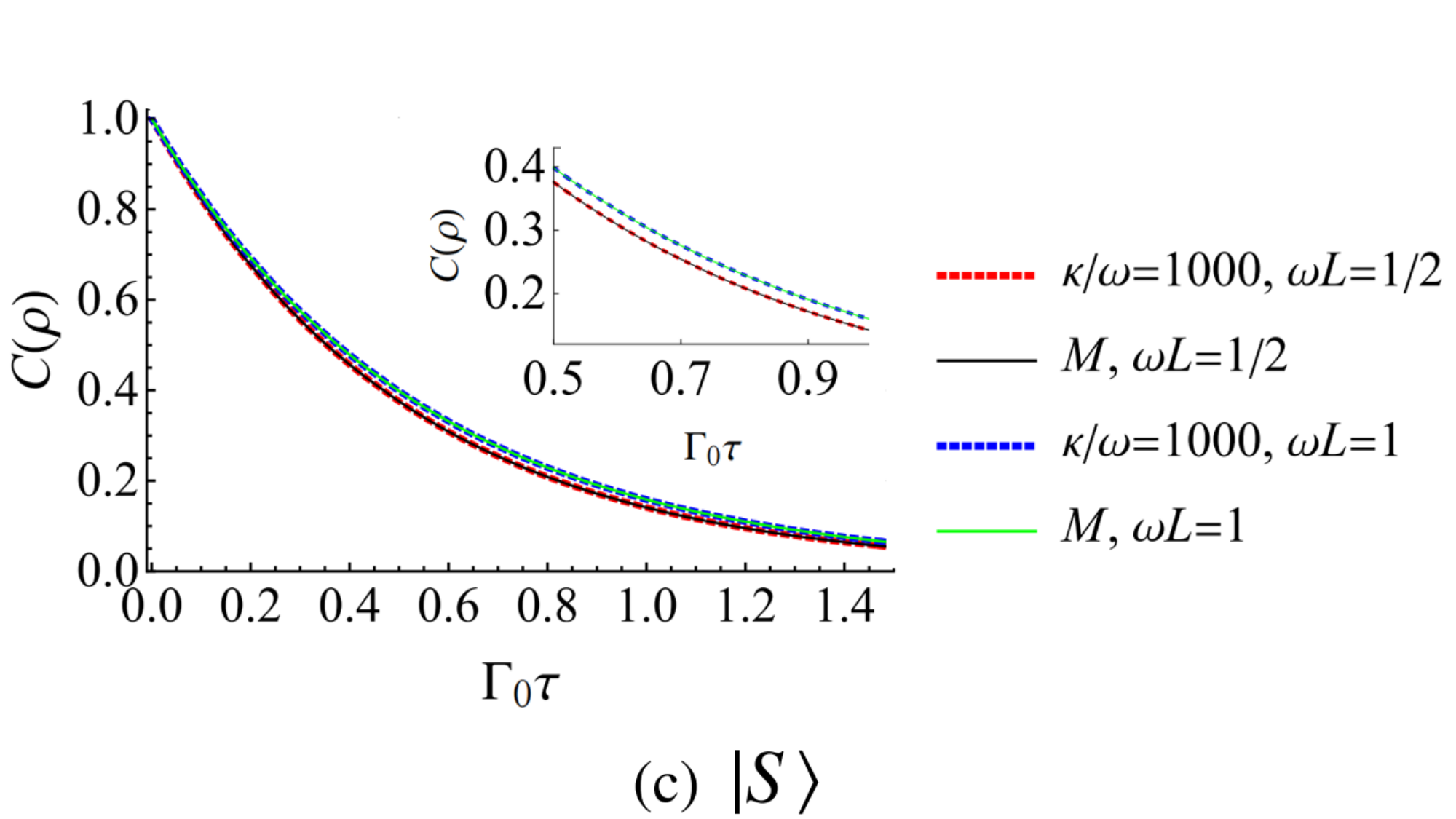}}
{\includegraphics[height=1.65in,width=2.65in]{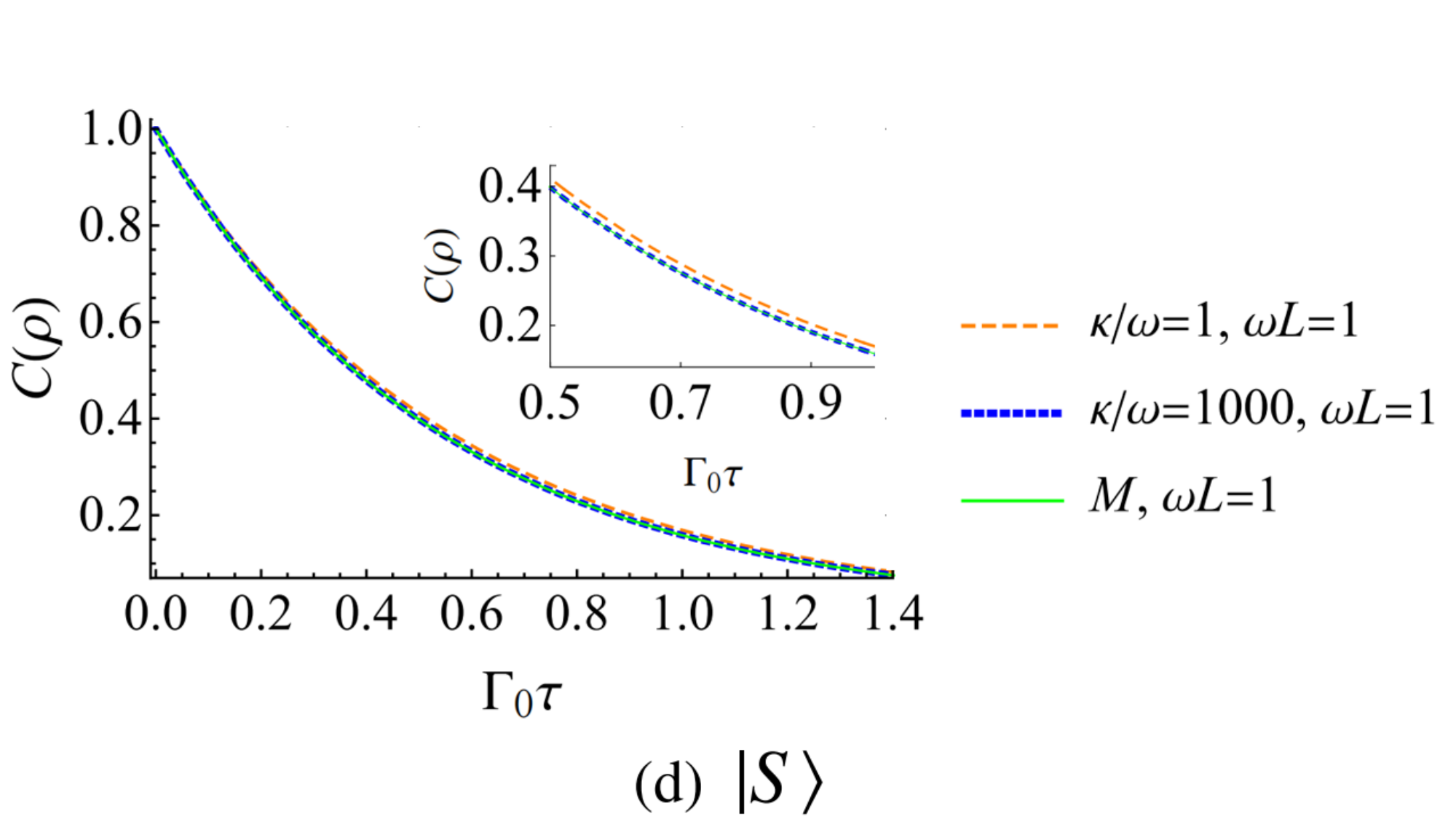}}
{\includegraphics[height=1.65in,width=2.65in]{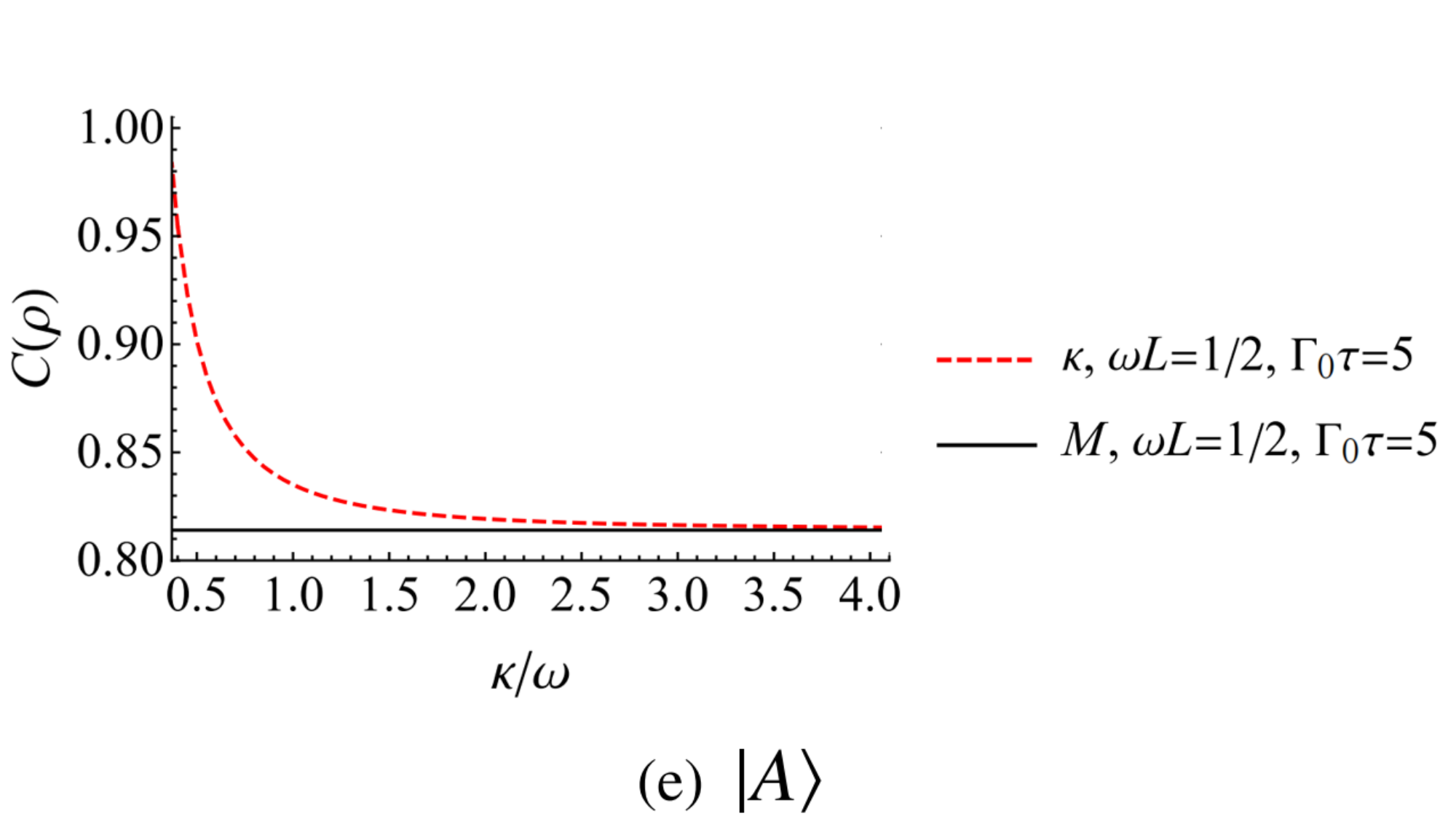}}
{\includegraphics[height=1.65in,width=2.65in]{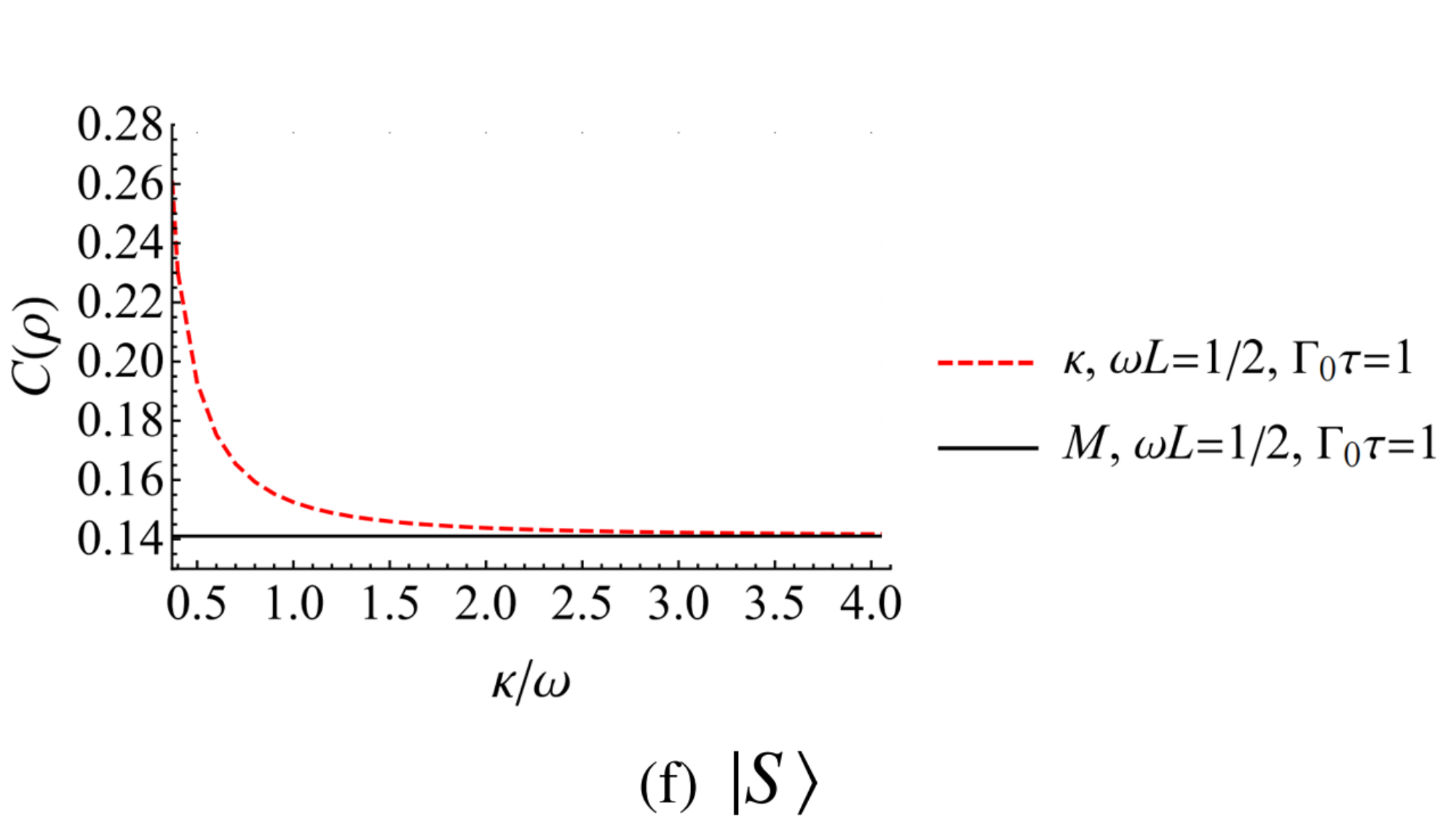}}
\caption{Time evolution of concurrence of static atoms initially prepared in $|A\rangle$ for different values of $\omega L=1/2,\;1$ (a) and $\kappa/\omega=1,\;1000$ (b) and $|S\rangle$ for different values of $\omega L=1/2,\;1$ (c) and $\kappa/\omega=1,\;1000$ (d). Concurrence is considered a function of deformation parameter $\kappa/\omega$ of static atoms initially prepared in $|A\rangle$ (e) and $|S\rangle$ (f).
}\label{R1-2}
\end{figure}

From Fig.~\ref{R1-2}, we find that concurrence decreases monotonically with time until it reaches zero in the infinite time limit for both of these universes. We also find that the response of entanglement magnitude to the interatomic separation behaves differently for different initial entangled states (i.e., antisymmetric and symmetric entangled states). Specifically, the entanglement magnitude of the initial antisymmetric entangled state case (initial symmetric entangled state) decreases (increases) with the increase in the interatomic separation [Figs.~\ref{R1-2} (a) and (c)]. In particular, for a fixed evolution time, concurrence decreases to an asymptotic value as the spacetime deformation parameter $\kappa$ increases [Figs.~\ref{R1-2} (e) and (f)]. In other words, when the deformation parameter $\kappa$ is large, the entanglement dynamics curves of two atoms in $\kappa$-deformed spacetime nearly coincide with that in Minkowski spacetime [also see Figs.~\ref{R1-2} (a), (b), (c), and (d)].

We provide a summary here. For the case of two static atoms initially prepared in the three specific states, when the spacetime deformation parameter $\kappa$ is large, the entanglement dynamics in $\kappa$-deformed spacetime are nearly indistinguishable from that in Minkowski spacetime regardless of the choice of interatomic separation. This finding indicates that we cannot distinguish these two spacetimes using the entanglement dynamics of static atoms with these initial states when the deformation parameter is large. However, the theoretical hypothesis states that the value of the deformation parameter involved in $\kappa$-deformed spacetime should be large~\cite{Xu2023,Harikumar2019,Harikumar2017,Harikumar2011,Amelino1998,Amelino2000,Amelino2001}; thus, the $\kappa$-deformed spacetime generally exhibits nearly the same properties as the Minkowski spacetime and obeys the Poincar\'{e} algebra. Hence, an issue arises: Can we identify some external auxiliary conditions, such as relativistic motion, to distinguish these two spacetimes using entanglement dynamics when the spacetime deformation parameter is large? This is what we will explore in the subsequent section.

\subsection{Entanglement dynamics of two uniformly moving atoms}
In this section, we will investigate the entanglement dynamics of two atoms moving with a constant velocity in $\kappa$-deformed and Minkowski spacetimes. We mainly focus on how velocity affects the entanglement behavior in these two different spacetimes.
The trajectories of the two inertial atoms, which are moving with a constant velocity and are separated from each other by a distance $L$, can be described as follows:
\begin{eqnarray}\label{trajectories1}
&&t_1(\tau)=\gamma\tau,\;\;\;x_1(\tau)=v\gamma\tau,\;\;\;y_1(\tau)=0,\;\;\;z_1(\tau)=0,\nonumber\\
&&t_2(\tau)=\gamma\tau,\;\;\;x_2(\tau)=v\gamma\tau,\;\;\;y_2(\tau)=0,\;\;\;z_2(\tau)=L,\nonumber\\
\end{eqnarray}
where $v$ is the velocity and $\gamma=1/\sqrt{1-v^2}$ is the usual Lorentz factor.

By substituting the trajectory~(\ref{trajectories1}) into Eq.~(\ref{deformed correlation function}), the two-point correlation functions in $\kappa$-deformed spacetime can be rewritten as follows:
\begin{eqnarray}\label{CF deformed0}
&&G^{11}(x,x')=G^{22}(x,x')=-\frac{1}{4\pi^2}\frac{1}{\triangle\tau^2}+\frac{1}{16\pi^2\kappa^2}
\frac{\gamma^2(3+v^2)}{\triangle\tau^4}\nonumber\\
&&\;\;\;\;\;\;\;\;\;\;\;\;\;\;\;\;\;\;\;\;\;\;\;\;\;\;\;\;\;\;\;\;\;\;\;\;\;\;\;\;\;\;\;\;
\;\;\;\;\;\;\;\;\;\;\;
-\frac{1}{4\pi^2\kappa^2}\frac{\gamma^4(v^2+1)}{\triangle\tau^4},\nonumber\\
&&\nonumber\\
&&G^{12}(x,x')=G^{21}(x,x')=-\frac{1}{4\pi^2}\frac{1}{\triangle\tau^2-L^2}+\frac{1}{4\pi^2\kappa^2}
\nonumber\\
&&\;\;\;\;\;\;\;\;\;\times\bigg\{\frac{\gamma^2(3+v^2)\triangle\tau^2+L^2}{4(\triangle\tau^2-L^2)^4}
-\frac{\gamma^4(v^2+1)\triangle\tau^4+L^2\gamma^2\triangle\tau^2}{(\triangle\tau^2-L^2)^4}
\bigg\}.\nonumber\\
\end{eqnarray}
Subsequently, the Fourier transforms of the aforementioned correlation functions can be calculated using the residue theorem method as follows:
\begin{eqnarray}\label{deformed Fourier0}
\mathcal{G}^{11}(\lambda)&=&\mathcal{G}^{22}(\lambda)=\frac{\lambda}{2\pi}\bigg[1-\frac{\lambda^2}{6\kappa^2}
\frac{1+v^2}{(1-v^2)^2}\nonumber\\
&&\;\;\;\;\;\;\;\;\;\;\;\;\;\;\;\;\;\;\;\;\;\;\;\;\;\;\;\;
+\frac{\lambda^2}{24\kappa^2}\frac{3+v^2}{1-v^2}\bigg]\theta(\lambda),\nonumber\\
\mathcal{G}^{12}(\lambda)&=&\mathcal{G}^{21}(\lambda)
=\frac{\lambda}{2\pi}\bigg[\frac{\sin \lambda L}{\lambda L}+\frac{ f(\lambda,L,v)}{24\lambda\kappa^2 L^3}\bigg]\theta(\lambda),
\end{eqnarray}
where $f(\lambda,L,v)=\frac{(3v^4\lambda L-\lambda^3L^3)\cos \lambda L -3v^2(v^2+2\lambda^2 L^2)\sin \lambda L }{(1-v^2)^2}$ and $\theta(\lambda)$ is the step function. By submitting the Fourier transforms (\ref{deformed Fourier0}) into Eq.~(\ref{AB}), we can derive the coefficients in Eq.~(\ref{coefficient matrix1}) for the $\kappa$-deformed spacetime case, as follows:
\begin{eqnarray}\label{uniformlyAB}
&&A_1=A_2=B_1=B_2=\frac{\Gamma_0}{4}\bigg[1-\frac{\omega^2}{6\kappa^2}
\frac{1+v^2}{(1-v^2)^2}\nonumber\\
&&\;\;\;\;\;\;\;\;\;\;\;\;\;\;\;\;\;\;\;\;\;\;\;\;\;\;\;\;\;\;\;\;\;\;\;\;\;
\;\;\;\;\;\;\;\;\;\;
+\frac{\omega^2}{24\kappa^2}\frac{3+v^2}{1-v^2}\bigg],\nonumber\\
&&A_3=A_4=B_3=B_4=\frac{\Gamma_0}{4}\bigg[\frac{\sin \omega L}{\omega L}+\frac{f(\omega,L,v)}{24\omega\kappa^2 L^3} \bigg].
\end{eqnarray}
Similarly, the coefficients in Eq.~(\ref{coefficient matrix1}) for two uniformly moving atoms in Minkowski spacetime can be derived as follows:
\begin{eqnarray}\label{uniformlyMAB}
&&A_1=A_2=B_1=B_2=\frac{\Gamma_0}{4},\nonumber\\
&&A_3=A_4=B_3=B_4=\frac{\Gamma_0}{4}\frac{\sin \omega L}{\omega L}.
\end{eqnarray}
Note that the coefficients in Eq.~(\ref{uniformlyMAB}) are similar to those of two static atoms in the Minkowski spacetime case (\ref{coefficient mrest}). That is, in Minkowski spacetime, all of the laws of physics for two uniformly moving atoms are the same for two static atoms.

From the aforementioned coefficients, we can infer that the dynamics of two uniformly moving atoms in $\kappa$-deformed spacetime are related to the velocity of atoms, whereas those in Minkowski spacetime are not. Therefore, we investigate how velocity affects the entanglement dynamics so that we can determine whether it is possible to distinguish these two spacetimes with the help of relativistic motion.

\subsubsection{Two uniformly moving atoms initially prepared in a separable state $|E\rangle$}
Let us start with the effects of velocity on the entanglement generation for two uniformly moving atoms initially prepared in a separable state $|E\rangle$. When the interatomic separation is small or nearly vanishes $(L\rightarrow 0)$, it can be directly inferred from Eqs.~(\ref{uniformlyAB}) and (\ref{uniformlyMAB}) that, in this limit, $A_i=B_i$, and thus, the evolution of the population $\rho_{AA}(\tau)$ remains zero. Therefore, the threshold factor $2\sqrt{\rho_{GG}(\tau)\rho_{EE}(\tau)}$ always outweighs the population $\rho_{SS}(\tau)$, indicating that, with the vanishing interatomic distance, there is no generated entanglement for the two atoms in both of these universes.

\begin{figure}[H]
\centering
{\includegraphics[height=1.65in,width=2.65in]{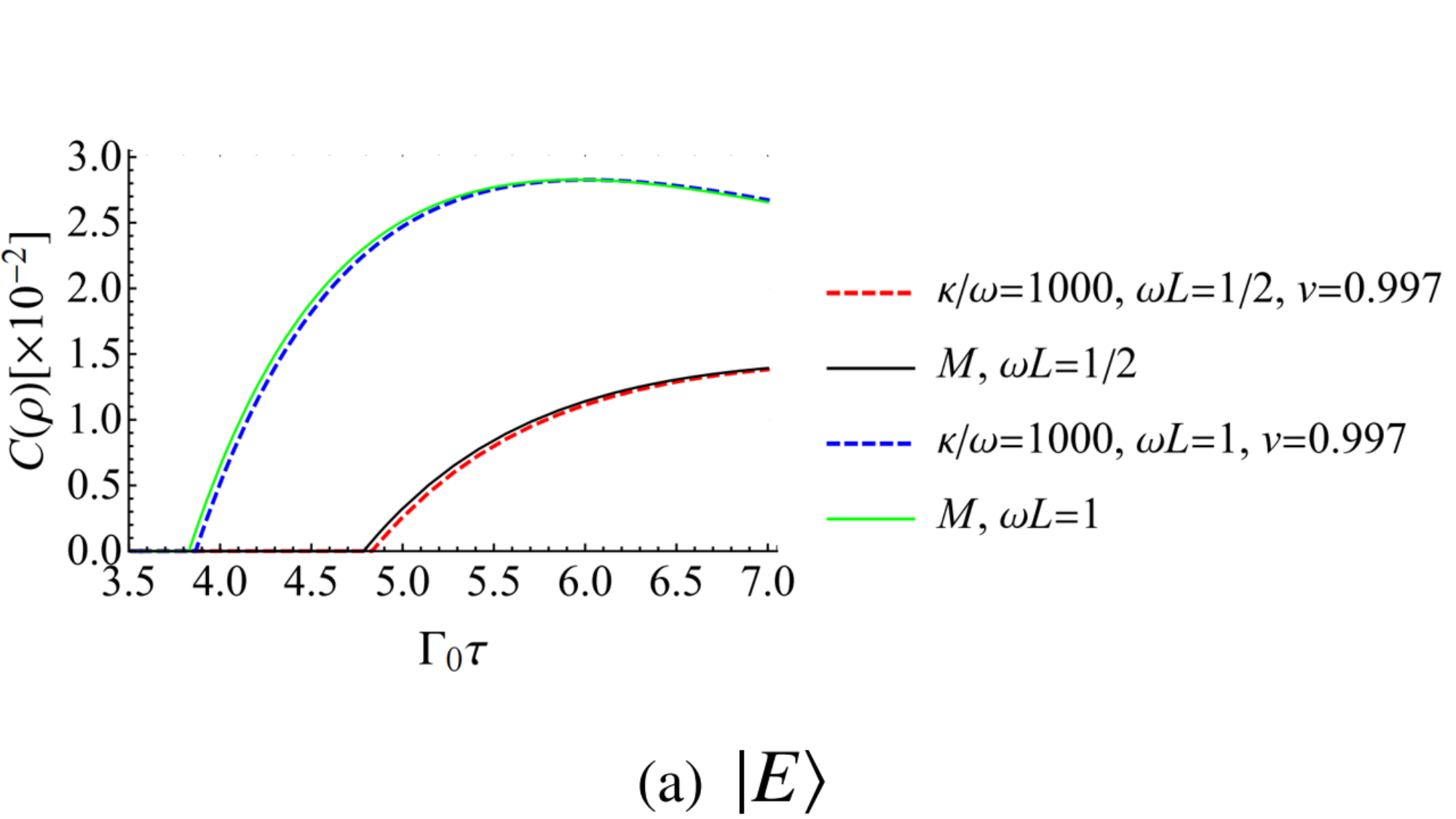}}
{\includegraphics[height=1.65in,width=2.65in]{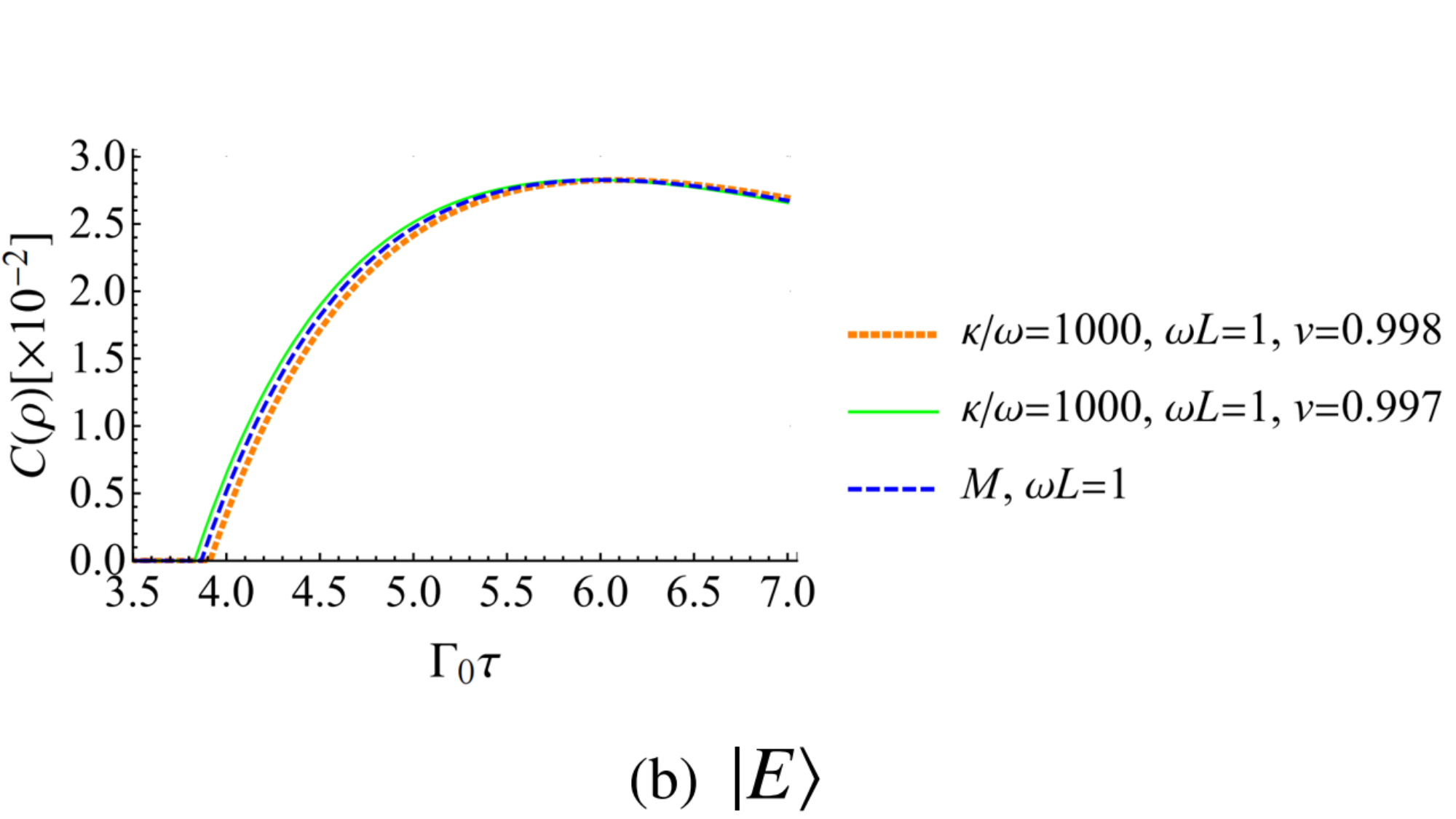}}
{\includegraphics[height=1.65in,width=2.65in]{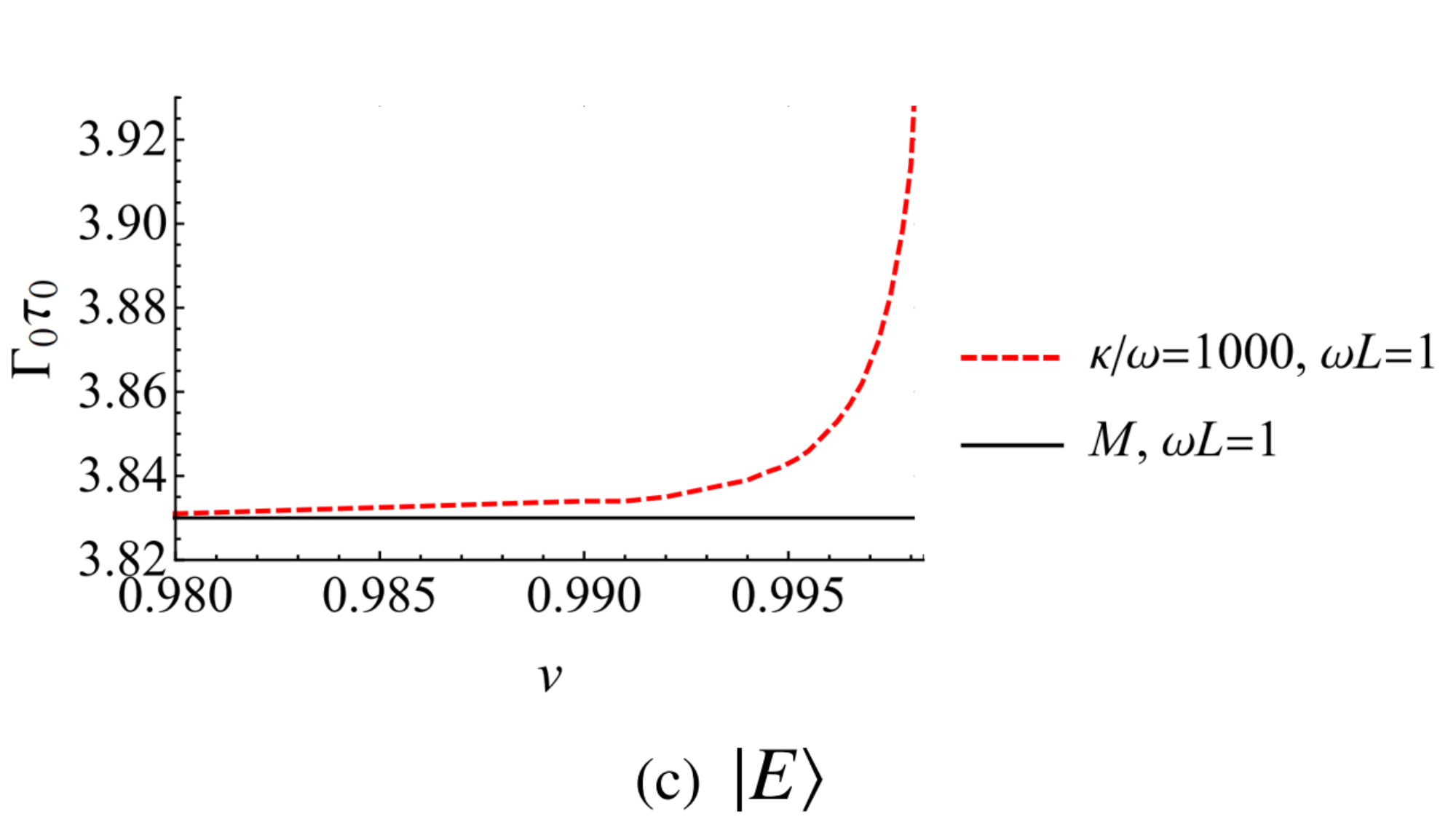}}
\caption{Time evolution of concurrence of two uniformly moving atoms initially prepared in $|E\rangle$ for different values of $\omega L=1/2,\;1$ with fixed $v=0.997$ (a) and $v=0.997,\;0.998$ with fixed $\omega L=1$ (b), and the waiting time to generate entanglement $\Gamma_0 \tau_0$ as a function of the velocity $v$ (c).
}\label{f2}
\end{figure}

\begin{figure}[H]
\centering
{\includegraphics[height=1.65in,width=2.65in]{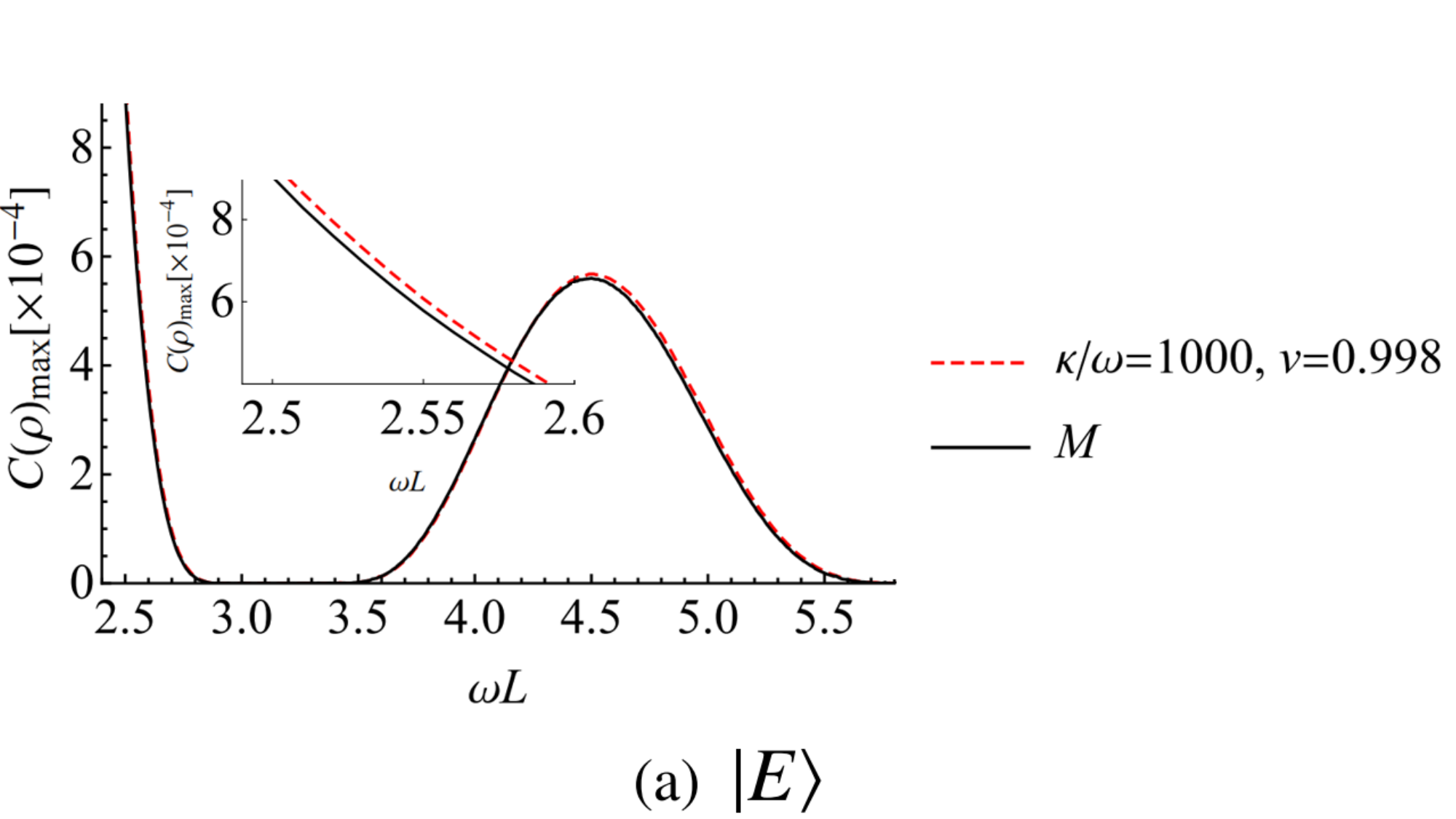}}
{\includegraphics[height=1.65in,width=2.65in]{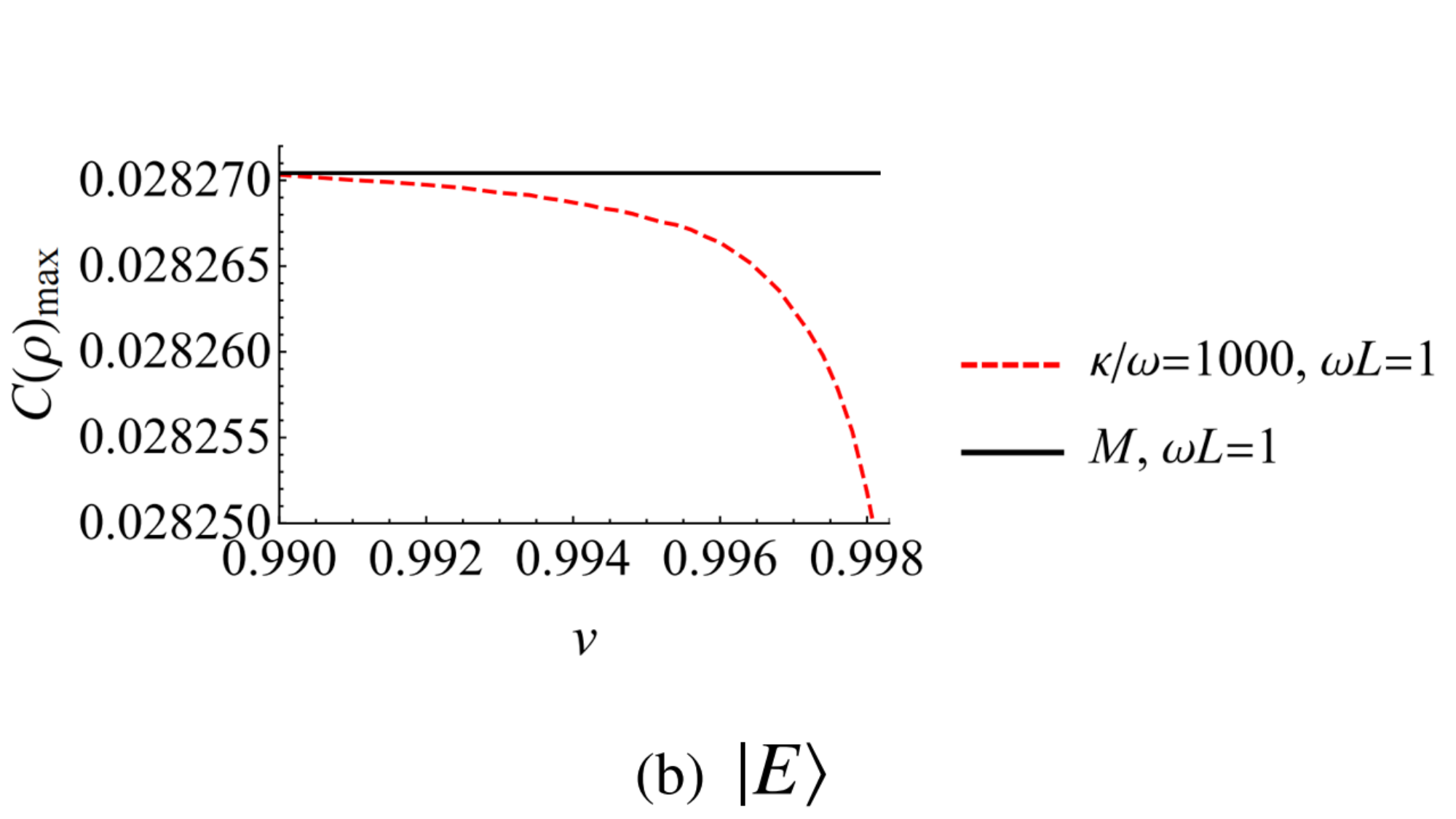}}
\caption{Maximum of concurrence during the evolution of two uniformly moving atoms initially prepared in $|E\rangle$, via interatomic distance $\omega L$ (a) and velocity $v$ (b).
}\label{f3}
\end{figure}

For the interatomic separation comparable to the transition wavelength ($L\sim\omega^{-1}$), we compare the entanglement dynamics of the $\kappa$-deformed and Minkowski spacetime cases shown in Fig.~\ref{f2}. As shown in Fig.~\ref{f2} (a), the waiting time to generate entanglement depends on the interatomic separation. In addition, with the increase in the velocity of atoms, the waiting time to generate entanglement becomes longer, i.e., the sudden creation of entanglement is delayed [Figs.~\ref{f2} (b) and (c)]. Therefore, here, we note that, with the increase in velocity, the differences in the entanglement dynamics between the $\kappa$-deformed and Minkowski spacetime cases become more distinguishable. That is, even when the spacetime deformation parameter $\kappa$ is large, in principle, we can still discriminate these two spacetimes with the help of relativistic motion. This result is different from that of the two static atoms discussed previously. We also note that the amplitude of the maximum entanglement generated depends on the interatomic separation. This result is shown in detail in Fig.~\ref{f3} (a).
Fig.~\ref{f3} (b) shows the behavior of the maximum entanglement generated during evolution with the assistance of atomic motion. We find that when the velocity of atoms is sufficiently large, the maximum of concurrence of the $\kappa$-deformed and Minkowski spacetime cases behaves differently even when the spacetime deformation parameter is relatively large. This result indicates that velocity will affect the difference in the maximum of concurrence between these two spacetimes compared with that shown in Fig.~\ref{R4} (b). Moreover, in the Minkowski spacetime, the maximum of concurrence is irrespective of velocity and remains a constant, see Fig.~\ref{f3} (b) and Eq.~(\ref{uniformlyMAB}). The aforementioned phenomenon indicates that, when the velocity is sufficiently large, in principle, the entanglement behavior of two atoms can be used to distinctly discriminate these two universes even when the spacetime deformation parameter $\kappa$ is relatively large.

\subsubsection{Two uniformly moving atoms initially prepared in entangled states $|A\rangle$ and $|S\rangle$}
\begin{figure}[H]
\centering
{\includegraphics[height=1.65in,width=2.65in]{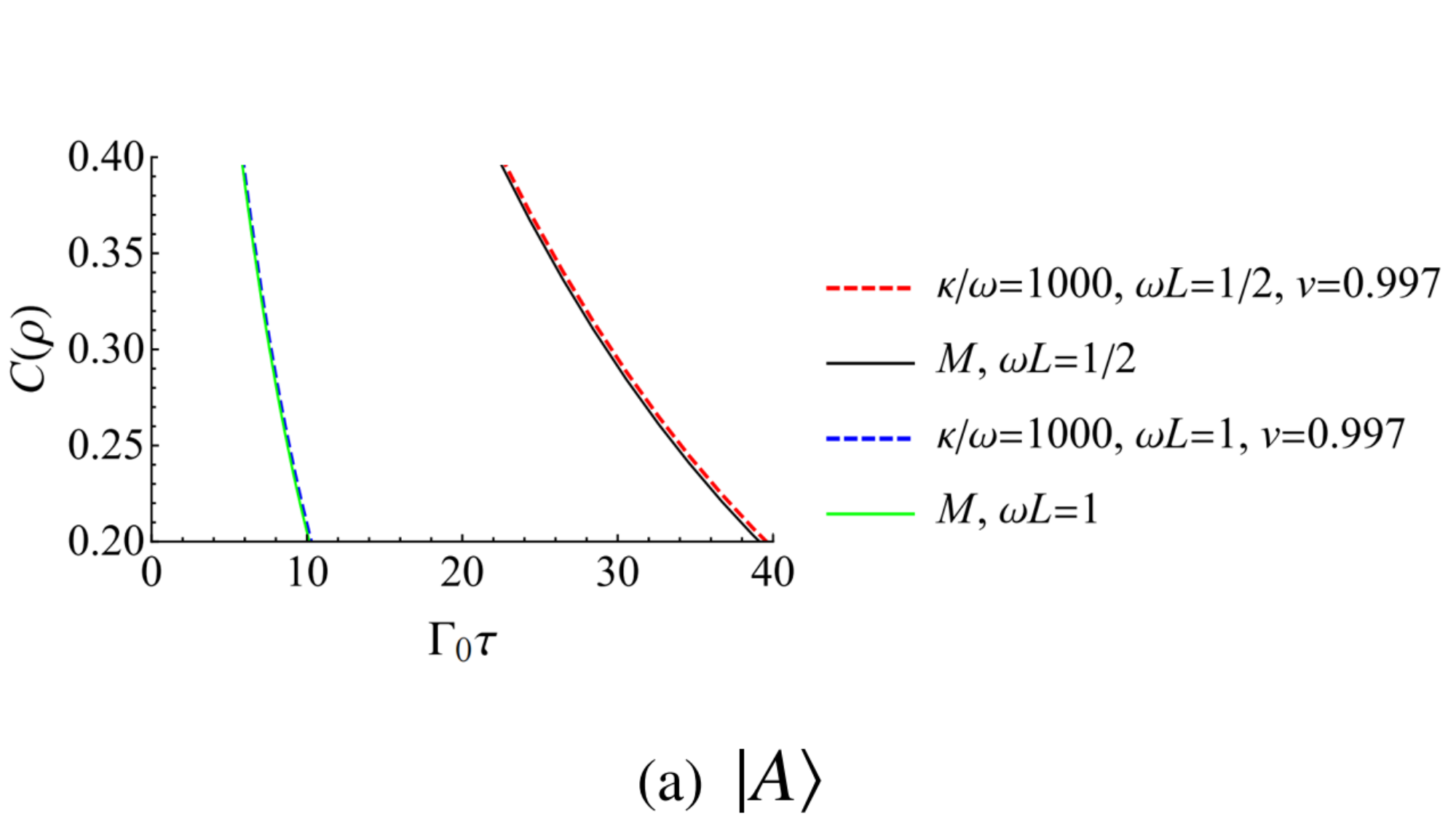}}
{\includegraphics[height=1.65in,width=2.65in]{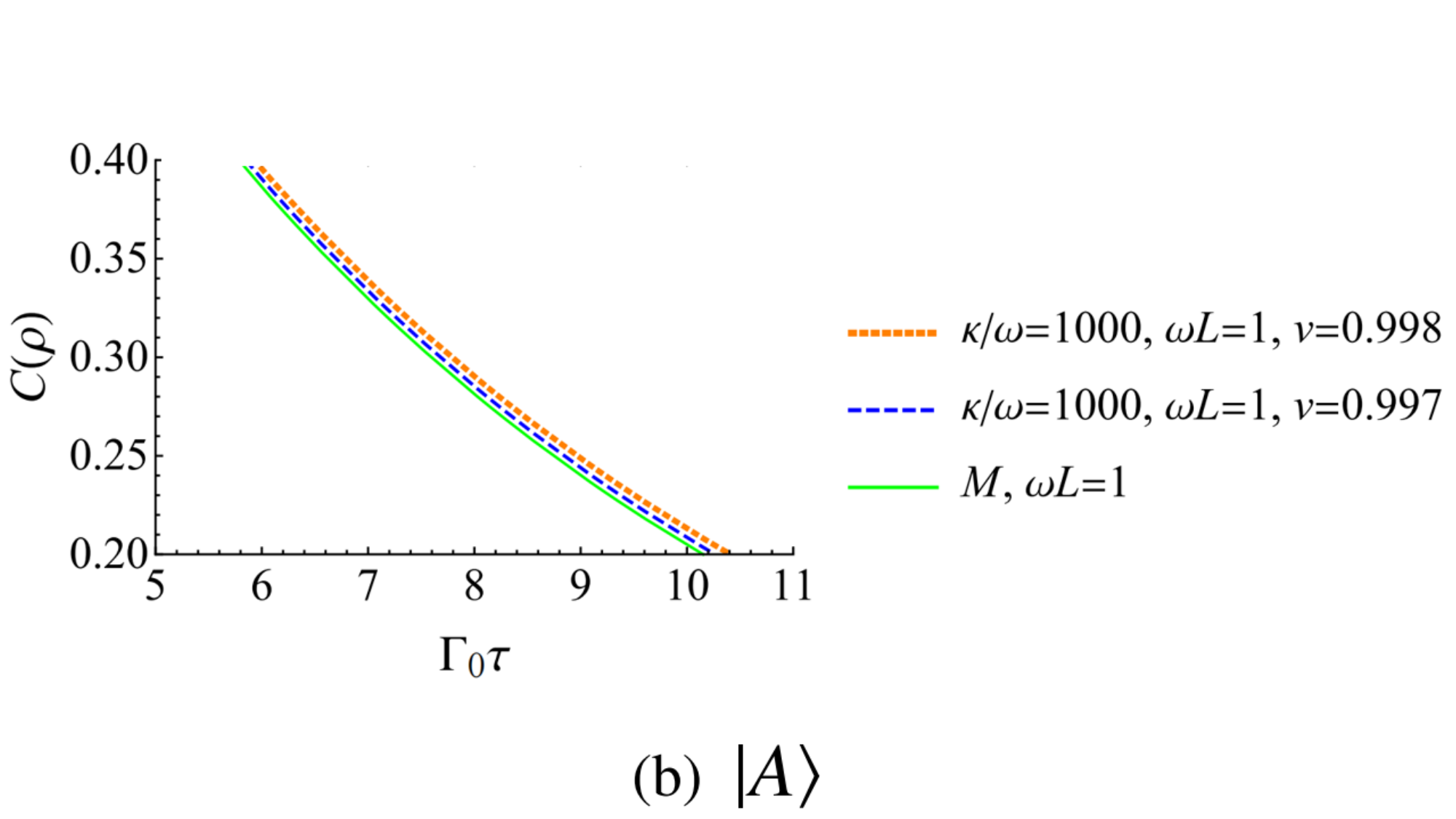}}\\
{\includegraphics[height=1.65in,width=2.65in]{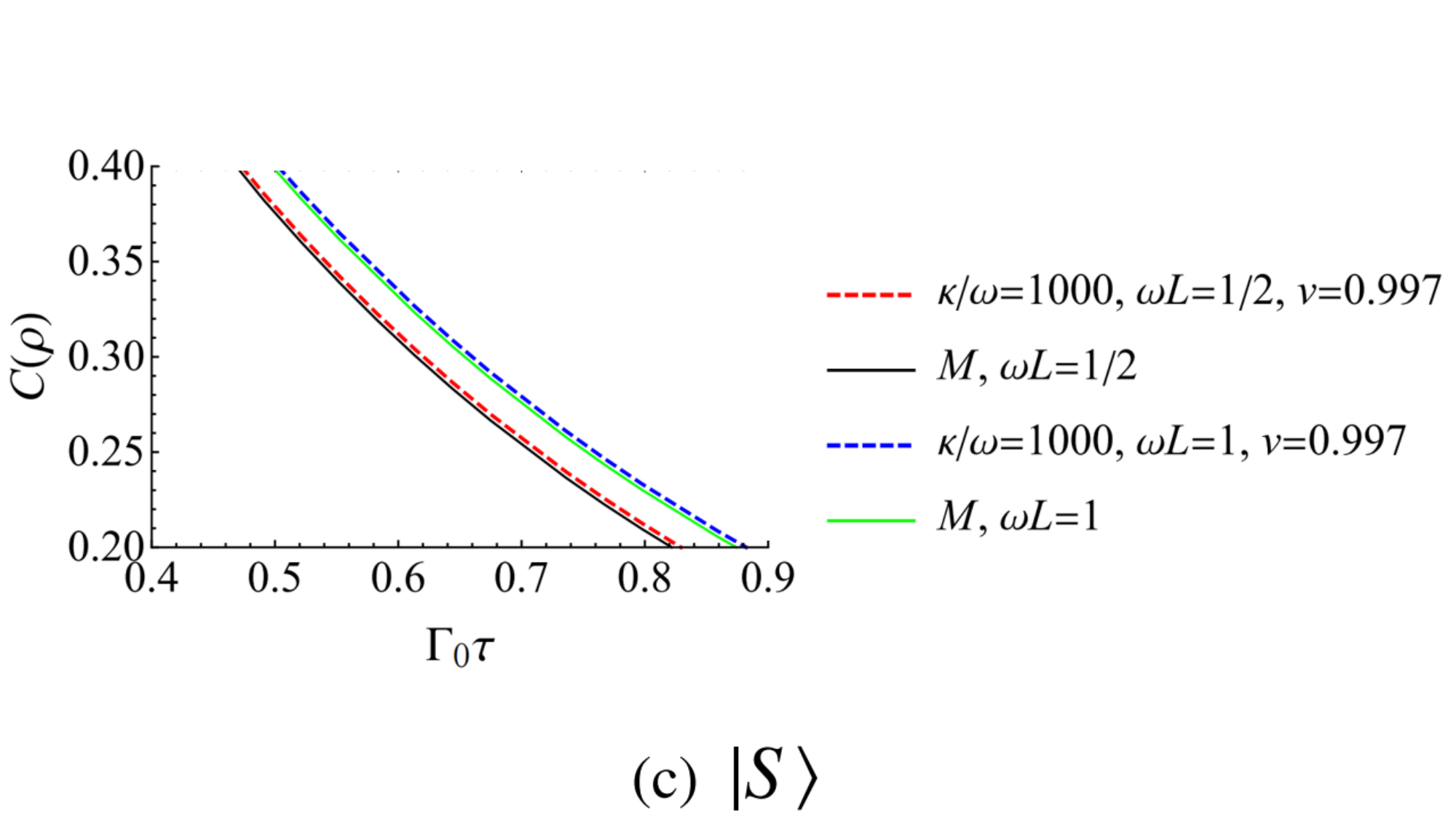}}
{\includegraphics[height=1.65in,width=2.65in]{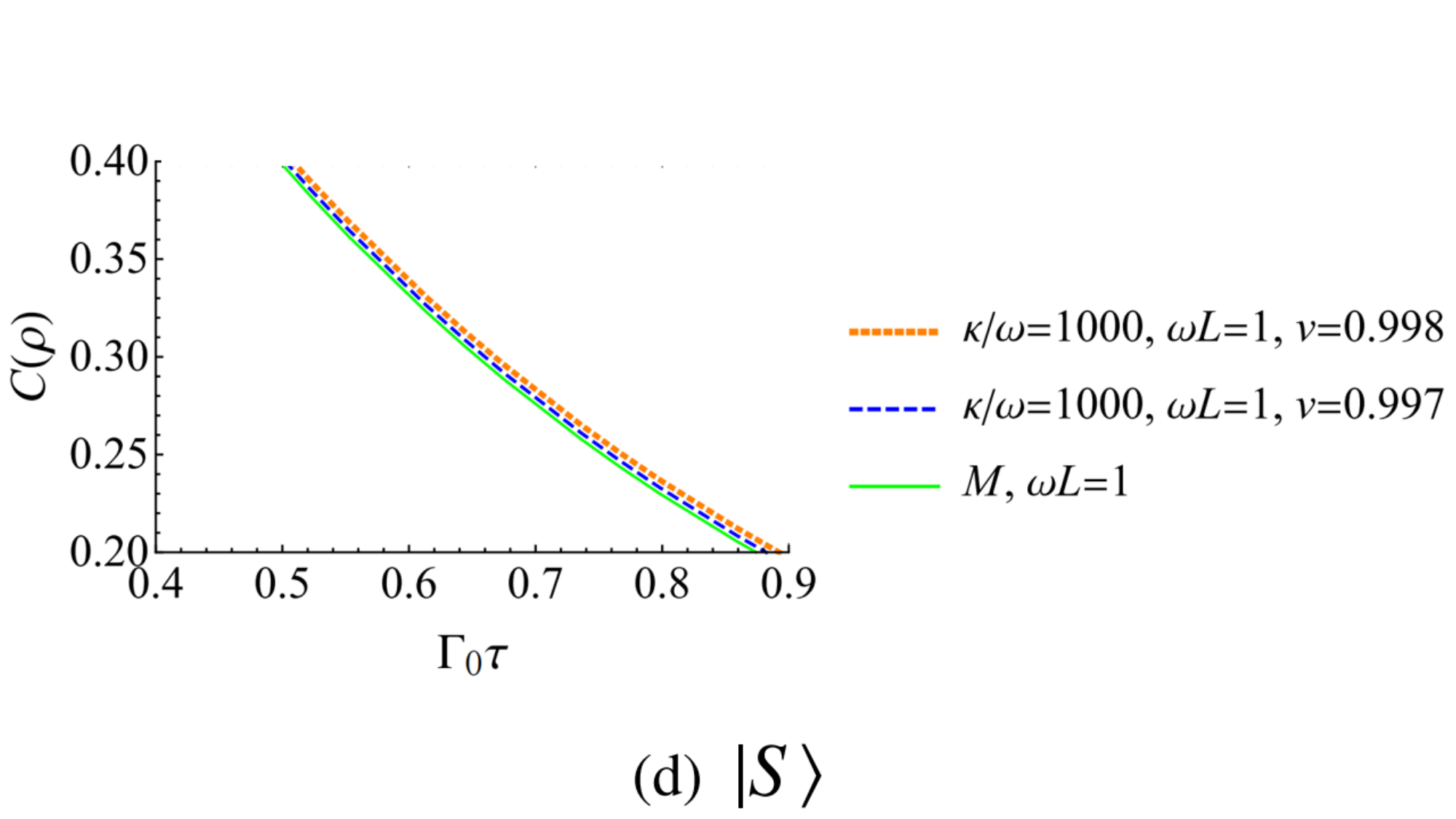}}
\caption{Time evolution of concurrence of two uniformly moving atoms initially prepared in $|A\rangle$ with different values of $\omega L=1/2,\;1$ (a) and different values of $v=0.997,\;0.998$ (b) and $|S\rangle$ with different values of $\omega L=1/2,\;1$ (c) and different values of $v=0.997,\;0.998$ (d).
}\label{f1}
\end{figure}

Now, we investigate the effects of velocity on the entanglement dynamics of two uniformly moving atoms initially prepared in two kinds of maximally entangled states, i.e., $|A\rangle$ and $|S\rangle$, as shown in Fig.~\ref{f1}. We find that, as the two-atom system evolves, their concurrence decreases monotonously and finally decays to zero in the infinite time limit both for the $\kappa$-deformed and Minkowski spacetime cases. However, we note that, although the spacetime deformation parameter $\kappa$ is relatively large, the atomic concurrence curve in the $\kappa$-deformed spacetime with an LDP still does not overlap with that in the Minkowski spacetime case, which is different from those shown in Fig.~\ref{R1-2} for two static atoms. This result originates from the influence of atomic velocity. Moreover, we find that as the velocity of the atoms increases, the difference in the entanglement dynamics between $\kappa$-deformed and Minkowski spacetimes becomes more distinct. For the initial sates, i.e., $|A\rangle$ and $|S\rangle$, their evolution responds differently to the interatomic distance. In $|A\rangle$, the entanglement decreases with the increase in the interatomic distance, while in $|S\rangle$, the entanglement increases with the increase in the interatomic distance. This finding indicates that the symmetry of the atomic entangled state may play an important role in distinguishing the two universes with the entanglement dynamics.

\subsection{Entanglement dynamics of two circularly accelerated atoms}
In this section, we will explore the entanglement dynamics of two circularly accelerated atoms (with an LDP) in $\kappa$-deformed and Minkowski spacetimes. We are interested in whether the uniform circular motion of the atoms is more readily able to tell us the difference between the $\kappa$-deformed and Minkowski spacetimes. We assume that these two atoms rotate synchronically with a separation $L$ perpendicular to the rotating plane, whose trajectories are described as follows:
\begin{eqnarray}\label{trajectories2}
&&t_1(\tau)=\gamma\tau, x_1(\tau)=R\cos\frac{\gamma v\tau}{R},
y_1(\tau)=R\sin\frac{\gamma v\tau}{R},z_1(\tau)=0,\nonumber\\
&&t_2(\tau)=\gamma\tau, x_2(\tau)=R\cos\frac{\gamma v\tau}{R},y_2(\tau)=R\sin\frac{\gamma v\tau}{R},
 z_2(\tau)=L,\nonumber\\
\end{eqnarray}
where $R$ is the radius of the circular orbit. In the rest frame of the atoms, the centripetal acceleration is $a=\frac{\gamma^2v^2}{R}$.

Now, by substituting the trajectory~(\ref{trajectories2}) into Eq.~(\ref{deformed correlation function}), we consider the ultrarelativistic limit, i.e., $v\rightarrow 1$. Then, the two-point correlation functions in $\kappa$-deformed spacetime become
\begin{eqnarray}\label{circular Green function1}
G^{11}(x,x')&=&G^{22}(x,x')
=-\frac{1}{4\pi^2}\frac{1}{\triangle\tau^2[1+\frac{1}{12}(a^2\triangle\tau^2)]}
\nonumber\\
&&\;\;\;\;\;\;\;\;\;\;\;\;\;\;\;\;\;\;\;-\frac{1}{16\pi^2\kappa^2} \frac{(4\gamma^2-1)-\frac{1}{12}a^2\triangle\tau^2}
{\triangle\tau^4(1+\frac{1}{12}a^2\triangle\tau^2)^3} \nonumber\\
&&\;\;\;\;\;\;\;\;\;\;\;\;\;\;\;\;\;\;\;-\frac{1}{4\pi^2\kappa^2} \frac{[(2\gamma^2-1)-\frac{1}{12}a^2\triangle\tau^2]\gamma^2}
{\triangle\tau^4(1+\frac{1}{12}a^2\triangle\tau^2)^4},\nonumber\\
\end{eqnarray}
and
\begin{eqnarray}\label{circular Green function2}
&&G^{12}(x,x')=G^{21}(x,x')=-\frac{1}{4\pi^2}\frac{1}{\triangle\tau^2
[1+\frac{1}{12}(a^2\triangle\tau^2)]-L^2}\nonumber\\
&&\;\;\;\;\;\;\;\;\;\;\;\;\;\;\;\;\;\;\;\;\;\;\;\;\;-\frac{1}{16\pi^2\kappa^2} \frac{(4\gamma^2-1)\triangle\tau^2-\frac{1}{12}a^2\triangle\tau^4+L^2}
{[\triangle\tau^2(1+\frac{1}{12}a^2\triangle\tau^2)-L^2]^3} \nonumber\\
&&\;\;\;\;\;\;\;\;\;\;\;\;\;\;\;\;
-\frac{1}{4\pi^2\kappa^2} \frac{[(2\gamma^2-1)\triangle\tau^2-\frac{1}{12}a^2\triangle\tau^4+L^2]
\gamma^2\triangle\tau^2}{[\triangle\tau^2(1+\frac{1}{12}a^2\triangle\tau^2)-L^2]^4}.\nonumber\\
\end{eqnarray}
Using the residue theorem method, the Fourier transforms of the correlation functions and the corresponding expressions of the coefficients $A_i$ and $B_i$ in $\kappa$-deformed spacetime can be derived, as shown in
Supplemental Material [Eqs.~(S1)--(S4)]. Note that, when $\kappa\rightarrow\infty$ for this case, we can recover the result obtained in Ref.~\cite{Hu20151} for a circularly accelerated two-atom system in Minkowski spacetime as expected [Eq.~(S5) in Supplemental Material].

\subsubsection{Two circularly accelerated atoms initially prepared in a separable state $|E\rangle$}
To analyze the entanglement generation for two circularly accelerated atoms in $\kappa$-deformed spacetime with an LDP, we assume that these two atoms are initially prepared in a separable state $|E\rangle$.

\begin{figure}[H]
\centering
{\includegraphics[height=1.65in,width=2.65in]{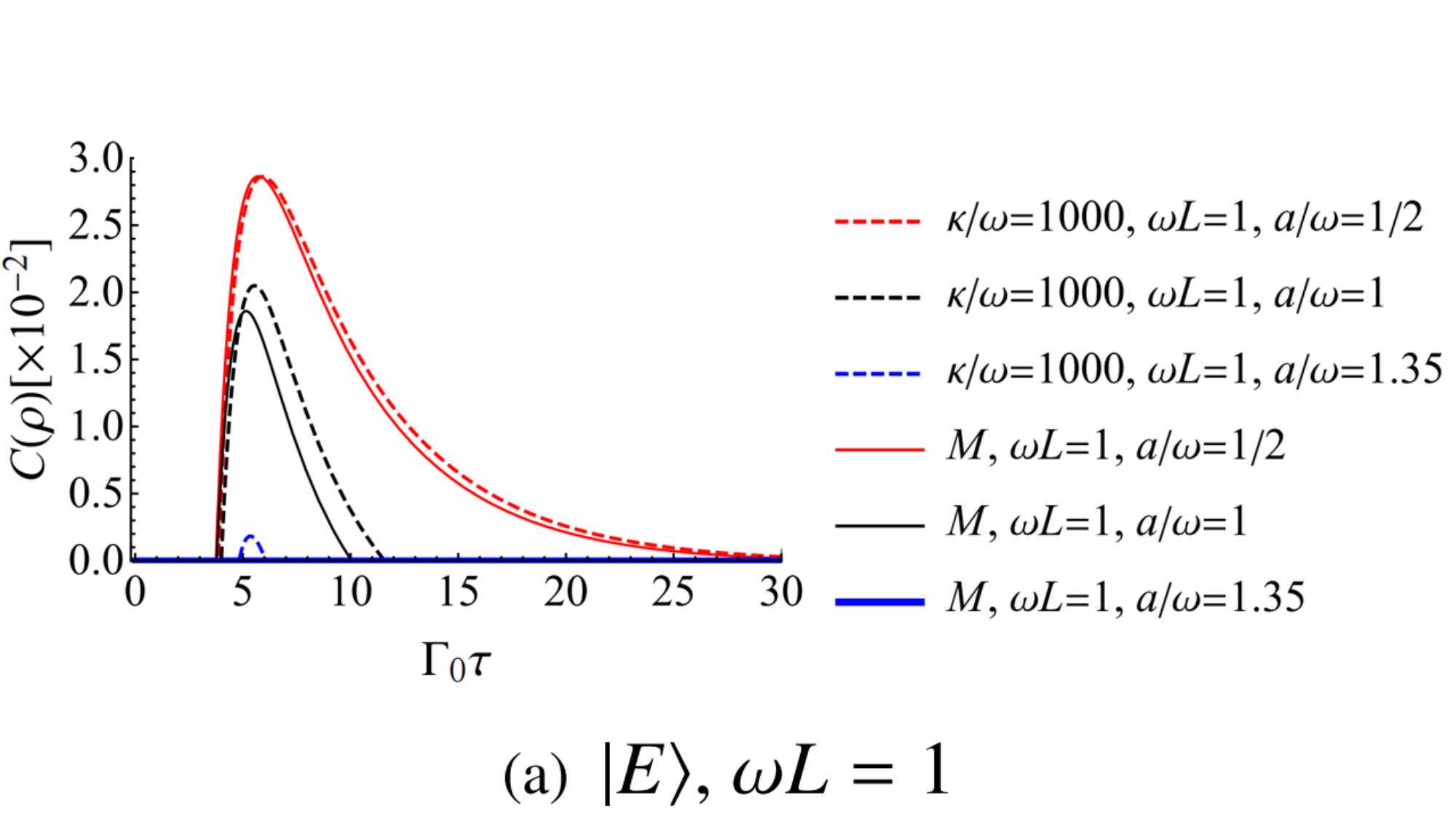}}
{\includegraphics[height=1.65in,width=2.65in]{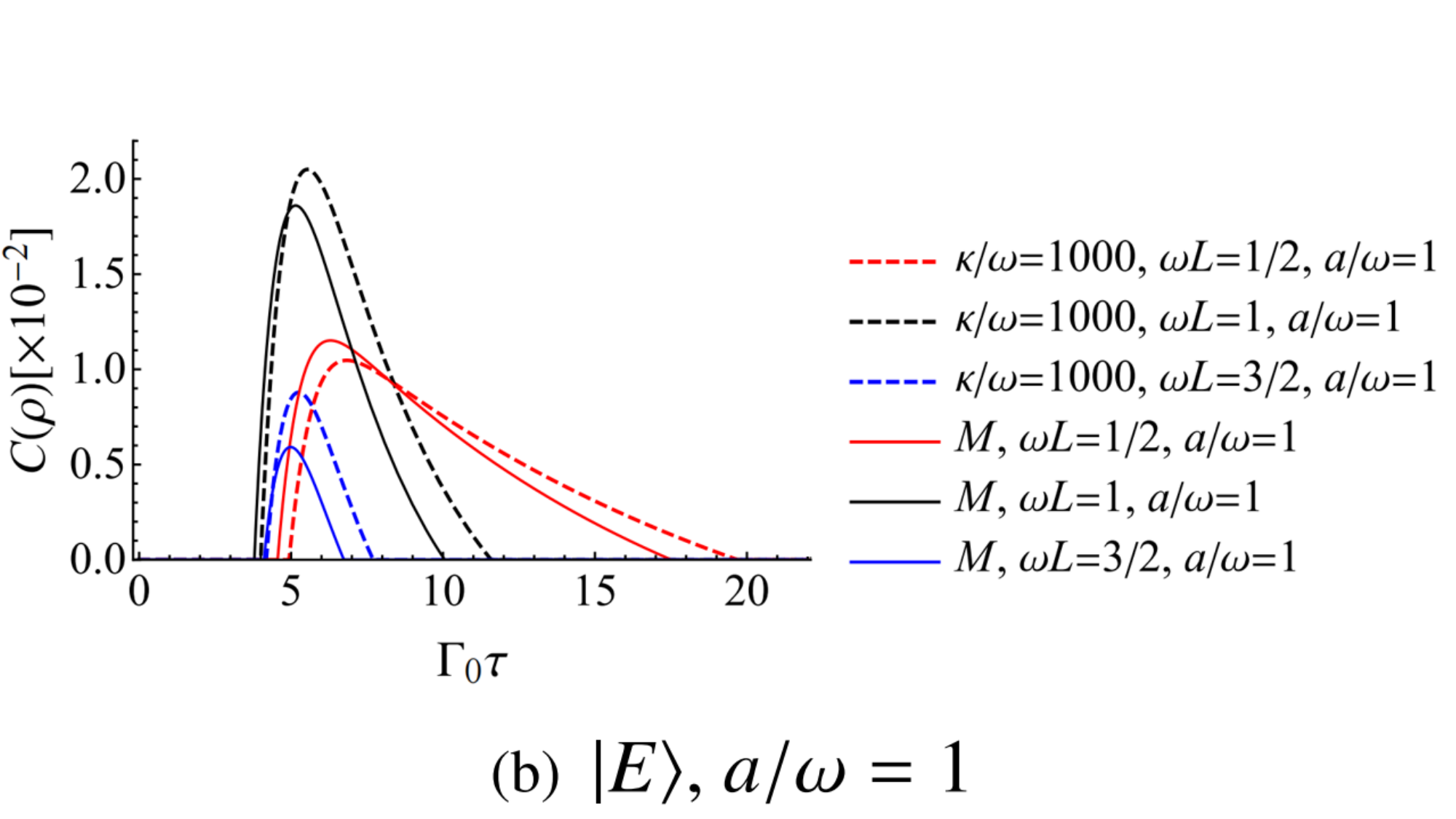}}
\caption{Time evolution of concurrence of circularly accelerated atoms initially prepared in $|E\rangle$, varying the centripetal acceleration $a/\omega$ (a) and the interatomic distance $\omega L$ (b).
}\label{circularly2}
\end{figure}

We focus on the case where the intermediate separations are comparable to the transition wavelength of the atoms ($L\sim \omega^{-1}$). There exists a delayed feature of the entanglement generation for two circularly accelerated atoms in $\kappa$-deformed and Minkowski spacetimes, as depicted in Fig.~\ref{circularly2}. We note that the waiting time to generate entanglement is not only related to the centripetal acceleration but also associated with the interatomic separation. Even when the spacetime deformation parameter is relatively large, the maximum entanglement generated and the waiting time are different for these two spacetime cases concerned. Specifically, for the case of circularly accelerated atoms with a fixed separation distance in Minkowski spacetime, there exists a critical value, i.e., $a/\omega \approx 1.35$ for the centripetal acceleration, beyond which entanglement generation does not occur. However, in $\kappa$-deformed spacetime, this critical value will be modified and increase with the decrease in the value of the spacetime deformation parameter. This finding indicates that, in some cases, entanglement can be generated in $\kappa$-deformed spacetime, whereas in Minkowski spacetime, entanglement cannot be generated. In principle, this presence/absence of entanglement provides us with a good criterion to check which universe we are living in (see the detailed discussion of this criterion in the subsequent paragraphs). Furthermore, we find
that the existing time of entanglement generated depends on the centripetal acceleration $a/\omega$ and interatomic distance $\omega L$.
Although the two-atom system is subjected to the same conditions, i.e., the same $a/\omega$ and $\omega L$, as a result of the spacetime deformation,
the existing time of entanglement generated is different for these two universes concerned.
This property can help us distinguish the $\kappa$-deformed and Minkowski spacetimes in principle.

\begin{figure}[H]
\centering
{\includegraphics[height=1.65in,width=2.65in]{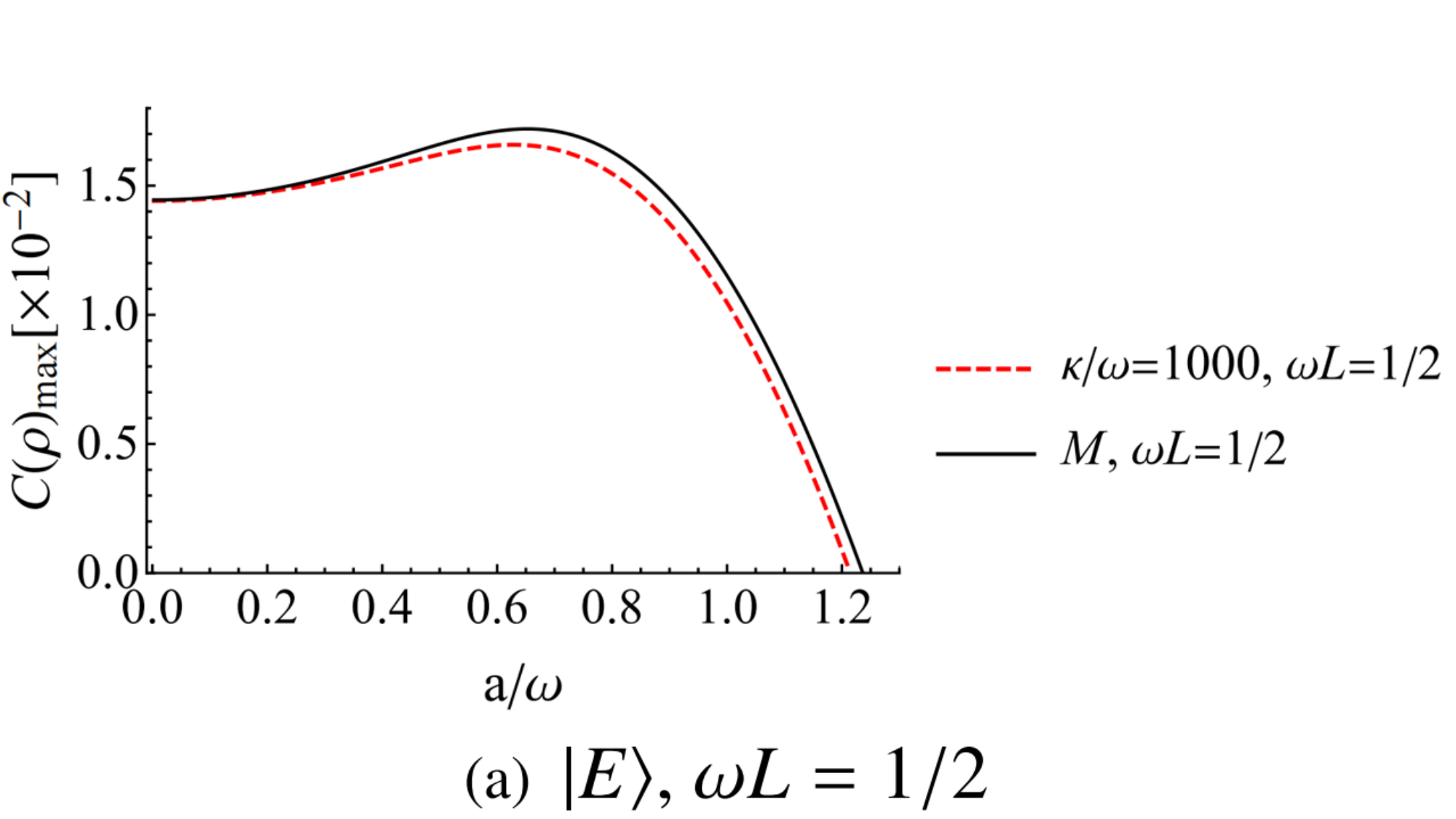}}
{\includegraphics[height=1.65in,width=2.65in]{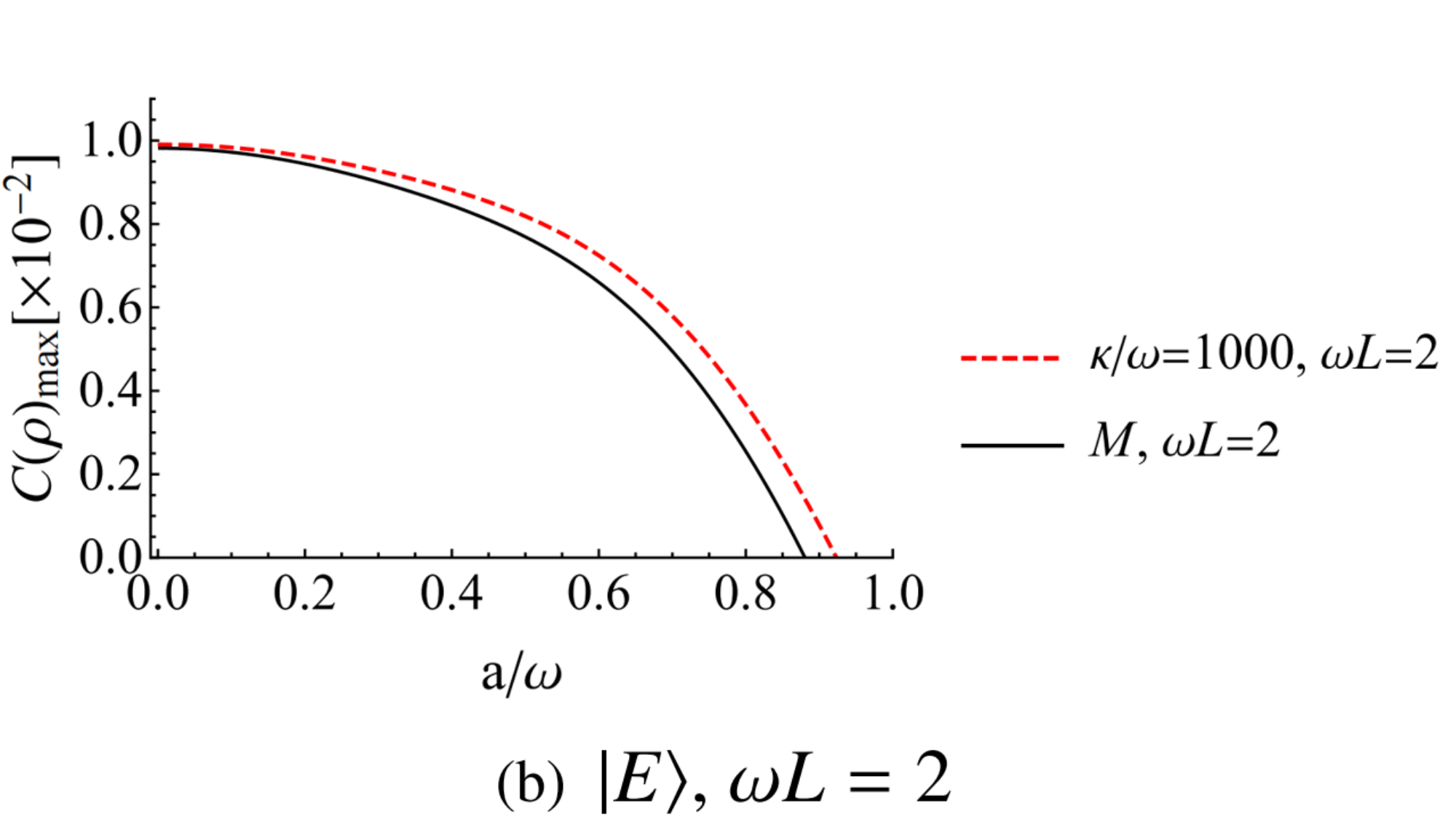}}\\
{\includegraphics[height=1.65in,width=2.65in]{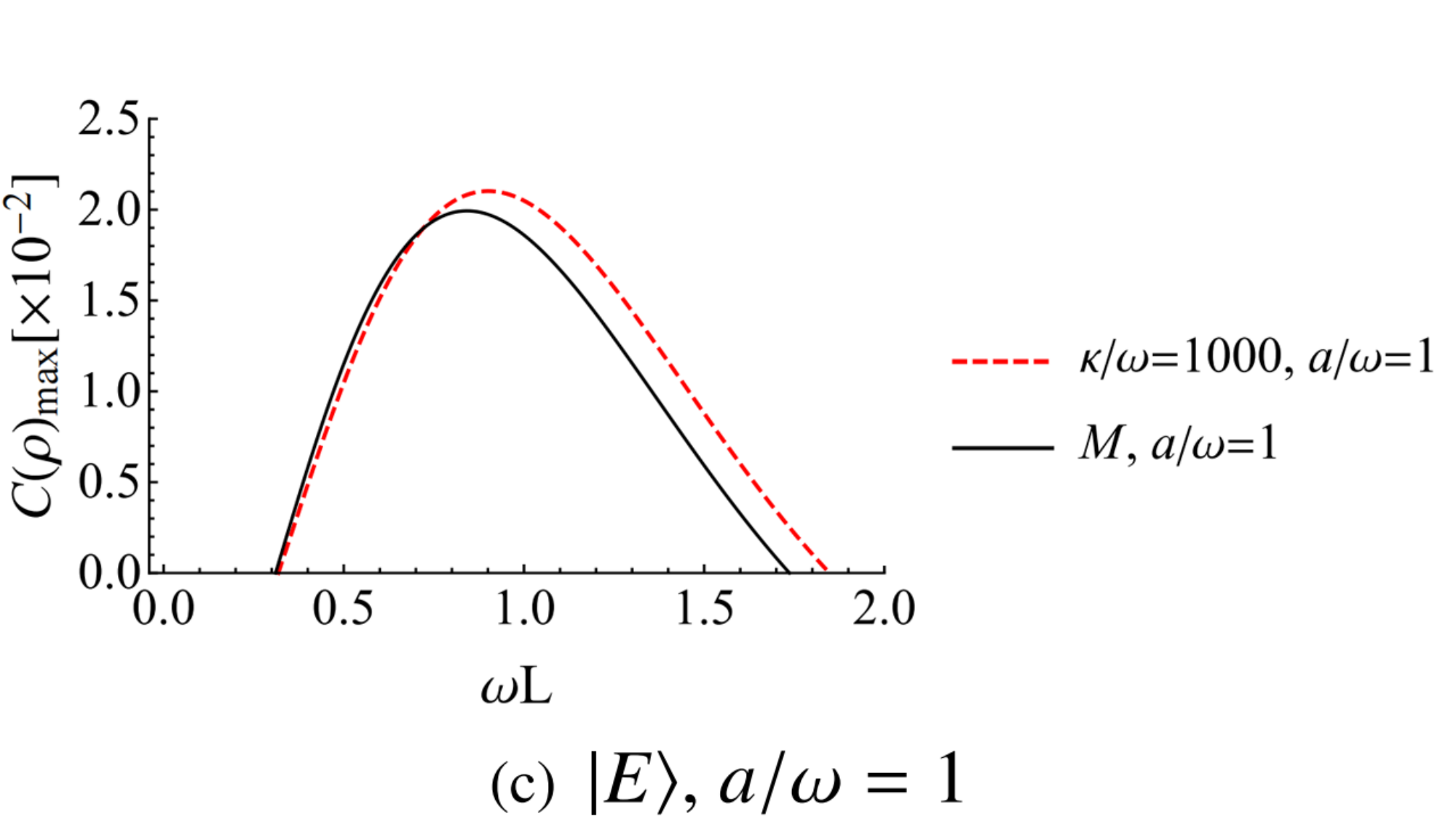}}
{\includegraphics[height=1.65in,width=2.65in]{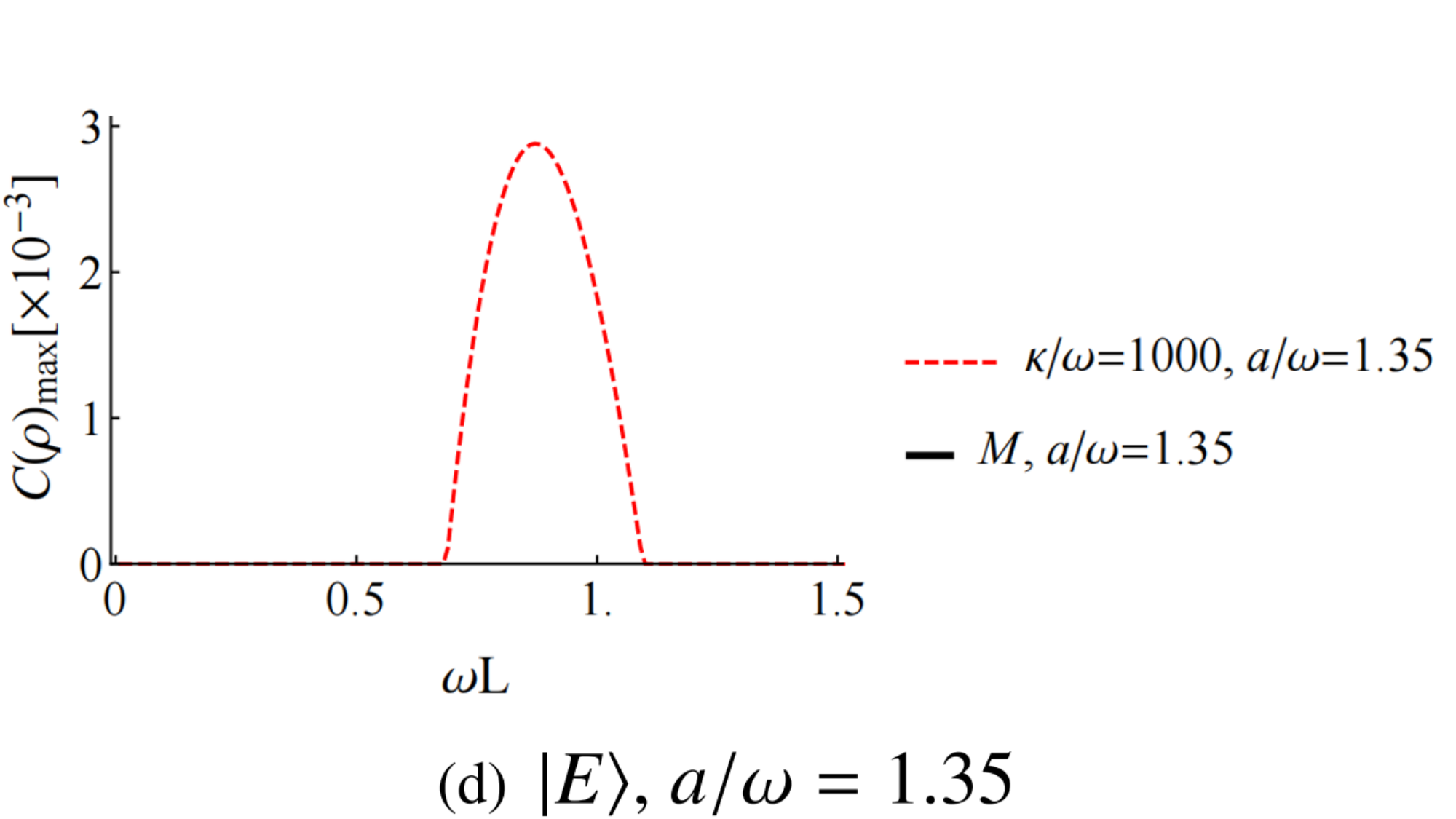}}
\caption{Comparison between the maximum of concurrence during the evolution of circularly accelerated atoms initially prepared in $|E\rangle$ via centripetal acceleration $a/\omega$ (a, b) and interatomic distance $\omega L$ (c, d).
}\label{circularly3-4}
\end{figure}

In Fig.~\ref{circularly3-4}, we plot how the maximum of concurrence during evolution is affected by the interatomic separation and centripetal acceleration of atoms. As shown in Figs.~\ref{circularly3-4} (a) and (b), when the two atoms are static ($a/\omega=0$), the maximum entanglement in the $\kappa$-deformed and Minkowski spacetime cases cannot be distinguished. However, one can find that they become more distinguishable with the increase in the centripetal acceleration. The maximum entanglement behavior via centripetal acceleration depends on the interatomic separation: it would first increase to a maximum and then decay to zero with the increase in the centripetal acceleration when the interatomic separation is relatively small, or it would decay monotonically to zero when the interatomic separation is relatively far. Furthermore, Figs.~\ref{circularly3-4} (c) and (d) show how the maximum entanglement generated depends on the interatomic separation with fixed centripetal acceleration. Remarkably, the interatomic distance regime where entanglement can be generated is dependent on acceleration. Therefore, we find, even when the spacetime deformation parameter $\kappa$ is relatively large, a spatial region where entanglement can be created in the $\kappa$-deformed spacetime. Meanwhile, entanglement cannot be created in the Minkowski spacetime even with the assistance of the centripetal acceleration (Fig.~\ref{circularly3-4} (d)).

\begin{figure}[H]
\centering
\includegraphics[height=2.05in,width=3.65in]{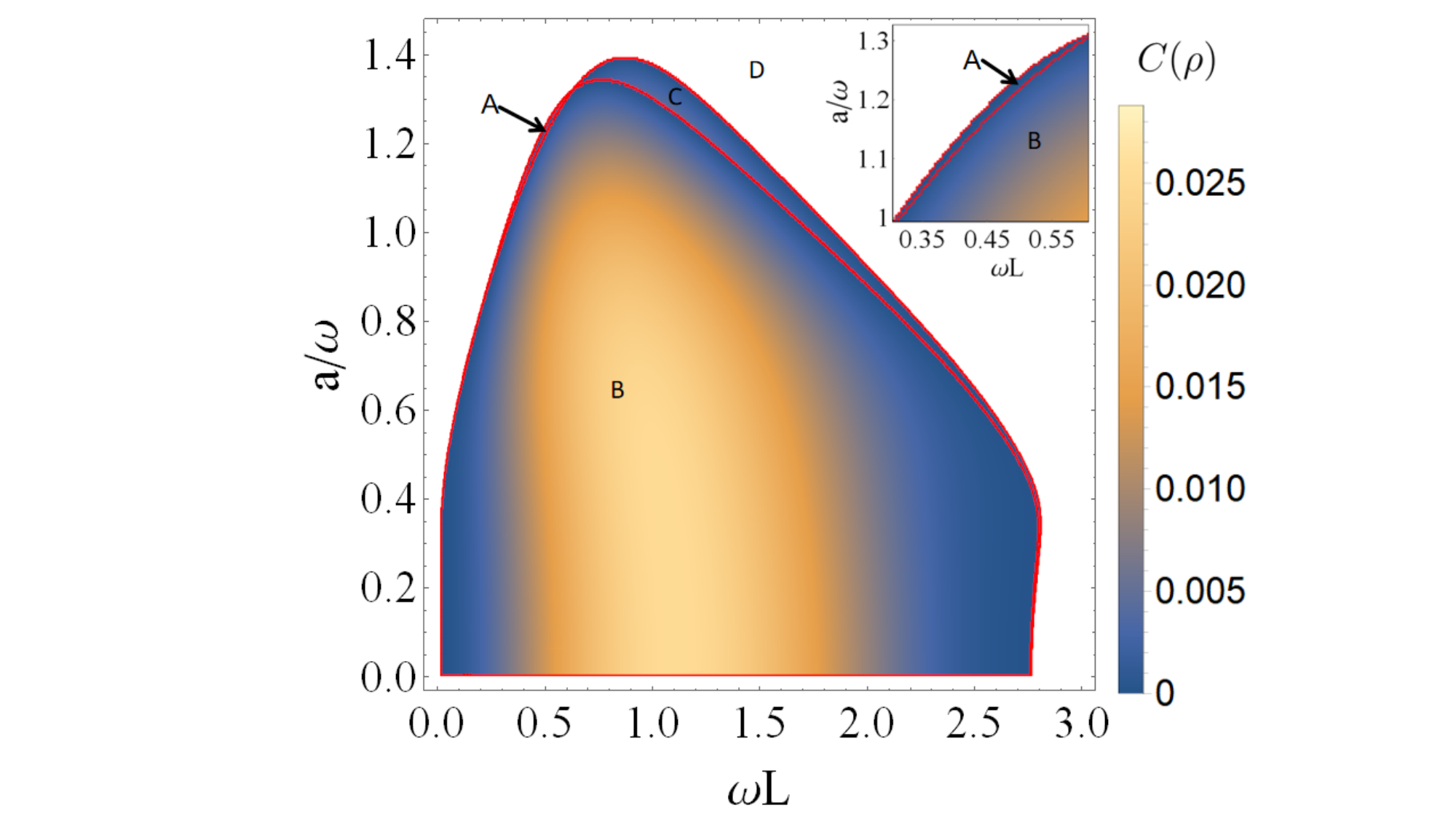}
\caption{Entanglement profile of the two-atom system initially prepared in $|E\rangle$. Region A: two atoms in $\kappa$-deformed spacetime with an LDP cannot get entangled, whereas two atoms in Minkowski spacetime can get entangled. Region B: two atoms in both of these universes can get entangled. Region C: two atoms in $\kappa$-deformed spacetime with an LDP can get entangled, whereas two atoms in Minkowski spacetime cannot get entangled. Region D: two atoms in both of these universes cannot get entangled. Here, we fixed $\kappa/\omega=1000$, and the legend for $C(\rho)$ has been applied to the Minkowski spacetime case.
}\label{circularly5}
\end{figure}

From the previously presented analysis, we find that, in the presence of centripetal acceleration, the entanglement generation for the two-atom system behaves differently in $\kappa$-deformed and Minkowski spacetimes. Therefore, an interesting issue arises: In which parameter regions can we discriminate these two universes concerned? In Fig.~\ref{circularly5}, we show in detail the parameter regions of the centripetal acceleration and interatomic separation where the entanglement can/cannot be generated in $\kappa$-deformed and Minkowski spacetimes. Notably, there exist different regions that indicate different properties of entanglement in these two universes. We infer from this diagram that the two atoms could get entangled only in some special regime of centripetal acceleration and interatomic separation. There exist upper bounds of centripetal acceleration and interatomic separation larger than which entanglement cannot be generated. Another fact shown in Fig.~\ref{circularly5} is that the possible region of entanglement generation for the two atoms in $\kappa$-deformed spacetime with an LDP does not completely overlap with that for the two atoms in Minkowski spacetime. Thus, using the different properties of entanglement generation as a criterion, one can, in principle, distinguish these two universes. Now, we estimate the order of magnitude. Taking hydrogen atoms as an example, we choose $\omega \sim 10^{15}s^{-1}$, $c\sim10^8m/s$, $\kappa\sim 1000\omega/c= 10^{10} m^{-1}$, and the centripetal acceleration $a\sim1.35 c\omega \;m/s^2=1.35\times10^{23} m/s^2$; thus, the accuracy of quantum entanglement estimation only need to reach $10^{-3}$ to distinguish the $\kappa$-deformed spacetime from the Minkowski spacetime. As reported in Ref.~\cite{Harikumar2011}, with the experimental data, the most optimistic bound on the spacetime deformation parameter is $\kappa>10^{12}\omega/c=10^{19} m^{-1}$. Thus, in future experimental setups, we take the centripetal acceleration $a\sim1.3430005 c\omega \;m/s^2=1.3430005\times10^{23} m/s^2$, $L\sim 0.7703 c/\omega=0.7703\times10^{-7} m$, and $\kappa\sim10^{12}\omega/c=10^{19} m^{-1}$. In this case, the entanglement in $\kappa$-deformed spacetime is $C(\rho)\sim 10^{-7}$, while the entanglement in Minkowski spacetime is zero; hence, we can distinguish these two spacetimes. Moreover, if the accuracy of quantum entanglement estimation can reach higher precision, such as $10^{-8}$, then, in future experimental setups, the bound of the deformation parameter can be improved to $\kappa\sim 10^{15}\omega/c= 10^{22} m^{-1}$.

\subsubsection{Two circularly accelerated atoms initially prepared in entangled states $|A\rangle$ and $|S\rangle$}
\begin{figure}[H]
\centering
{\includegraphics[height=1.65in,width=2.65in]{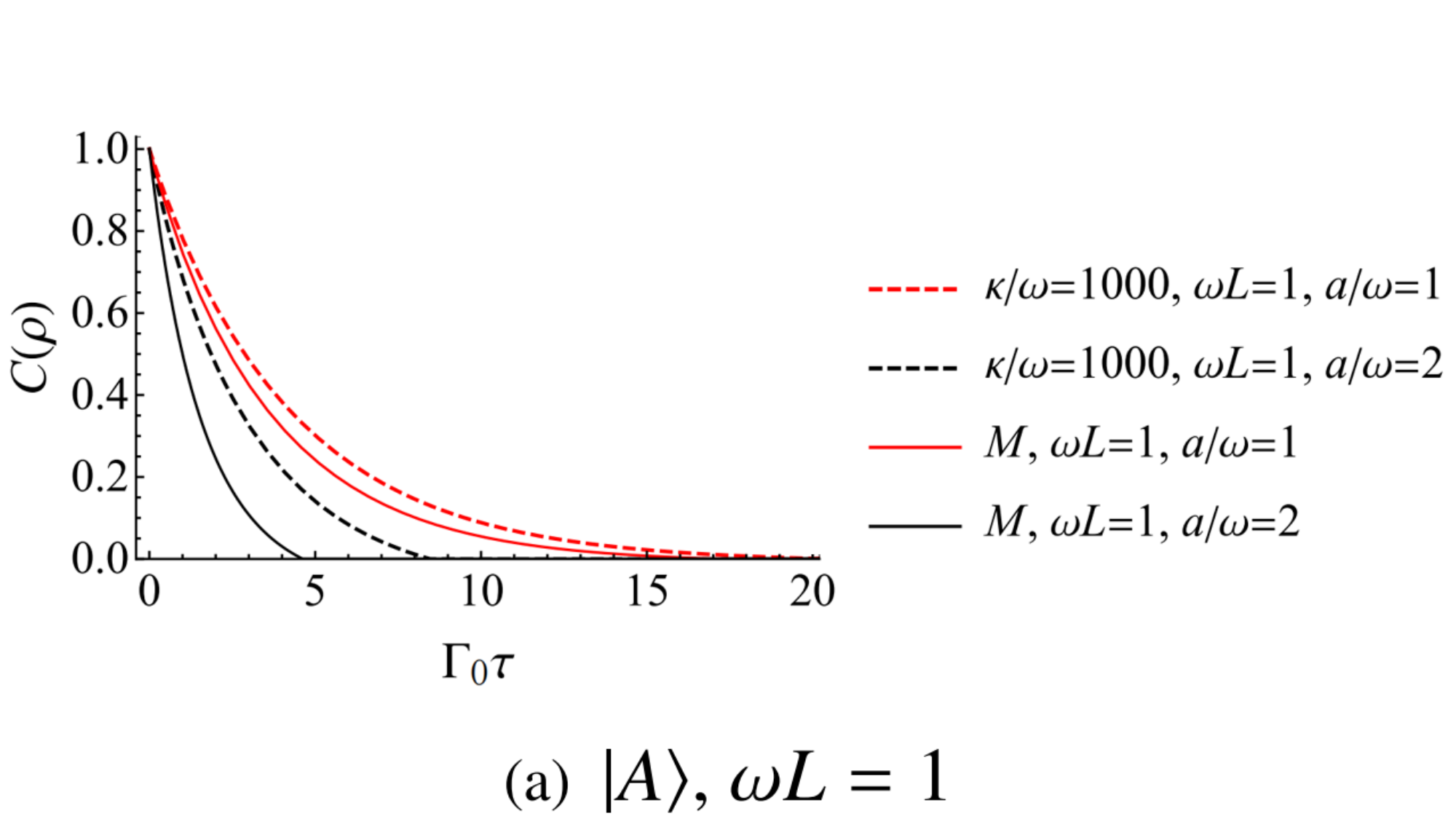}}
{\includegraphics[height=1.65in,width=2.65in]{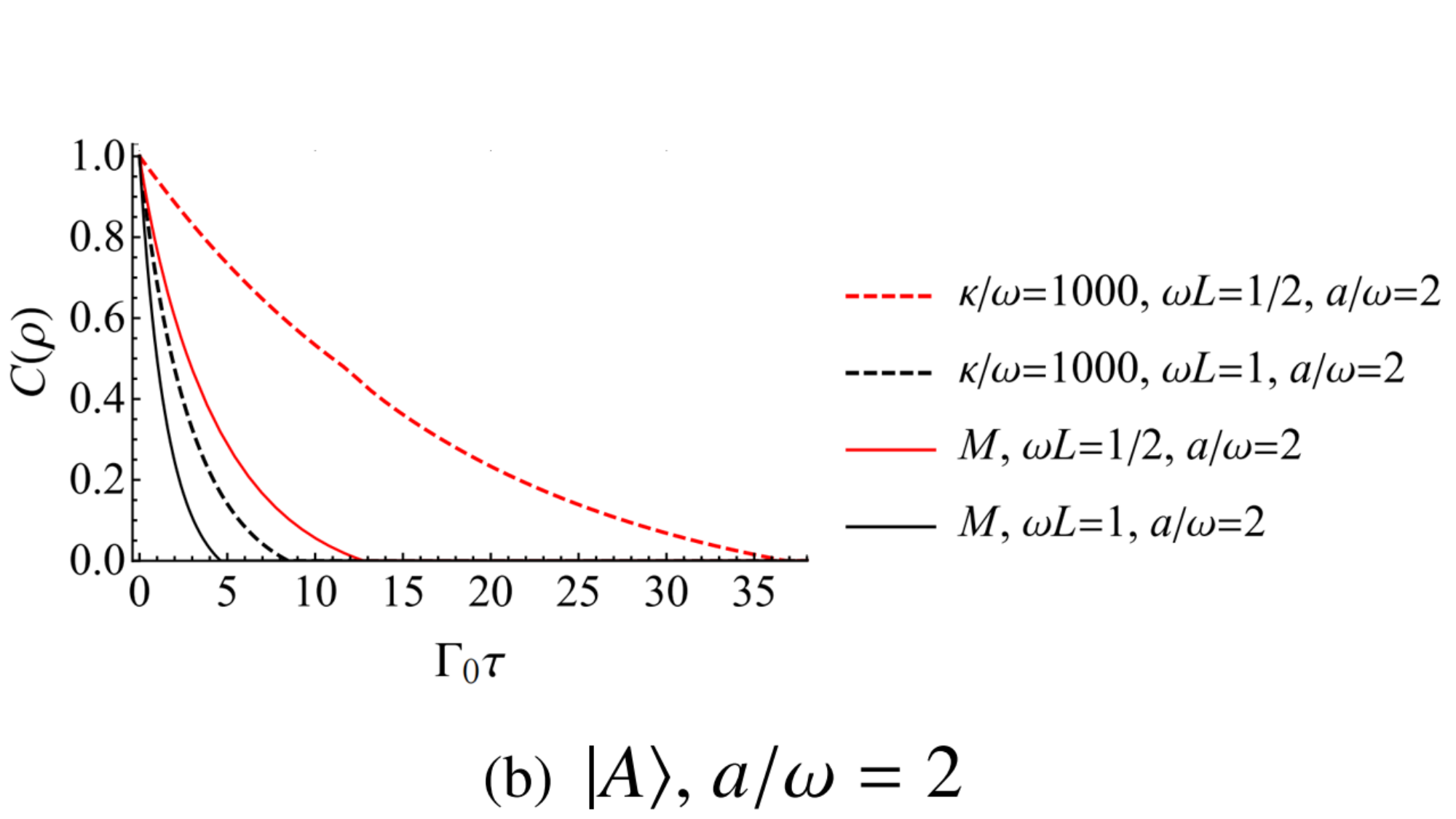}}\\
{\includegraphics[height=1.65in,width=2.65in]{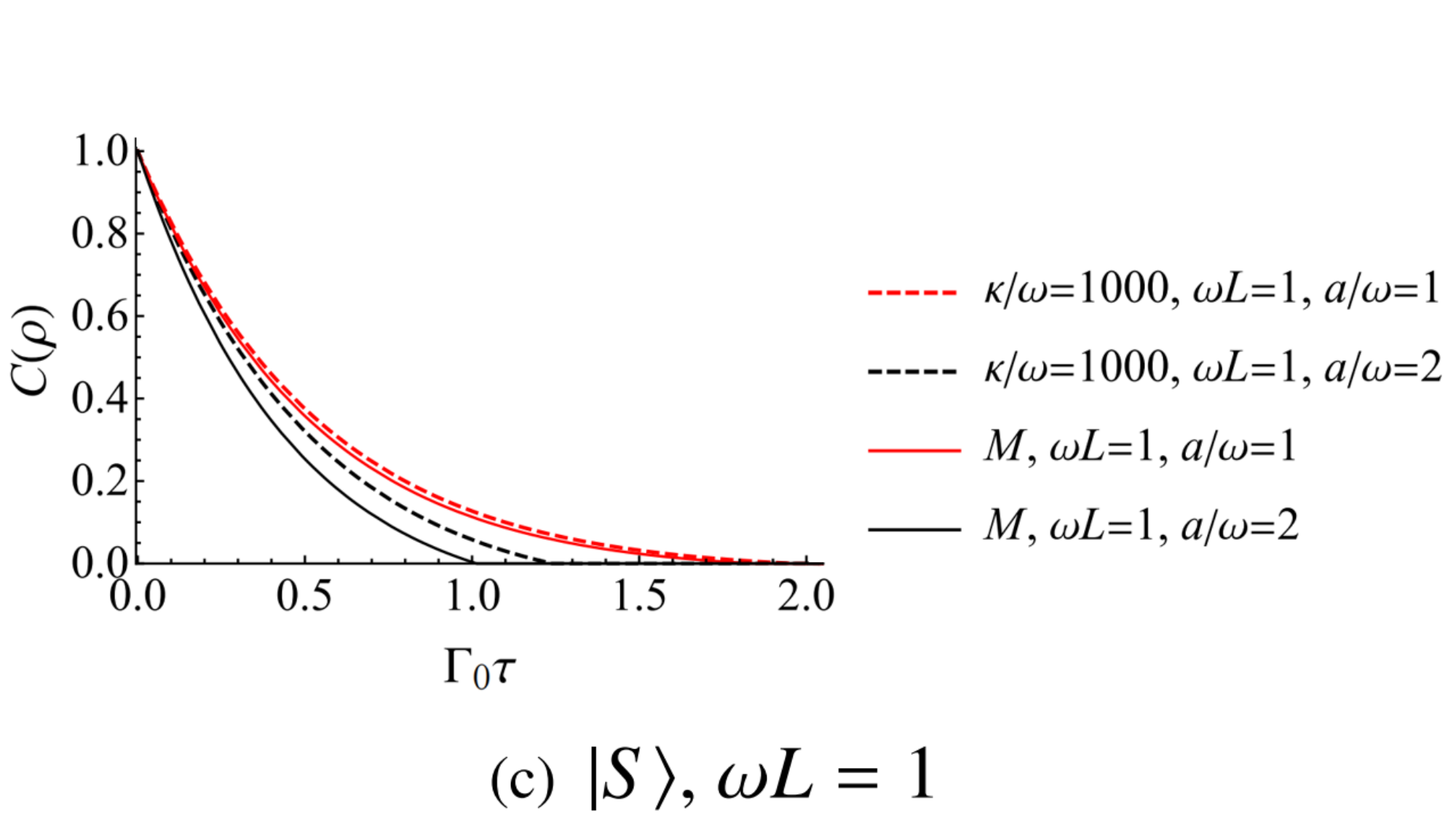}}
{\includegraphics[height=1.65in,width=2.65in]{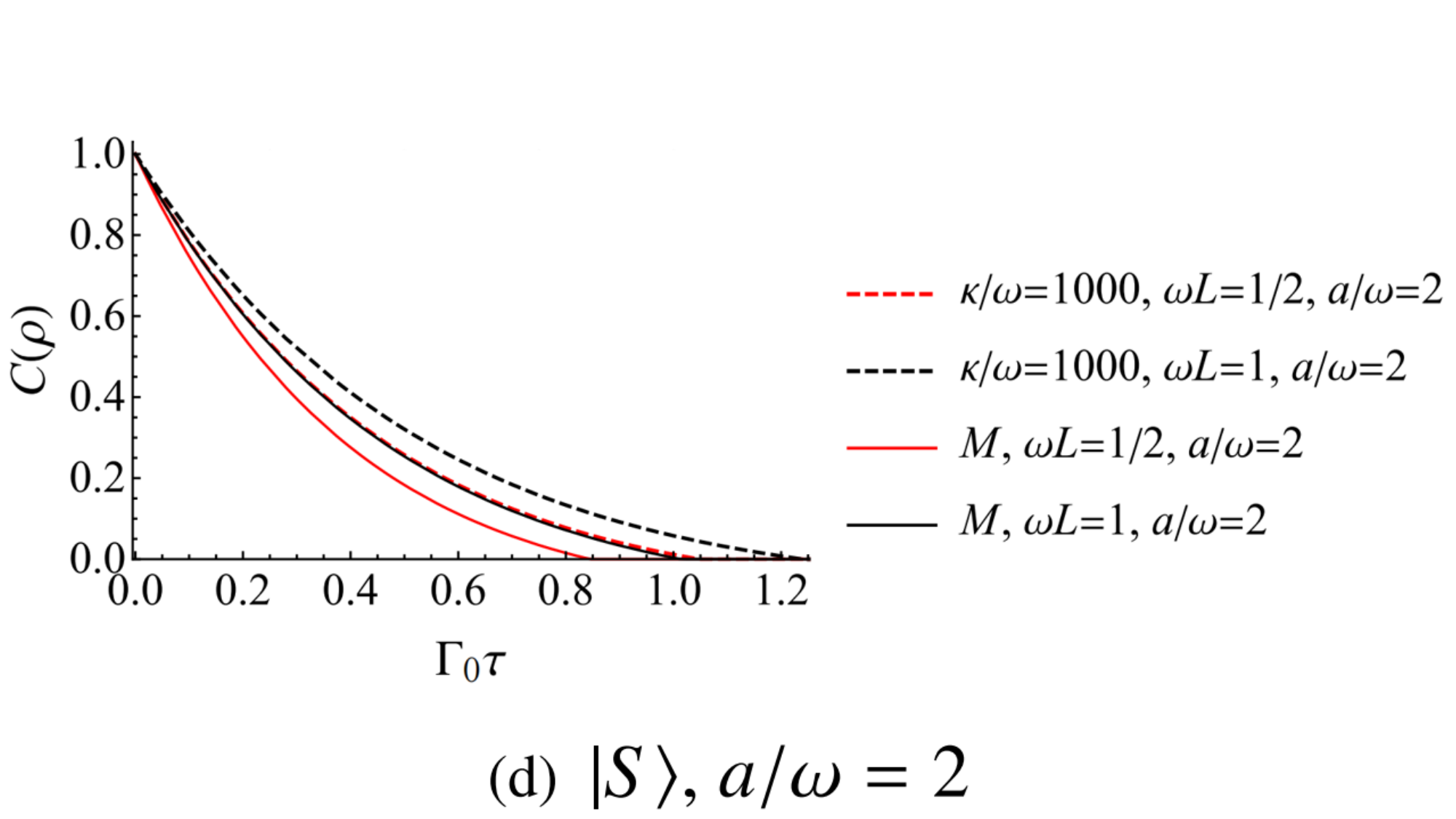}}
\caption{Time evolution of concurrence of two circularly accelerated atoms, varying the centripetal acceleration $a/\omega$ (left) and the interatomic distance $\omega L$ (right), initially prepared in $|A\rangle$ (a, b) and $|S\rangle$ (c, d).
}\label{circularly1-2}
\end{figure}

We analyze the entanglement degradation when the two-atom system is initially prepared in two kinds of maximally entangled states, i.e., $|A\rangle$ and $|S\rangle$. In Fig.~\ref{circularly1-2}, we show the concurrence as a function of the evolution time by fixing the centripetal acceleration and interatomic separation in $\kappa$-deformed and Minkowski spacetimes.

For the initial antisymmetry entangled state case, we infer from Figs.~\ref{circularly1-2} (a) and (b) that, with a fixed interatomic separation, the entanglement decays rapidly with the evolution time of the two circularly accelerated atoms in these two spacetimes. The larger the centripetal acceleration is, the faster the entanglement decays, as well as the variation of entanglement via interatomic separation with a fixed centripetal acceleration. For the initial symmetry entangled state case shown in Figs.~\ref{circularly1-2} (c) and (d), we find that, with a fixed interatomic separation, the two-atom entanglement decays rapidly with the evolution time in these two spacetimes. The larger the centripetal acceleration is, the faster the entanglement decays. However, the variation of entanglement via interatomic separation with a fixed centripetal acceleration is the opposite, which is different from the initial antisymmetry entangled state case. Furthermore, for both of the initial entangled state cases, the entanglement of two circularly accelerated atoms in Minkowski spacetime decays more quickly than that in $\kappa$-deformed spacetime. In this sense, with the help of centripetal acceleration, one can also exploit the entanglement behavior of two atoms to discriminate $\kappa$-deformed and Minkowski spacetimes in principle.

\section{Entanglement dynamics of two atoms with the environment-induced interatomic interaction}\label{section4}
In this section, we will consider the two atoms initially prepared in a separable state $|10\rangle$ and the superposition state of $|A\rangle$ and $|S\rangle$, i.e., $|\psi\rangle=\sqrt{p}|A\rangle+\sqrt{1-p}|S\rangle$ $(0<p<1, p\neq 1/2)$, to address how the environment-induced interatomic interaction affects the entanglement dynamics. More precisely, we determine whether the environment-induced interatomic interaction can help us distinguish the $\kappa$-deformed spacetime with an LDP and the Minkowski spacetime through the atomic entanglement dynamics.

To address the aforementioned issue, we analyze the evolution process of the entanglement of two atoms with $\rho_{AS}(0)=\rho_{SA}(0)\neq 0$. According to Eq.~(\ref{state1}), only the density elements, i.e., $\rho_{AS}$ and $\rho_{SA}$, are affected by the environment-induced interatomic interaction in the coupled basis. According to the previously presented calculation, the evolution time of the density matrix elements $\rho_{AS}(\tau)$ and $\rho_{SA}(\tau)$ can be calculated as follows:
\begin{eqnarray}\label{elements AS}
&&\rho_{AS}(\tau)=\rho_{AS}(0)e^{-4(A_1+iD)\tau},\;\;\;\;\;\;
\nonumber\\
&&\nonumber\\
&&\rho_{SA}(\tau)=\rho_{SA}(0)e^{-4(A_1-iD)\tau}.
\end{eqnarray}
By inserting Eq.~(\ref{elements AS}) into the definition of concurrence (\ref{concurrence0}), we derive the following expression:
\begin{eqnarray}\label{concurrence2}
C[\rho(\tau)]=\textrm{max}\{0,K_1(\tau)\},
\end{eqnarray}
where
\begin{eqnarray}\label{concurrence3}
K_1(\tau)&=&-2\sqrt{\rho_{GG}(\tau)\rho_{EE}(\tau)}+\bigg\{[\rho_{AA}(\tau)-\rho_{SS}(\tau)]^2\nonumber\\
&&+
[\rho_{AS}(0)e^{-4(A_1+iD)\tau}-\rho_{SA}(0)e^{-4(A_1-iD)\tau}]^2\bigg\}^{\frac{1}{2}}.
\end{eqnarray}
We find that there exists an extra term, i.e., $[\rho_{AS}(0)e^{-4(A_1+iD)\tau}-\rho_{SA}(0)e^{-4(A_1-iD)\tau}]^2$, in Eq.~(\ref{concurrence3}) because of the effects of the environment-induced interatomic interaction. Notably, the presence of this extra term may result in several intriguing physical properties, which will be discussed in detail in the subsequent sections.

\subsection{Entanglement dynamics of two static atoms}
To obtain the term of the environment-induced coupling between two static atoms, we insert Eqs.~(\ref{trajectories0})--(\ref{rest deformed Fourier0}) into Eq.~(\ref{Hilbert transform}). The Hilbert transforms of the correlation functions of two static atoms in $\kappa$-deformed spacetime are expressed as follows:
\begin{eqnarray}
&&\mathcal{K}^{12}(\omega)
=\frac{P}{\pi i}\int^\infty_{-\infty}d\omega'\frac{1}{\omega'-\omega}\frac{\omega'}{2\pi}\nonumber\\
&&\nonumber\\
&&\;\;\;\;\;\;\;\;\;\;\;\;\;\;\;\;\;\;\;\;\;\;\;\;\;\;\;\;\;\;\;\;\;
\times
\bigg[\frac{\sin \omega' L}{\omega' L}-\frac{\omega'^2 \cos \omega' L}{24\kappa^2}\bigg],\nonumber\\
&&\nonumber\\
&&\mathcal{K}^{12}(-\omega)
=\frac{P}{\pi i}\int^\infty_{-\infty}d\omega'\frac{1}{\omega'+\omega}\frac{\omega'}{2\pi}\nonumber\\
&&\nonumber\\
&&\;\;\;\;\;\;\;\;\;\;\;\;\;\;\;\;\;\;\;\;\;\;\;\;\;\;\;\;\;\;\;\;\;
\times\bigg[\frac{\sin \omega' L}{\omega' L}-\frac{\omega'^2 \cos \omega' L}{24\kappa^2}\bigg].
\end{eqnarray}
Then, from Eqs.~(\ref{D}) and (\ref{D1}), we derive the following expression:
\begin{eqnarray}\label{DK1}
&&D=\Gamma_0\frac{1}{2\pi}P\int^\infty_{0}dx \bigg[\frac{x}{x
-1}+\frac{x}{x+1}\bigg]\nonumber\\
&&\nonumber\\
&&\;\;\;\;\;\;\;\;\;\;\;\;\;\;\;\;\;\;\;\;\;\;\;\;\;\;\;\;\;\;\;\;\;
\times\bigg[\frac{\sin x \omega L}{x \omega L}-\frac{x^2 \cos x \omega L}{24(\frac{\kappa}{\omega})^2}\bigg].
\end{eqnarray}
Similarly, for the case of two static atoms in Minkowski spacetime, we derive the following expression:
\begin{eqnarray}\label{DM1}
D=\Gamma_0\frac{1}{2\pi}P\int^\infty_{0}dx \bigg[\frac{x}{x
-1}+\frac{x}{x+1}\bigg]\frac{\sin x \omega L}{x \omega L}.
\end{eqnarray}
Note that, for the case $\kappa\rightarrow\infty$, the result obtained using Eq.~(\ref{DK1}) recovers to that obtained using Eq.~(\ref{DM1}) for the Minkowski spacetime case.

\subsubsection{Two static atoms initially prepared in a separable state $|10\rangle$}
We analyze how the environment-induced interatomic interaction affects the entanglement dynamics of two atoms initially prepared in state $|10\rangle$. In this case, the extra term in Eq.~(\ref{concurrence3}) can be written as follows:
\begin{eqnarray}
[\rho_{AS}(0)e^{-4(A_1+iD)\tau}-\rho_{SA}(0)e^{-4(A_1-iD)\tau}]^2
=\sin^2(4D\tau)e^{-8A_1\tau}.\nonumber\\
\end{eqnarray}

\begin{figure}[H]
\centering
{\includegraphics[height=1.65in,width=2.65in]{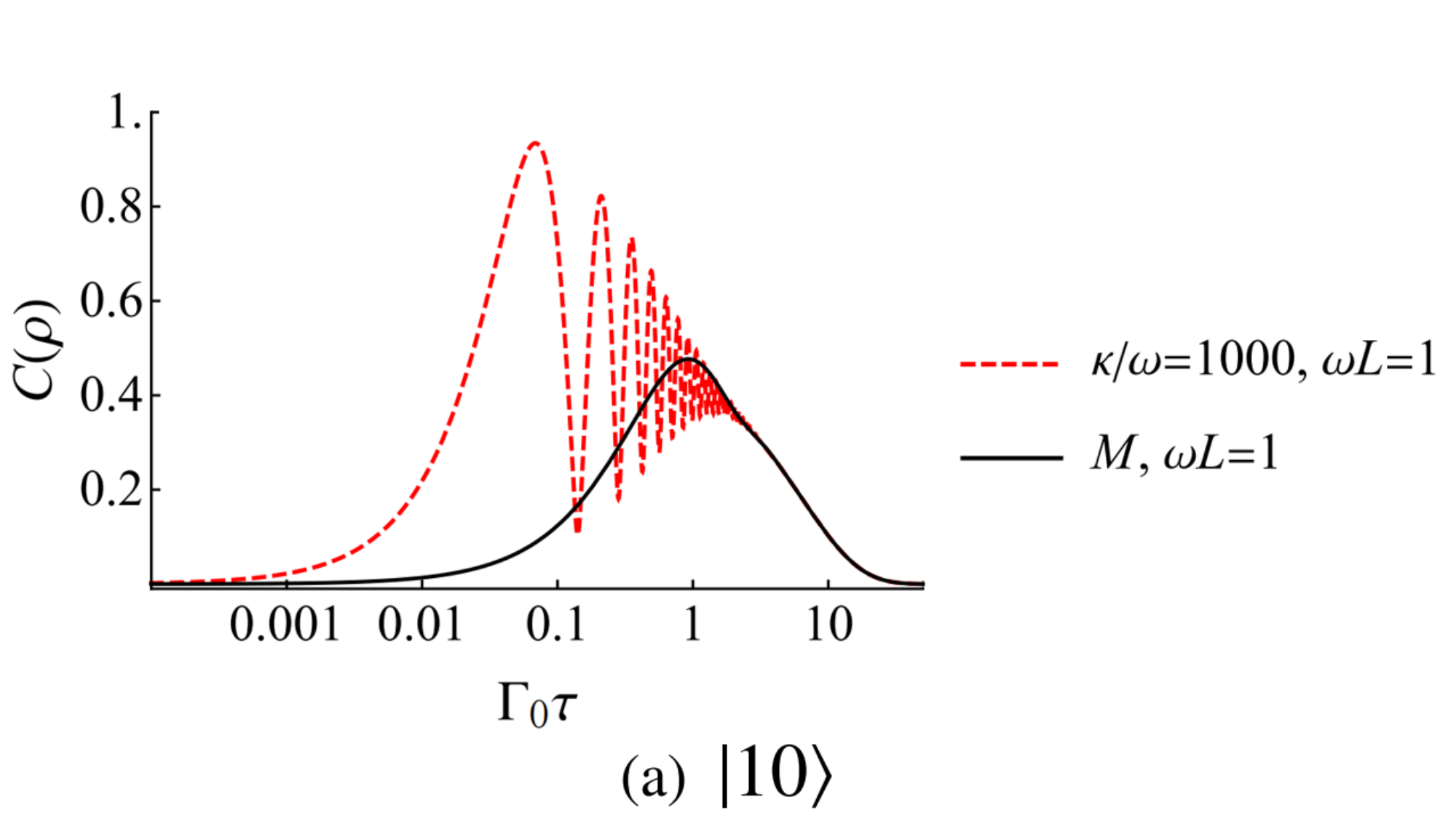}}
{\includegraphics[height=1.65in,width=2.65in]{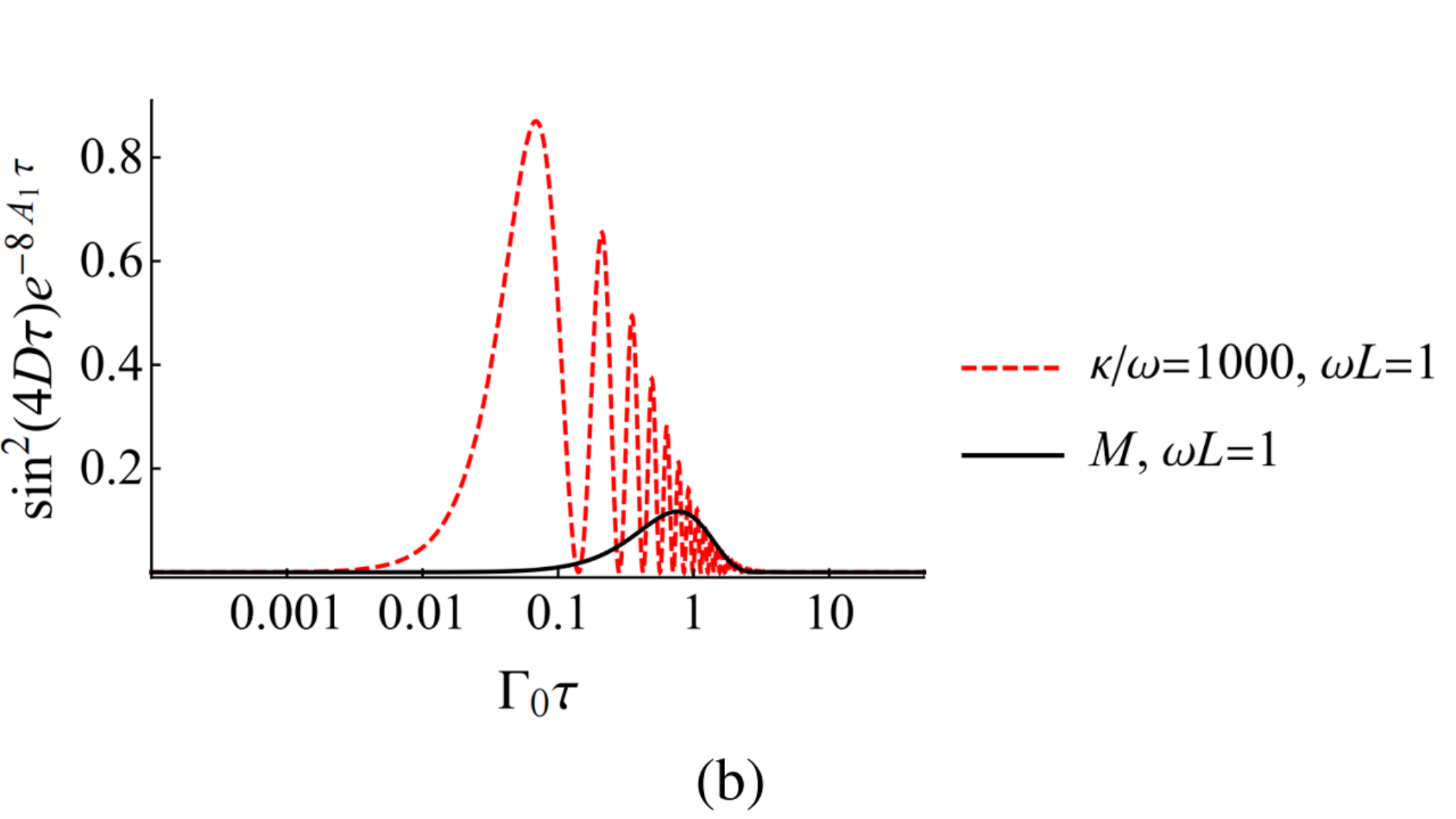}}
\caption{(a) Time evolution of concurrence of two static atoms initially prepared in $|10\rangle$. (b) Factor $\sin^2(4D\tau)e^{-8A_1\tau}$ as a function of time.
}\label{kerest1}
\end{figure}

In Fig.~\ref{kerest1}, we plot the evolution process of concurrence of two static atoms initially prepared in state $|10\rangle$ in $\kappa$-deformed spacetime with an LDP and Minkowski spacetime with fixed $\kappa/\omega=1000$. As shown in Fig.~\ref{kerest1} (a), the environment-induced interatomic interaction has a significant impact on concurrence during the initial period in $\kappa$-deformed spacetime. However, after a long time, the asymptotic concurrence of the $\kappa$-deformed spacetime case nearly coincides with that of the Minkowski spacetime case. This interesting behavior is caused by the parameter term $\sin^2(4D\tau)e^{-8A_1\tau}$ expressed in Eq.~(\ref{concurrence3}), which is dominated by the trigonometric term $\sin^2(4D\tau)$ in a short initial time and is determined by the exponential term $e^{-8A_1\tau}$ after a sufficiently long time [Fig.~\ref{kerest1} (b)]. Therefore, as a result of the environment-induced interatomic interaction, we find that, for the $\kappa$-deformed spacetime case, the entanglement generated evolves periodically in the short initial time and eventually decays to zero asymptotically. This interesting phenomenon is different from that in the Minkowski spacetime case. In this sense, the environment-induced interatomic interaction between two static atoms can help us distinguish these two universes in principle.

\subsubsection{Two static atoms initially prepared in entangled state $|\psi\rangle=\sqrt{p}|A\rangle+\sqrt{1-p}|S\rangle$}
We will investigate the entanglement behavior of two static atoms initially prepared in the entangled state $|\psi\rangle=\sqrt{p}|A\rangle+\sqrt{1-p}|S\rangle$ $(0<p<1, p\neq 1/2)$ under the effects of the environment-induced interatomic interaction. We note here that different $p$ denote the different weights of the symmetric and antisymmetric entangled states to construct the initial entangled state. In this case, we can calculate the extra term in Eq.~(\ref{concurrence3}) as follows:
\begin{eqnarray}
&&[\rho_{AS}(0)e^{-4(A_1+iD)\tau}-\rho_{SA}(0)e^{-4(A_1-iD)\tau}]^2\nonumber\\
&&\nonumber\\
&&\;\;\;\;\;\;\;\;\;\;\;\;\;\;\;\;\;\;\;\;\;\;\;\;\;\;\;\;
=4p(1-p)\sin^2(4D\tau)e^{-8A_1\tau}.
\end{eqnarray}

\begin{figure}[H]
\centering
{\includegraphics[height=1.45in,width=2.25in]{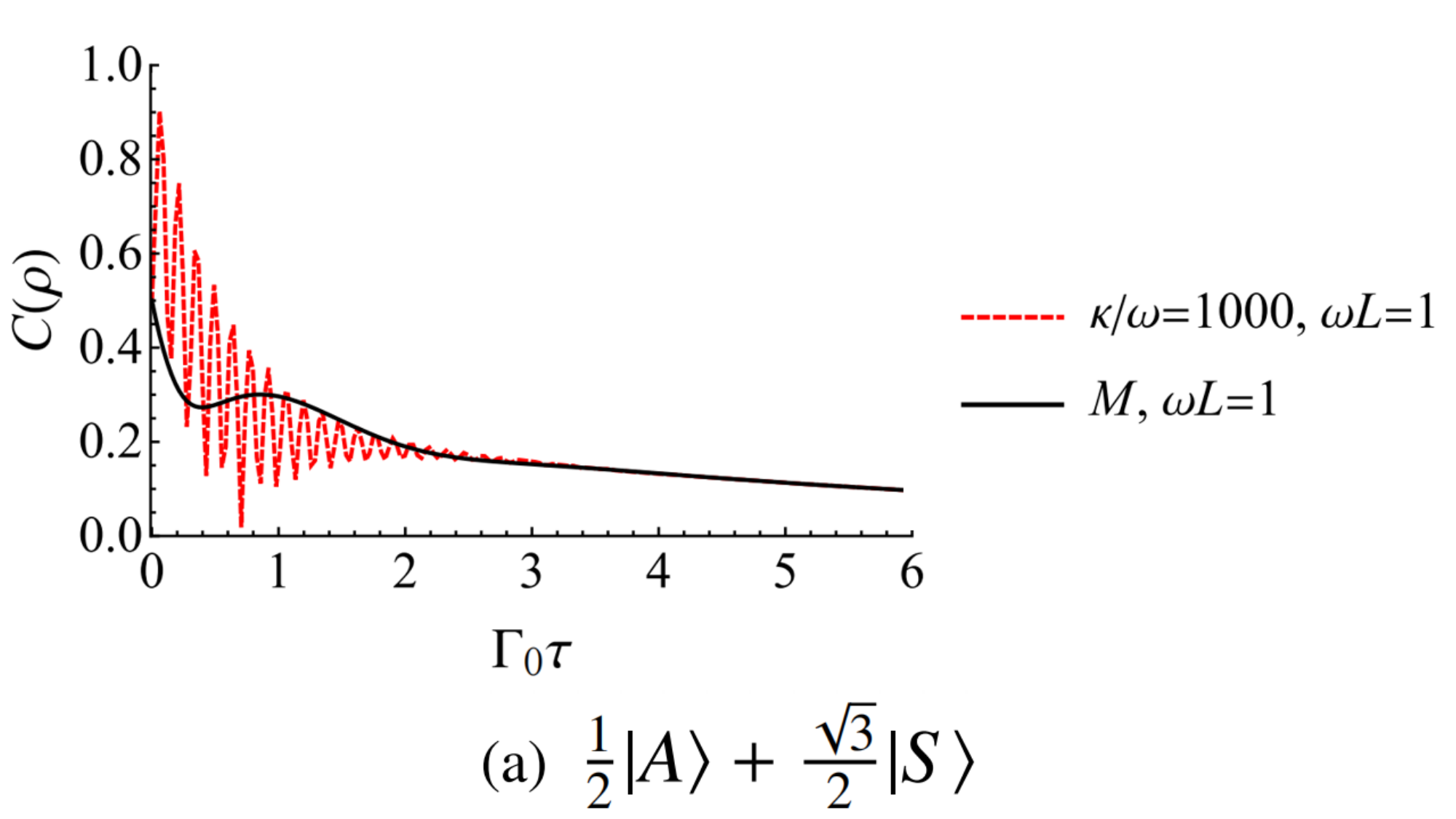}}
{\includegraphics[height=1.45in,width=2.25in]{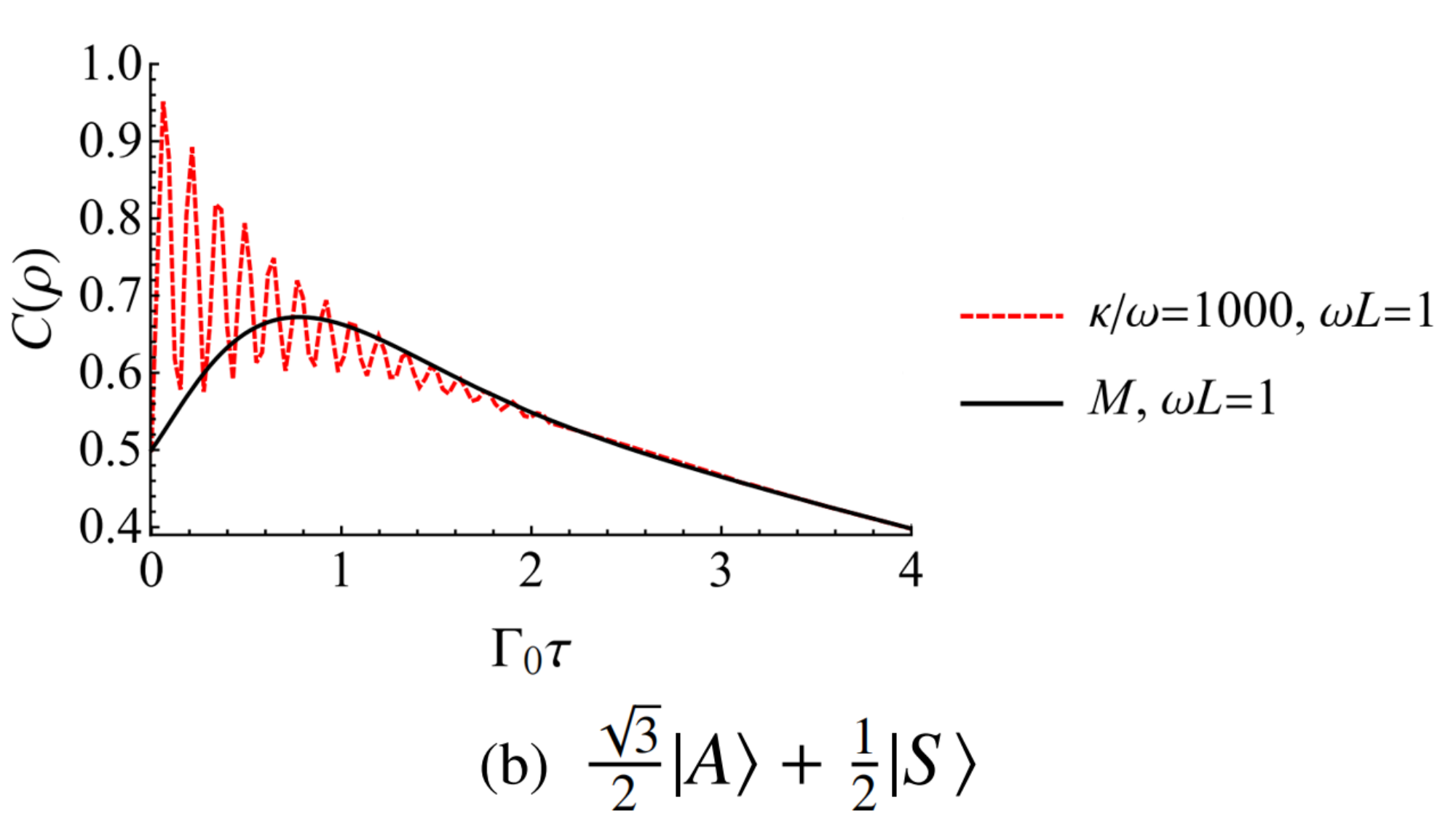}}
{\includegraphics[height=1.45in,width=2.25in]{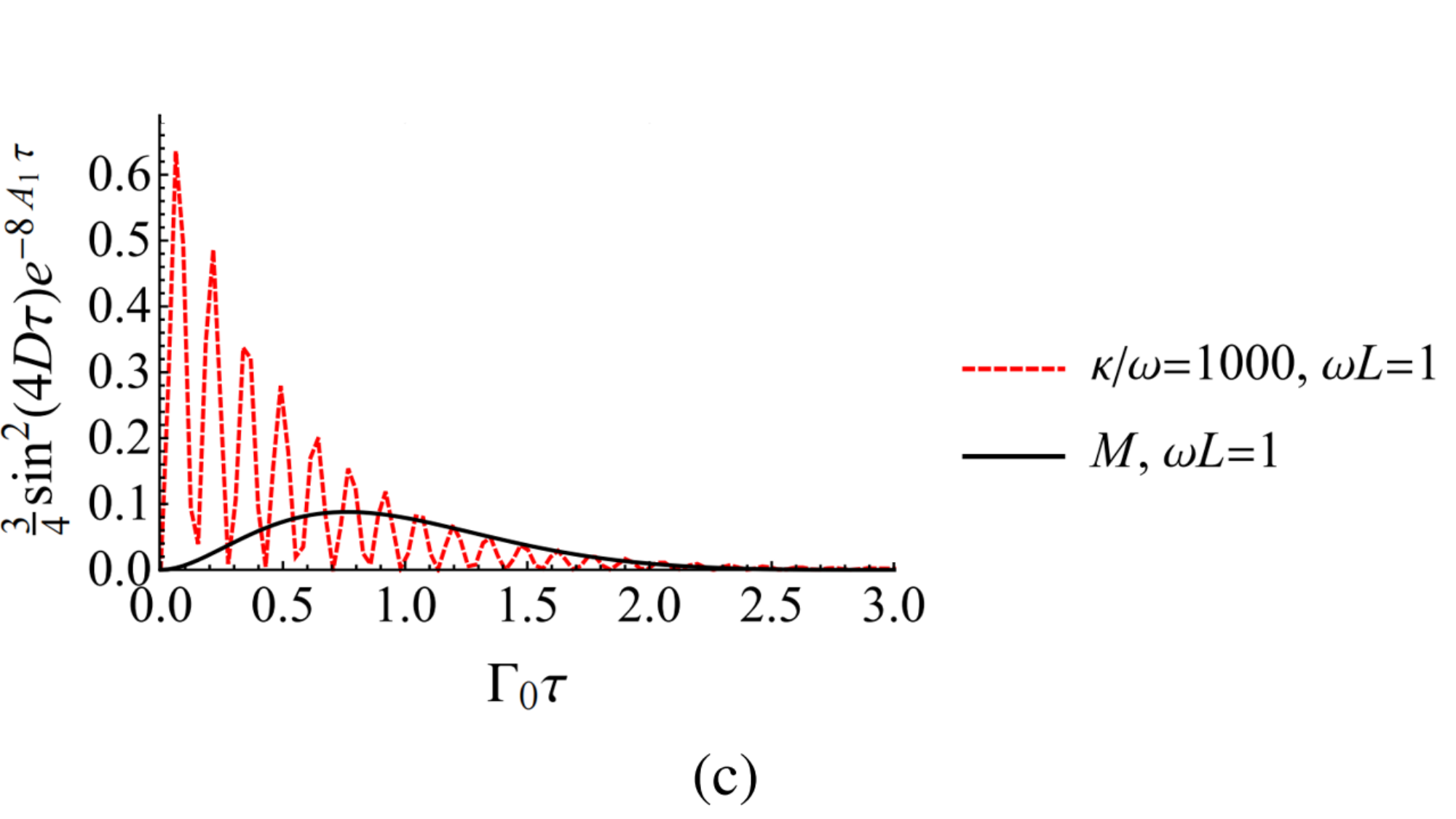}}
\caption{Time evolution of concurrence of static atoms initially prepared in $\frac{1}{2}|A\rangle+\frac{\sqrt{3}}{2}|S\rangle$ (a) and $\frac{\sqrt{3}}{2}|A\rangle+\frac{1}{2}|S\rangle$ (b). (c) The factor $\frac{3}{4}\sin^2(4D\tau)e^{-8A_1\tau}$ is a function of time.
}\label{R5}
\end{figure}
For $p=1/4$ [Fig.~\ref{R5} (a)], which means that the symmetric entangled state mainly contributes to the initial entangled state, we find that the concurrence of the Minkowski spacetime case behaves differently compared with those of the initial entangled states $|A\rangle$ and $|S\rangle$ cases (discussed previously). Although this entanglement decays with the increase in time, it is non-monotonous as a result of the environment-induced interatomic interaction. Moreover, in the early period, this concurrence shows an oscillatory behavior in $\kappa$-deformed spacetime with an LDP compared with that in Minkowski spacetime under the effects of environment-induced interatomic interaction. However, with a sufficiently long time limit, the entanglement behavior of atoms in the $\kappa$-deformed spacetime case nearly coincides with that in the Minkowski spacetime case, which indicates that all of the laws of physics in $\kappa$-deformed spacetime with an LDP recover to that in flat spacetime. For $p=3/4$ [Fig.~\ref{R5} (b)], which means that the antisymmetric entangled state mainly contributes to the initial entangled state. We also find that, with the environment-induced interatomic interaction, the entanglement dynamics of $\kappa$-deformed spacetime with an LDP behaves differently from that of Minkowski spacetime. We note that this difference in behavior is ultimately due to the trigonometric term $\sin^2(4D\tau)$ in a short initial time. Meanwhile, after a sufficiently long time, the entanglement behavior is dominated by the exponential term $e^{-8A_1\tau}$ [Fig.~\ref{R5} (c)].

\subsection{Entanglement dynamics of two uniformly moving atoms}
Now, we reconsider the aforementioned case while assuming that the two atoms move with a uniform speed. When the two-atom system moves uniformly in $\kappa$-deformed spacetime, by inserting Eqs.~(\ref{trajectories1})--(\ref{deformed Fourier0}) into Eq.~(\ref{Hilbert transform}), we can calculate the corresponding Hilbert transform of the correlation function as follows:
\begin{eqnarray}\label{KKD1}
\mathcal{K}^{12}(\omega)
&=&\frac{P}{\pi i}\int^\infty_{-\infty}d\omega'\frac{1}{\omega'-\omega}\frac{\omega'}{2\pi}\nonumber\\
&&\nonumber\\
&&\;\;\;\;\;\;\;\;\;\;\;\;\;\;\;\;\times\bigg[\frac{\sin \omega' L}{\omega' L}+\frac{ f(\omega',L,v)}{24\omega'\kappa^2 L^3}\bigg],\nonumber\\
&&\nonumber\\
\mathcal{K}^{12}(-\omega)
&=&\frac{P}{\pi i}\int^\infty_{-\infty}d\omega'\frac{1}{\omega'+\omega}\frac{\omega'}{2\pi}\nonumber\\
&&\nonumber\\
&&\;\;\;\;\;\;\;\;\;\;\;\;\;\;\;\;\times\bigg[\frac{\sin \omega' L}{\omega' L}+\frac{ f(\omega',L,v)}{24\omega'\kappa^2 L^3}\bigg].
\end{eqnarray}
From Eq.~\eqref{KKD1}, we derive the following expression:
\begin{eqnarray}\label{Dku}
D&=&\Gamma_0\frac{1}{2\pi}P\int^\infty_{0}dx \bigg[\frac{x}{x
-1}+\frac{x}{x+1}\bigg]\nonumber\\
&&\nonumber\\
&&\;\;\;\;\;\;\;\;\;\;\;\;\;\;\;\;\times
\bigg[\frac{\sin x \omega L}{x \omega L}+\frac{ x^2f(x,\omega L,v)}{24(\frac{\kappa}{\omega})^2 (x\omega L)^3}\bigg].
\end{eqnarray}
Similarly, for two uniformly moving atoms in Minkowski spacetime, we derive the following expression:
\begin{eqnarray}\label{DMu}
D=\Gamma_0\frac{1}{2\pi}P\int^\infty_{0}dx \bigg[\frac{x}{x
-1}+\frac{x}{x+1}\bigg]\frac{\sin x \omega L}{x \omega L},
\end{eqnarray}
which is the same as that in the case of two static atoms in Minkowski spacetime and is completely unaffected by velocity.

\subsubsection{Two uniformly moving atoms initially prepared in a separable state $|10\rangle$}
\begin{figure}[H]
\centering
{\includegraphics[height=1.65in,width=2.65in]{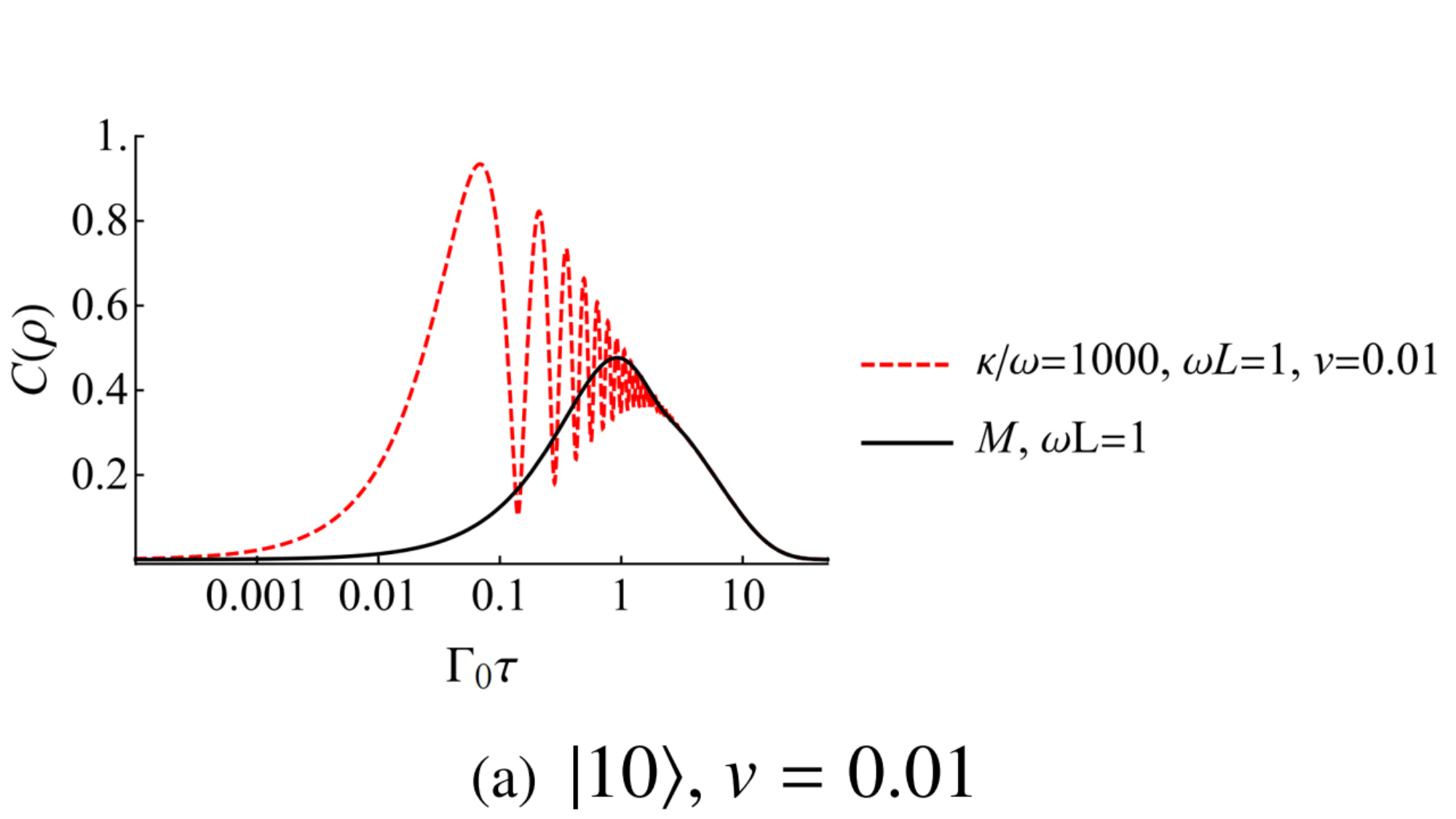}}
{\includegraphics[height=1.65in,width=2.65in]{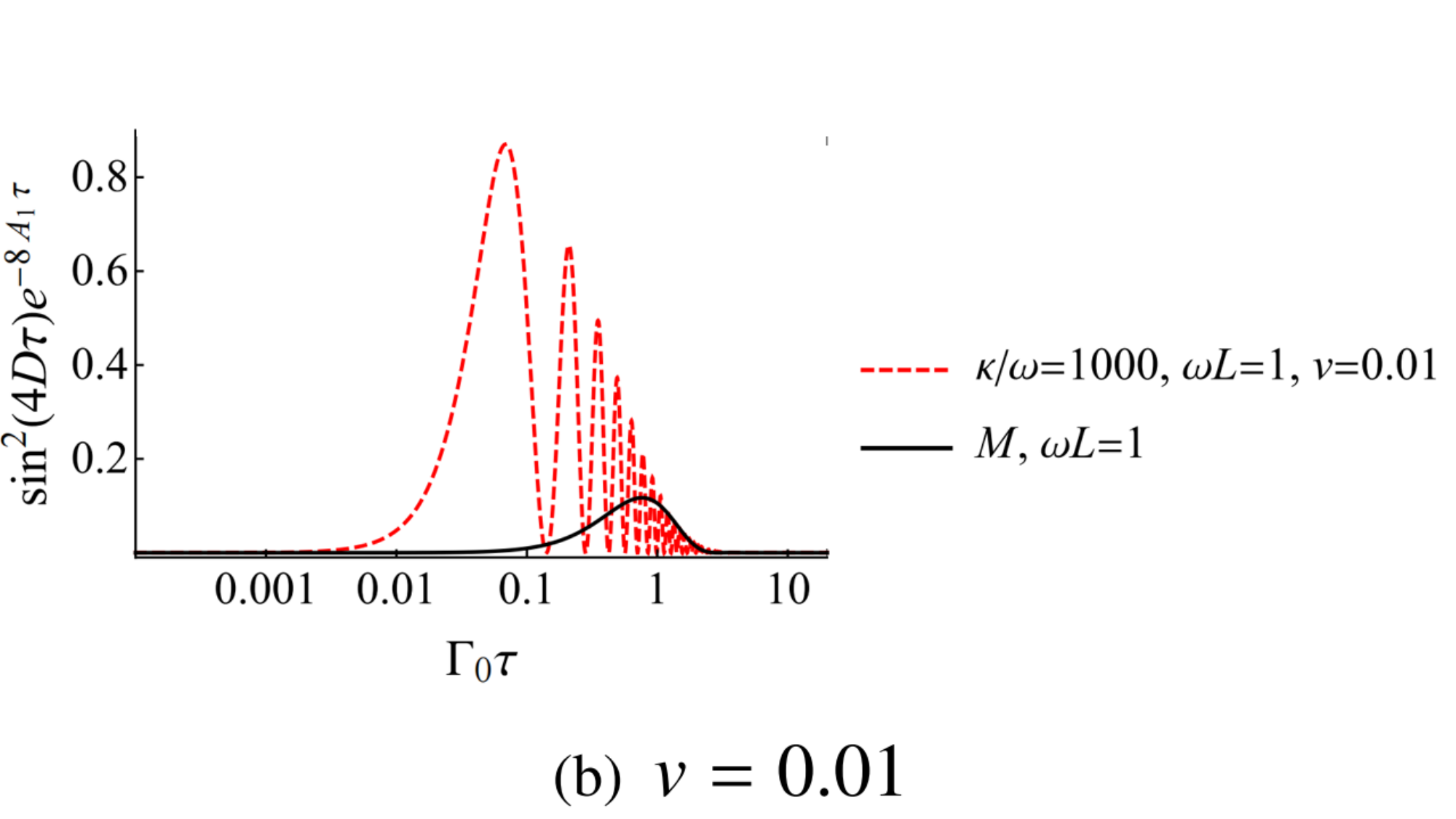}}\\
{\includegraphics[height=1.65in,width=2.65in]{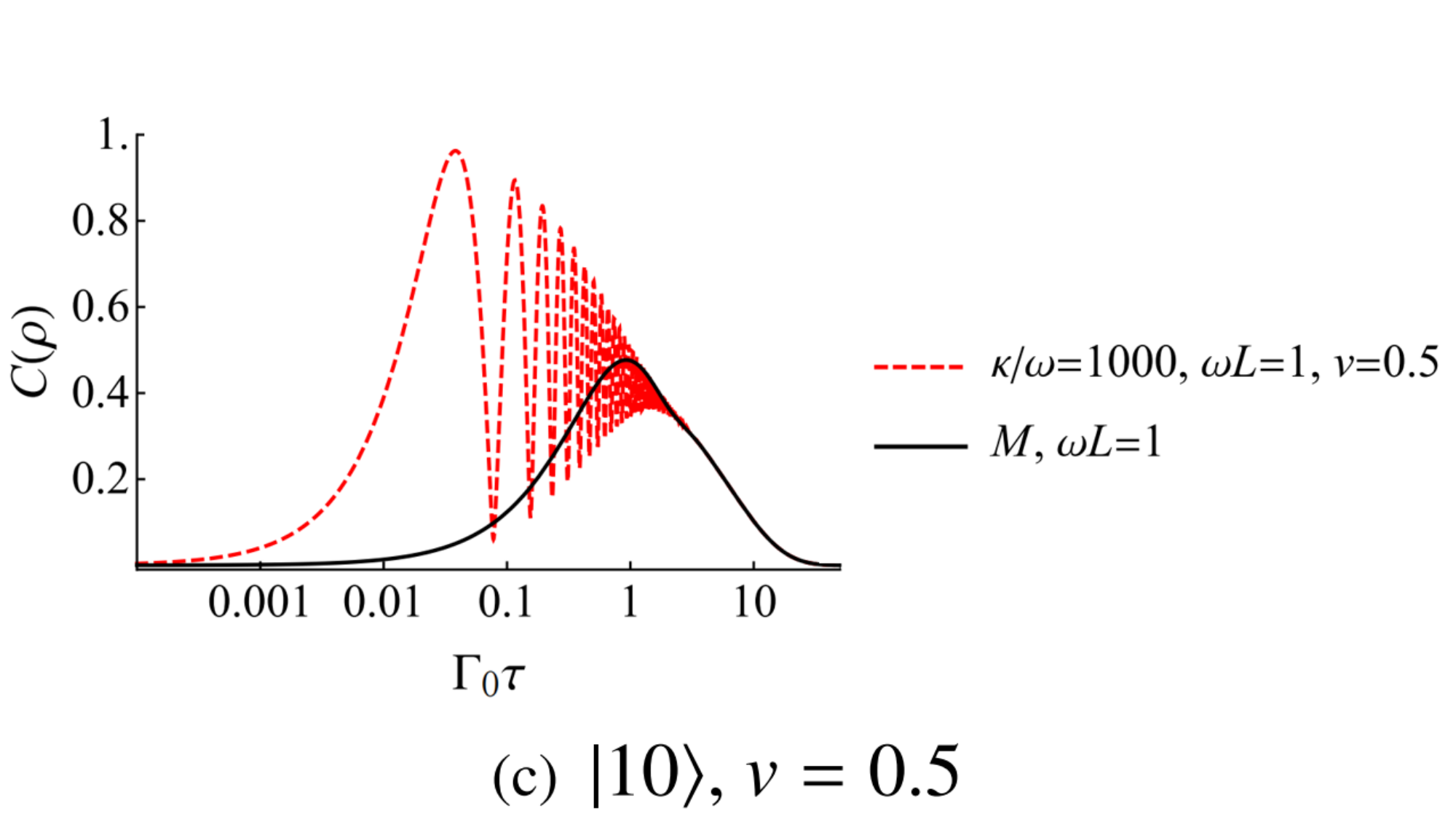}}
{\includegraphics[height=1.65in,width=2.65in]{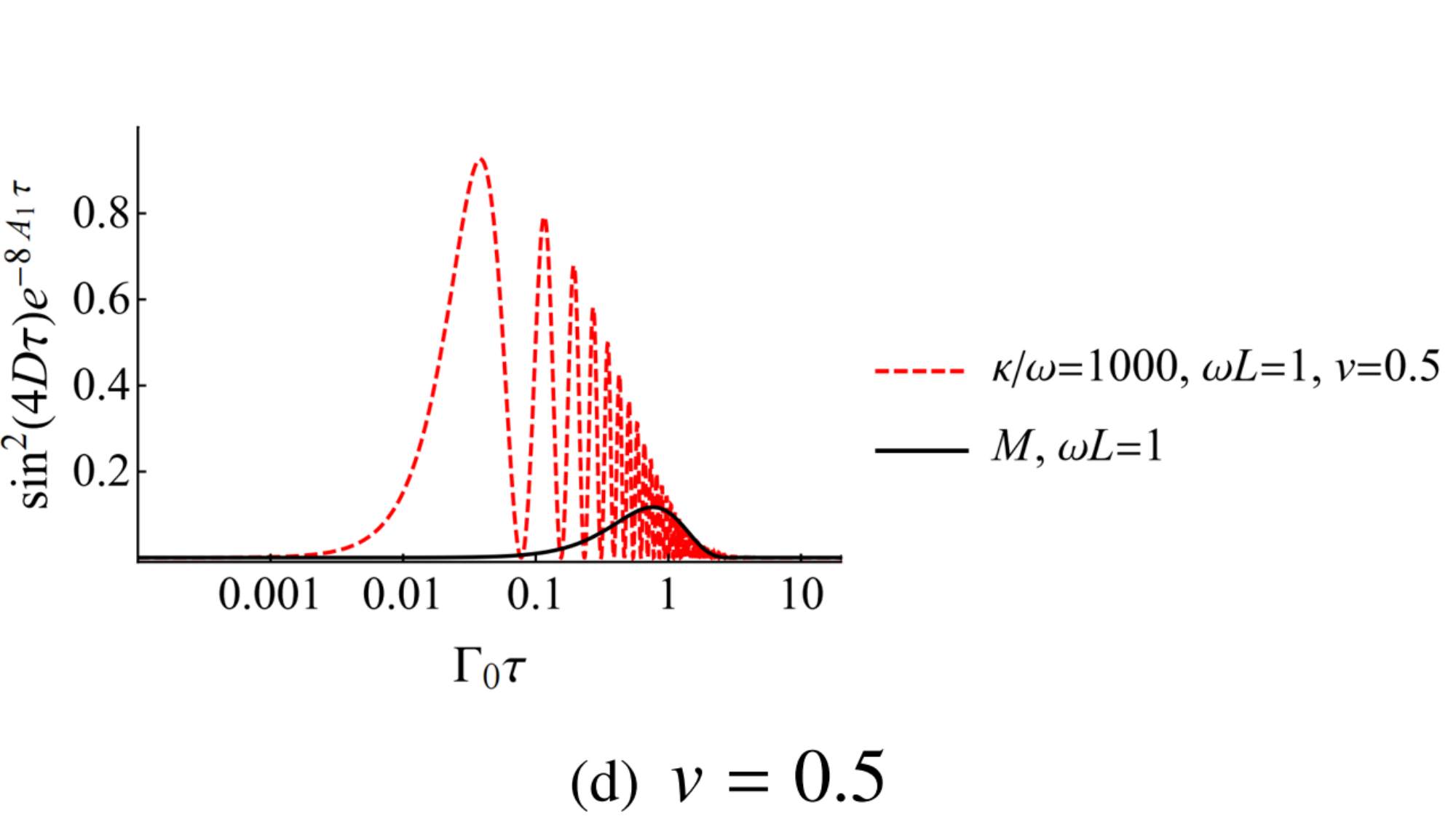}}
\caption{Time evolution of concurrence of atoms initially prepared in $|10\rangle$ with the velocities $v=0.01$ (a) and $v=0.5$ (c). The factor $\sin^2(4D\tau)e^{-8A_1\tau}$ is a function of time with the velocities $v=0.01$ (b) and $v=0.5$ (d).
}\label{kev1-2}
\end{figure}

To analyze how the environment-induced interatomic interaction of two uniformly moving atoms affects the entanglement generation in two different spacetimes, we show the dynamics of concurrence with different velocities in Fig.~\ref{kev1-2} for fixed $\kappa/\omega=1000$ and $\omega L=1$. When the environment-induced interatomic interaction is introduced, the oscillatory manner of the time evolution of concurrence during the initial period in $\kappa$-deformed spacetime is observed, and this oscillation frequency increases as the velocity of the atoms increases. In addition, the oscillation is damped during evolution; thus, the asymptotic concurrence will be consistent with the entanglement behavior of the Minkowski spacetime case. We note that this oscillatory behavior is dominated by the trigonometric term $\sin^2(4D\tau)$ during the initial stage, is determined by the exponential term $e^{-8A_1\tau}$, and decays to the Minkowski spacetime case after a long time. Hence, when the environment-induced interatomic interaction is considered, the differences in the entanglement dynamics between the $\kappa$-deformed spacetime with an LDP and Minkowski spacetime are more pronounced during the initial period. Therefore, the environment-induced interatomic interaction between two uniformly moving atoms can assist us in distinguishing these two universes.

\subsubsection{Two uniformly moving atoms initially prepared in entangled state $|\psi\rangle=\sqrt{p}|A\rangle+\sqrt{1-p}|S\rangle$}
\begin{figure}[H]
\centering
{\includegraphics[height=1.65in,width=2.65in]{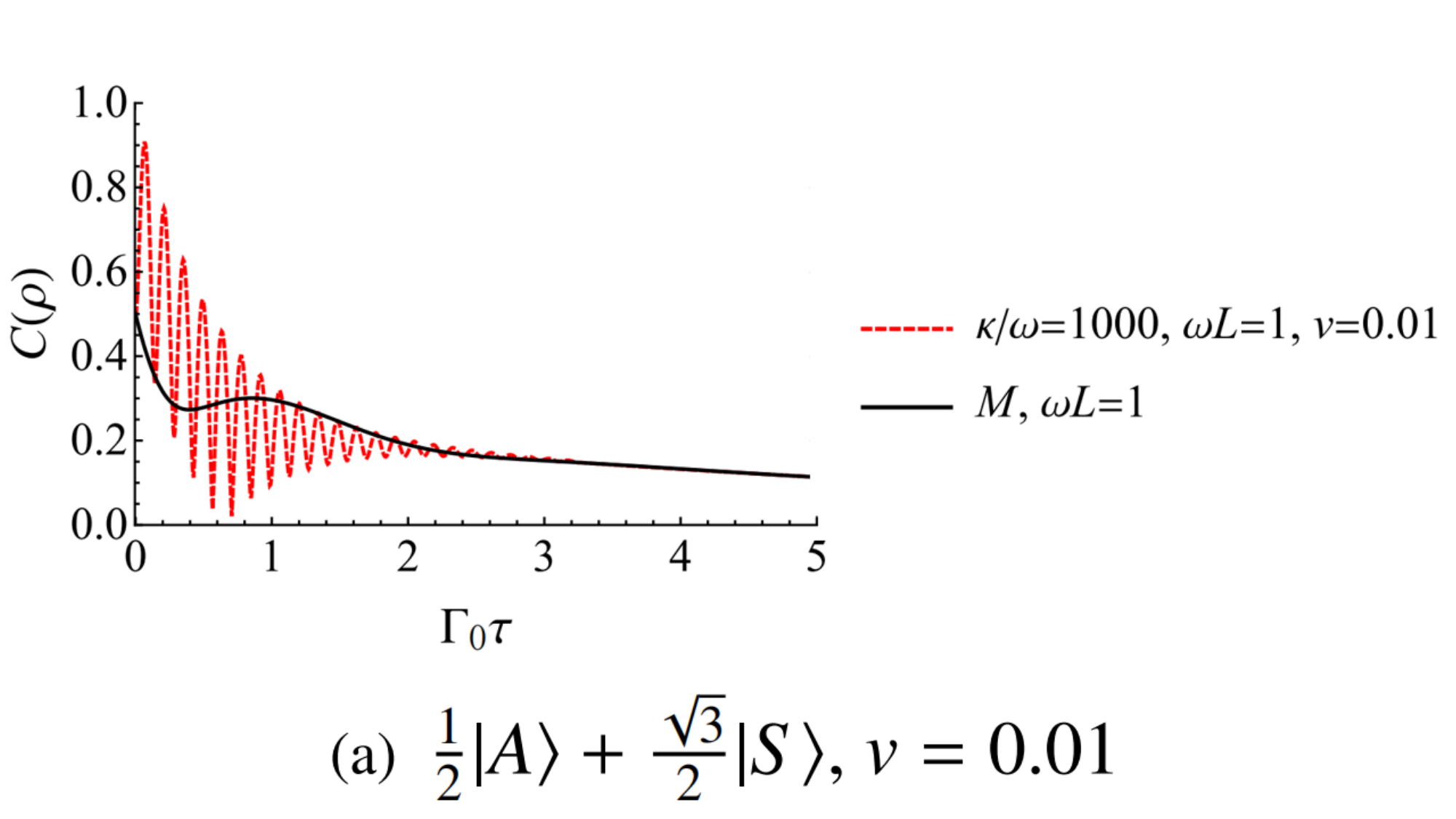}}
{\includegraphics[height=1.65in,width=2.65in]{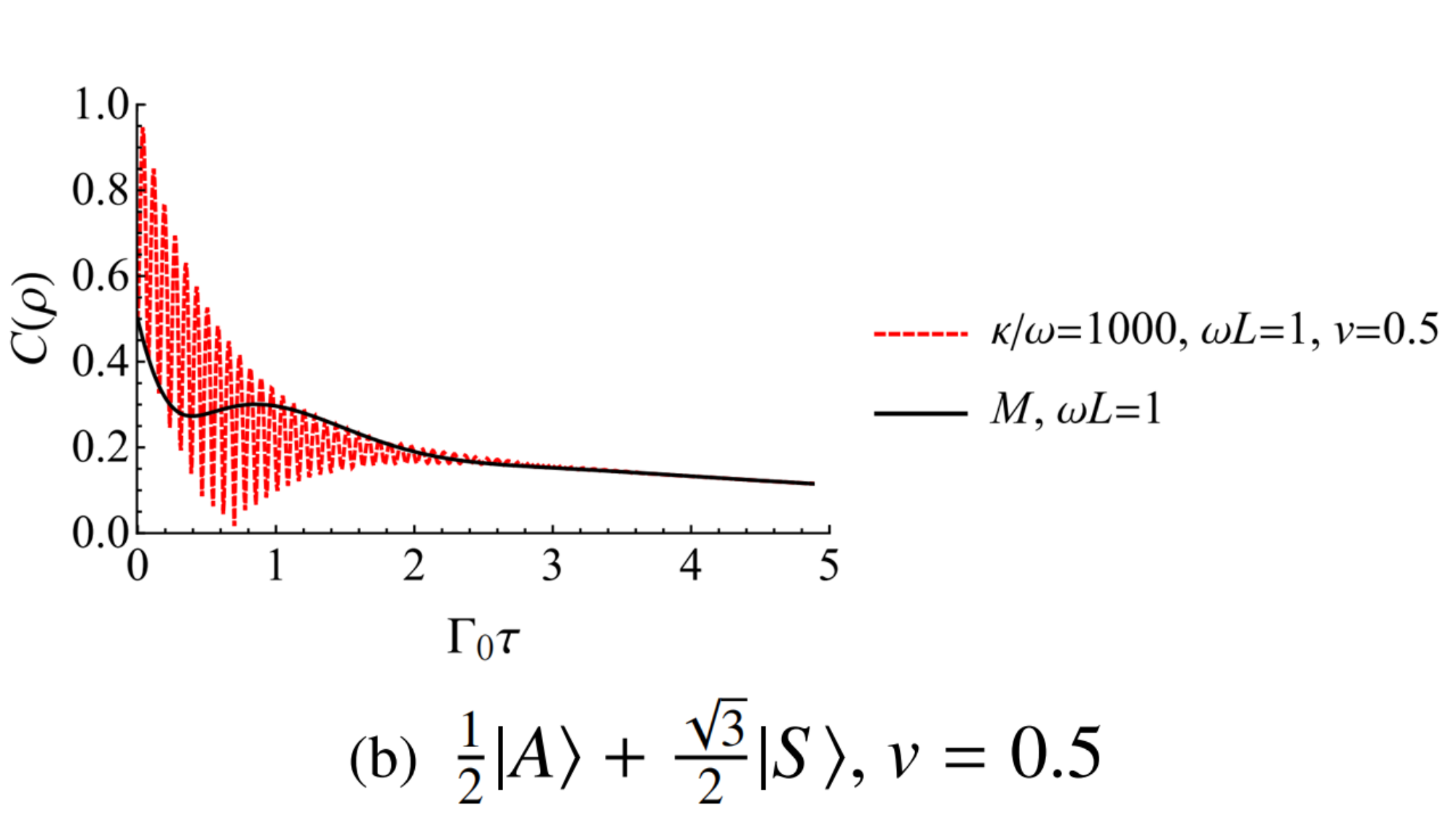}}
{\includegraphics[height=1.65in,width=2.65in]{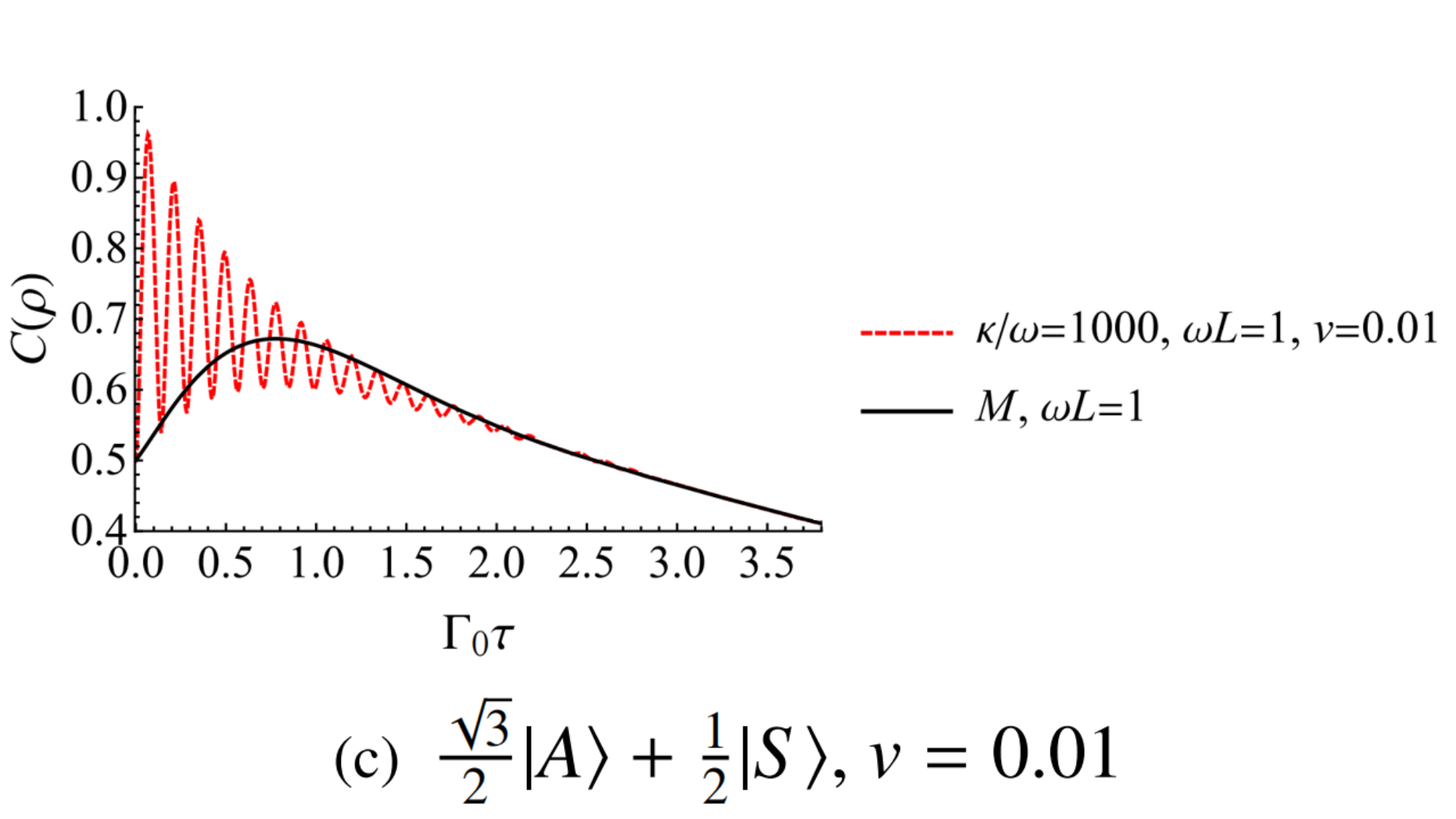}}
{\includegraphics[height=1.65in,width=2.65in]{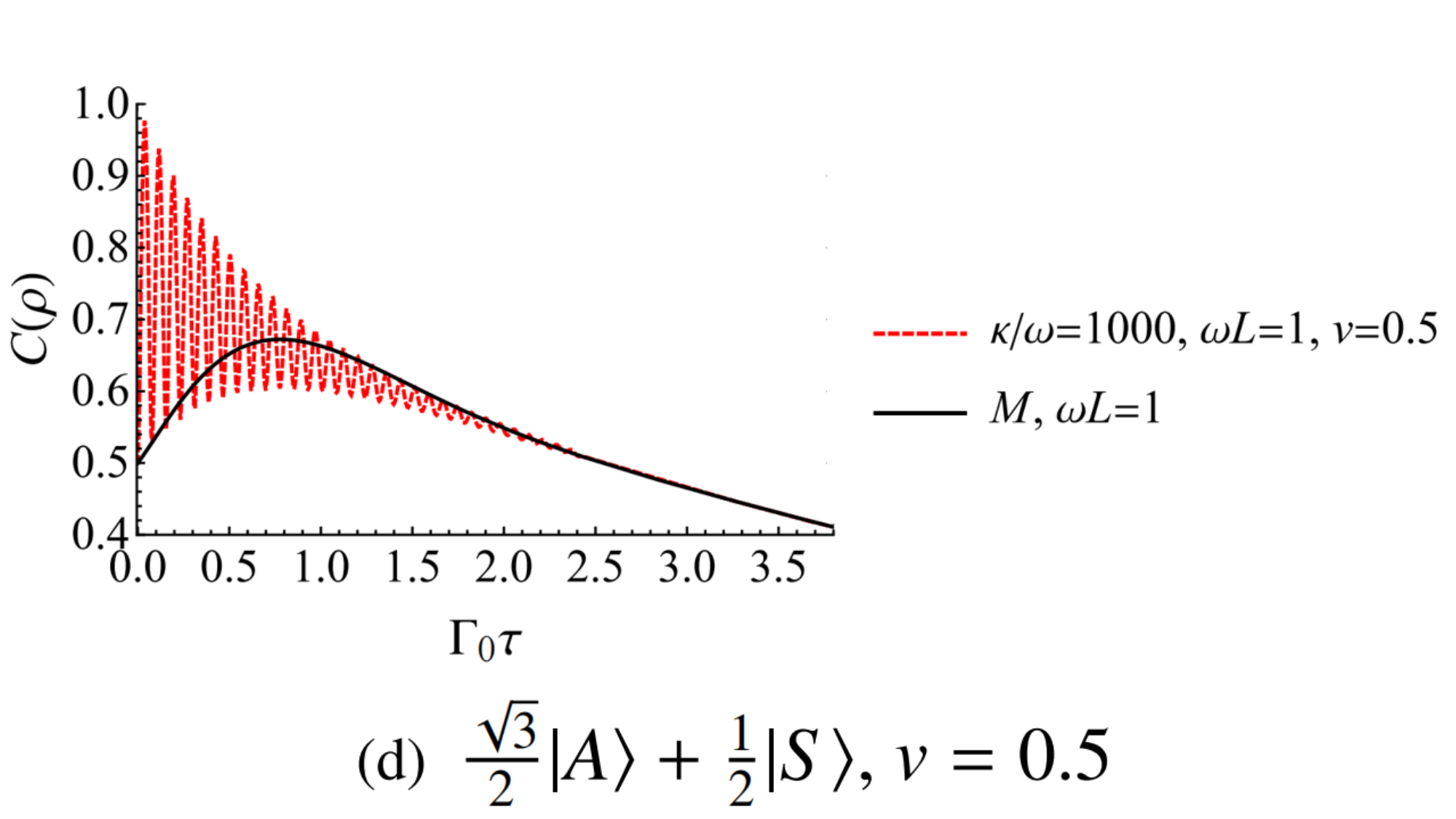}}
{\includegraphics[height=1.65in,width=2.65in]{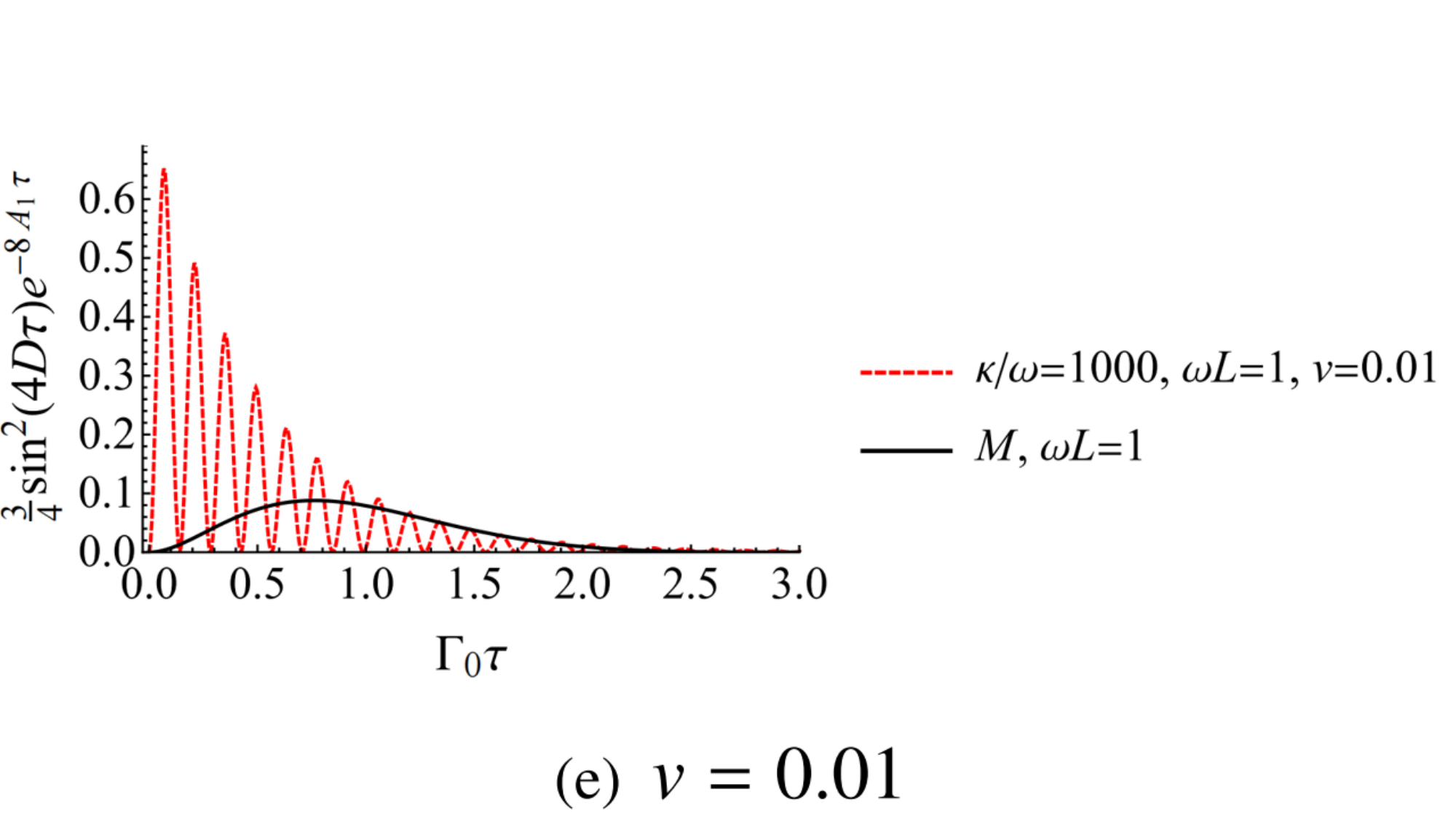}}
{\includegraphics[height=1.65in,width=2.65in]{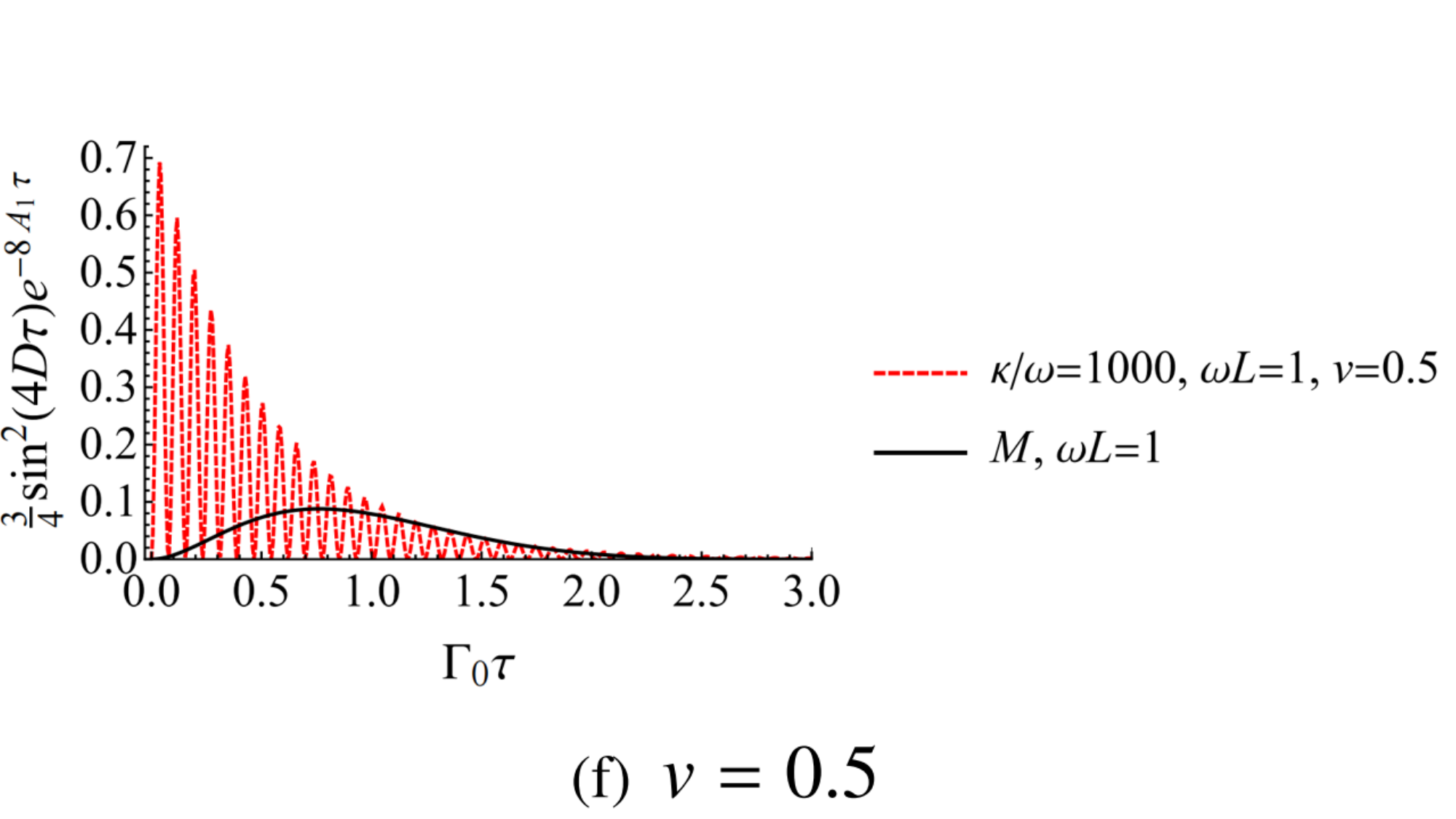}}
\caption{Time evolution of concurrence of two uniformly moving atoms initially prepared in $\frac{1}{2}|A\rangle+\frac{\sqrt{3}}{2}|S\rangle$ for different values $v=0.01$ (a) and $v=0.5$ (b), and $\frac{\sqrt{3}}{2}|A\rangle+\frac{1}{2}|S\rangle$ for different values $v=0.01$ (c) and $v=0.5$ (d). The factor $\sin^2(4D\tau)e^{-8A_1\tau}$ is a function of time with the velocities $v=0.01$ (e) and $v=0.5$ (f).
}\label{f4}
\end{figure}

We consider the scenario where the initial state of two uniformly moving atoms is prepared in the superposition state $|\psi\rangle=\sqrt{p}|A\rangle+\sqrt{1-p}|S\rangle$ $(0<p<1, p\neq 1/2)$, which causes the environment-induced interatomic interaction. In Figs.~\ref{f4} (a) and (b), we take $p=1/4$ and plot the relevant entanglement dynamics of atoms with different fixed velocities. One can see that the concurrence decays with the increase in evolution time, and the concurrence decays differently for these two spacetimes. Moreover, the oscillatory decay still dominates the entanglement behavior of the $\kappa$-deformed spacetime case in the early stage of evolution. Meanwhile, the oscillatory frequency is dependent on velocity, which increases with the increase in the atomic velocity. In the late stage of evolution, the entanglement in the $\kappa$-deformed spacetime case shares the same behavior as that in the Minkowski spacetime case. Furthermore, we prepare the $p=3/4$ initial state case and plot its entanglement dynamics in Figs.~\ref{f4} (c) and (d). Notably, in the early stage of evolution, the oscillatory manner of entanglement is also dependent on the atomic velocity, but they are different from the $p=1/4$ case. Similarly, in the late stage of evolution, one cannot distinguish these two spacetimes through the atomic entanglement behavior. Remarkably, as shown in Figs.~\ref{f4} (e) and (f), the oscillatory behavior of entanglement appears as a consequence of the environment-induced interatomic interaction, which is dominated by the trigonometric term $\sin^2(4D\tau)$ during the initial stage and is determined by the exponential term $e^{-8A_1\tau}$ after a long time.

Compared with that shown in Fig.~\ref{R5}, Fig.~\ref{f4} shows that, under the influence of velocity, the behavior of entanglement in $\kappa$-deformed spacetime is different from that in Minkowski spacetime: the higher the velocity is, the greater the difference in the entanglement behavior between these two universes. This finding indicates that, even when the spacetime deformation parameter $\kappa$ is relatively large, in principle, we can more easily distinguish these two spacetimes with the help of environment-induced interatomic interaction between two uniformly moving atoms.

\subsection{Entanglement dynamics of two circularly accelerated atoms}
In this section, we investigate how the entanglement dynamics between two circularly accelerated atoms are dependent on the environment-induced interatomic interaction in $\kappa$-deformed and Minkowski spacetimes. With the trajectories of the two circularly accelerated atoms~(\ref{trajectories2}) and the Fourier transform of the correlation function, we can straightforwardly derive an analytical expression for the corresponding Hilbert transform in Eq.~(\ref{D1}). The explicit forms are given in Supplemental Material in detail [Eqs.~(S6) and (S7)]. Similarly, when the two circularly accelerated atoms interact with a bath of a fluctuating massless scalar field in a Minkowski vacuum, we can also straightforwardly calculate the corresponding Hilbert transform for the uniform circular motion case [Eq.~(S8) in Supplemental Material].

\subsubsection{Two circularly accelerated atoms initially prepared in a separable state $|10\rangle$}
\begin{figure}[H]
\centering
{\includegraphics[height=1.45in,width=2.25in]{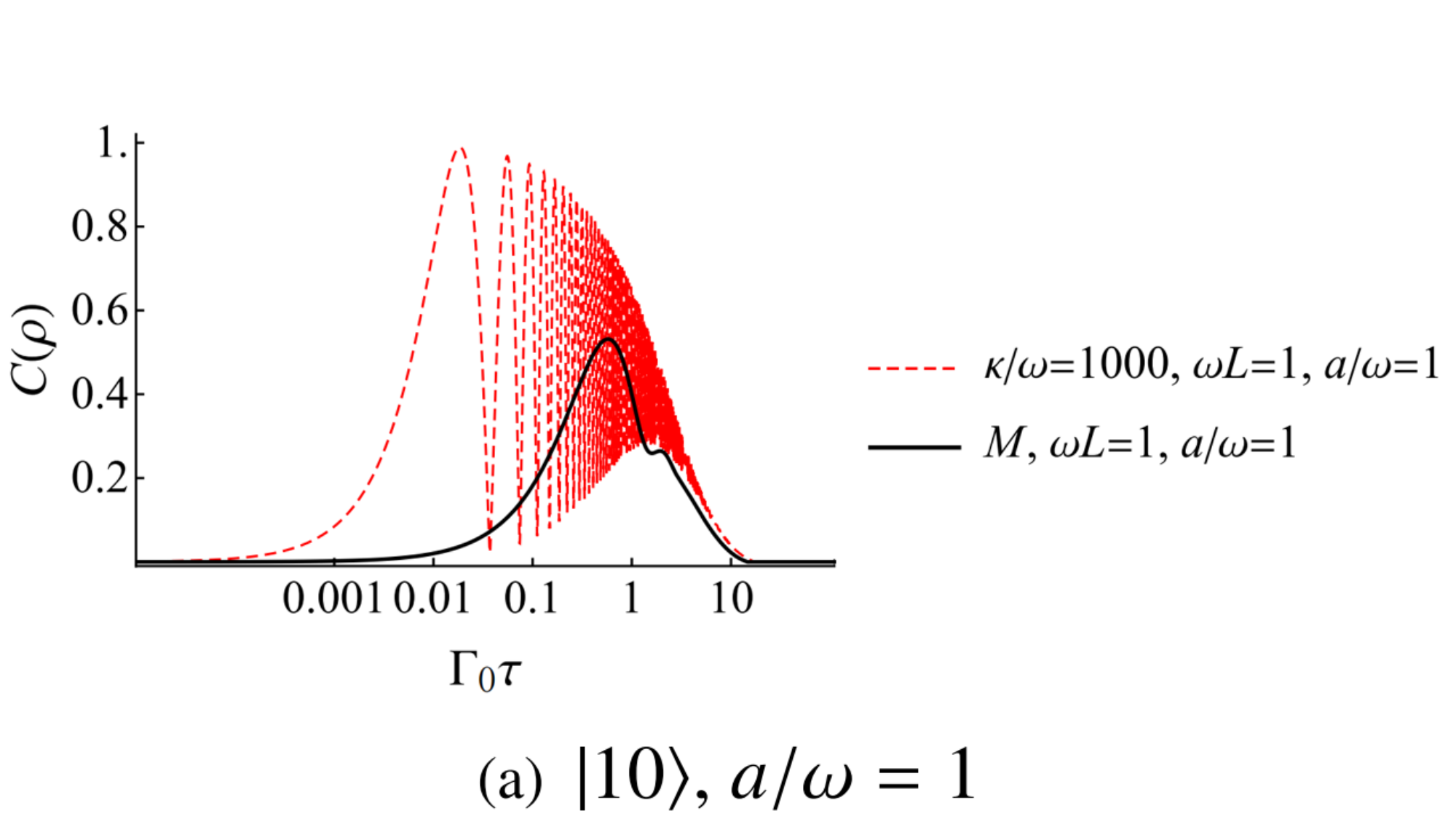}}
{\includegraphics[height=1.45in,width=2.25in]{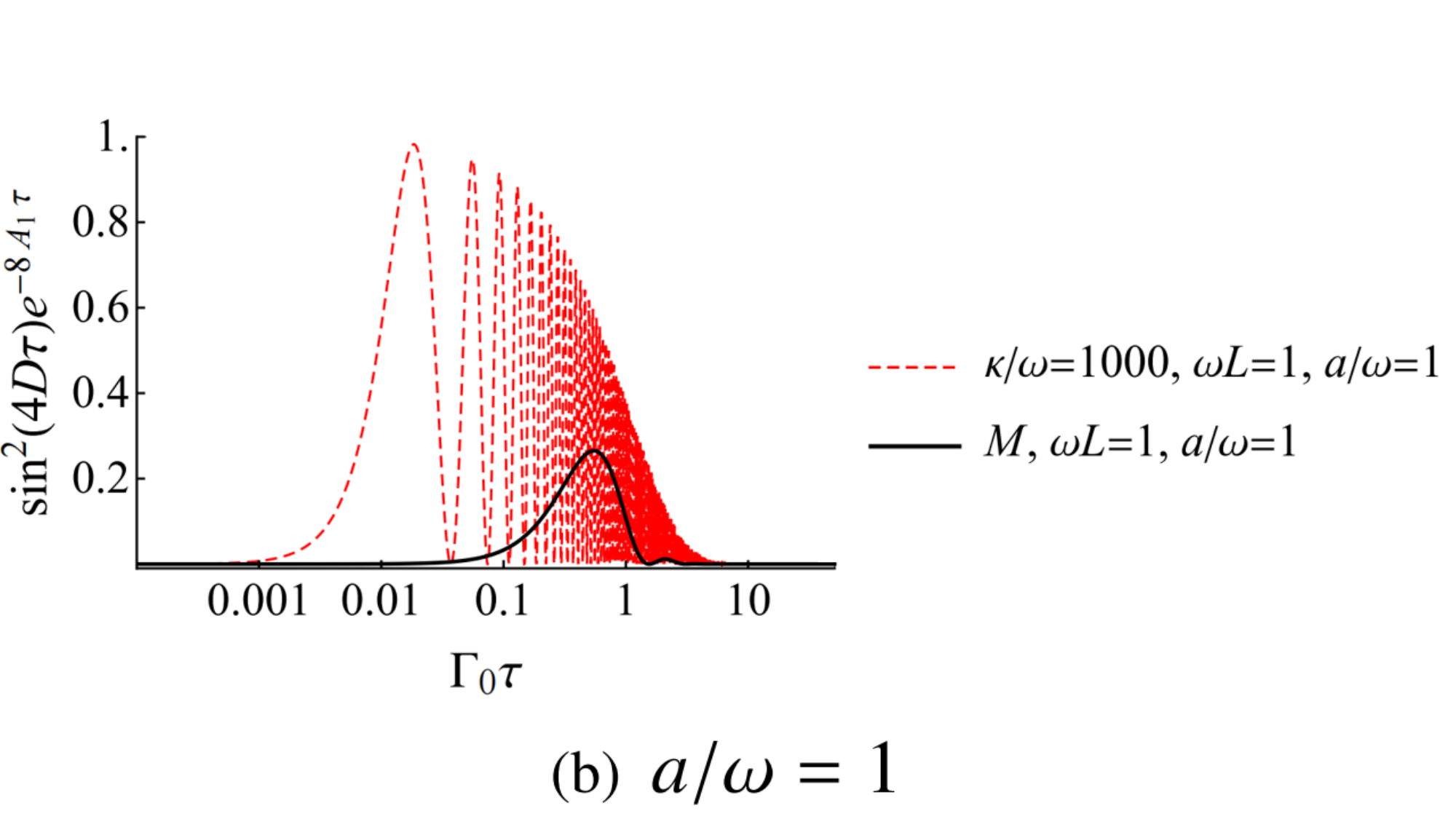}}\\
{\includegraphics[height=1.45in,width=2.25in]{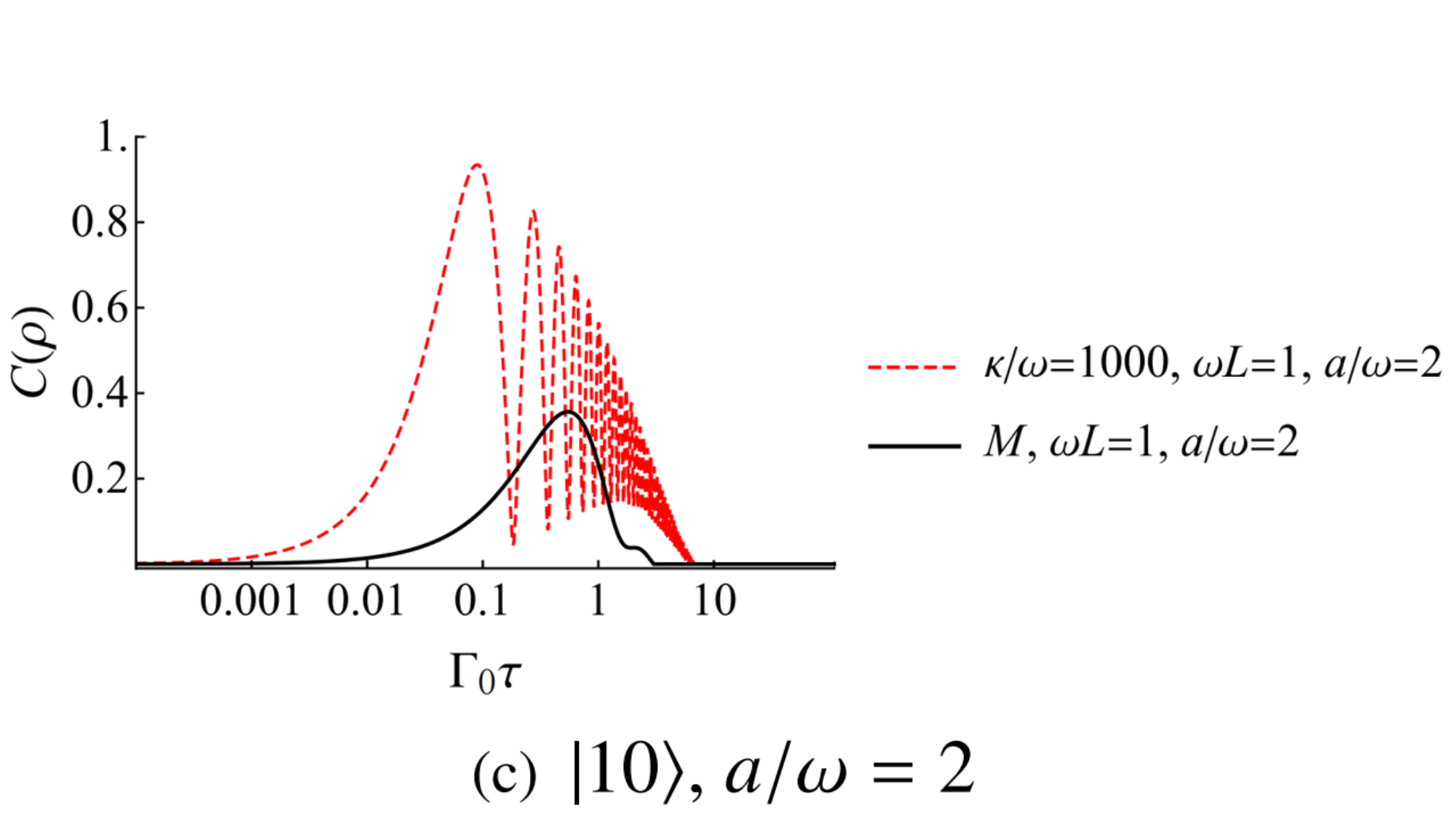}}
{\includegraphics[height=1.45in,width=2.25in]{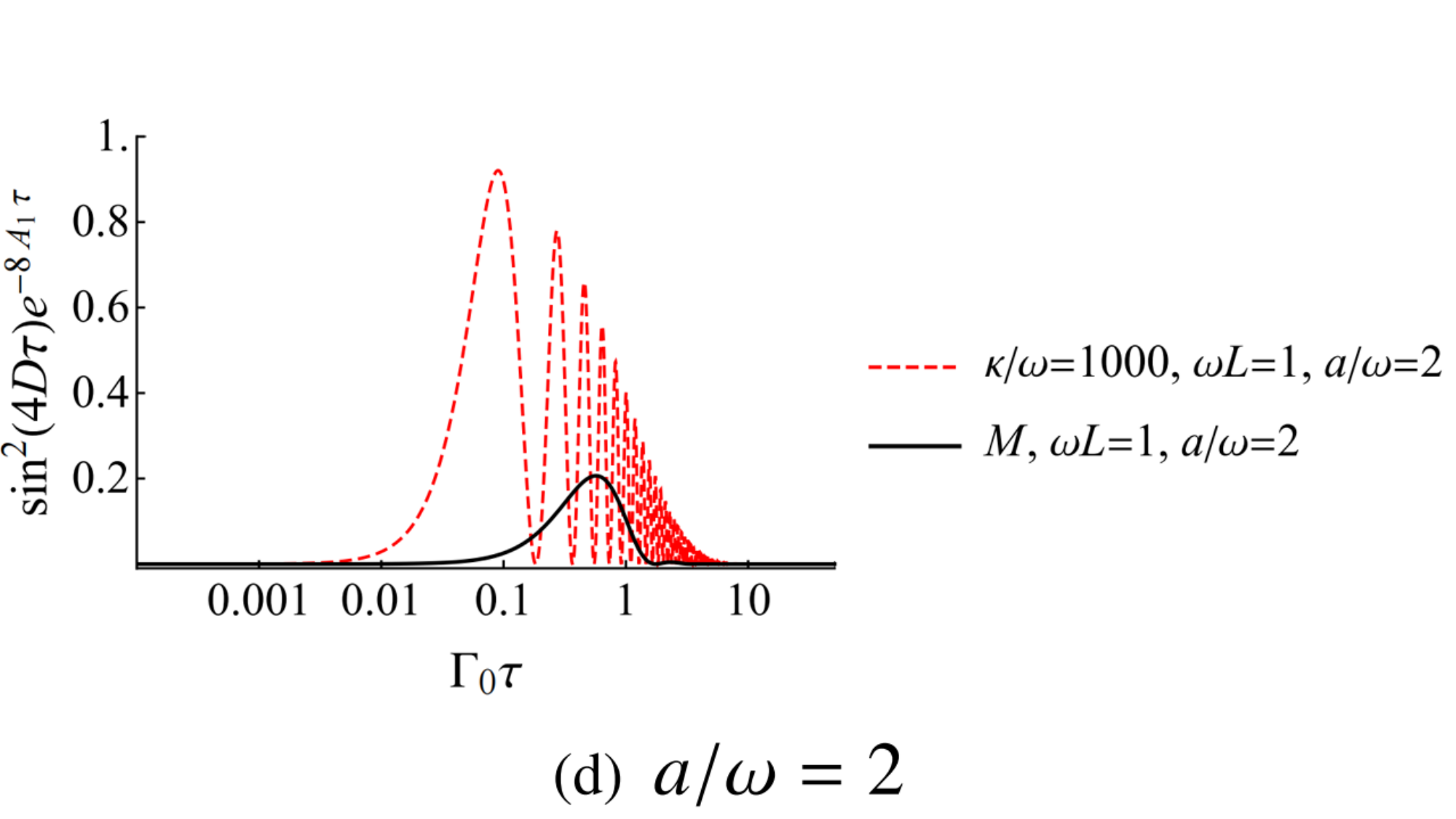}}
\caption{Time evolution of concurrence of two circularly accelerated atoms initially prepared in $|10\rangle$ with centripetal acceleration $a/\omega=1$ (a) and $a/\omega=2$ (c). The factor $\sin^2(4D\tau)e^{-8A_1\tau}$ is a function of time with centripetal acceleration $a/\omega=1$ (b) and $a/\omega=2$ (d).
}\label{kencir1-2}
\end{figure}

We assume that the two atoms are initially prepared in the separable state $|10\rangle$. As shown in Fig.~\ref{kencir1-2}, we take $\kappa/\omega=1000$ and $\omega L=1$ and plot the concurrence dynamics in $\kappa$-deformed and Minkowski spacetimes. This chosen initial state can cause the environment-induced interatomic interaction, which is embodied by the extra term $\sin^2(4D\tau)e^{-8A_1\tau}$ and significantly contributes to the two-atom entanglement dynamics.

We first note that entanglement will be generated as a result of the vacuum fluctuation of the quantum field and the motion of atoms. We also find that, as discussed previously, the oscillatory manner of the entanglement dynamics is observed during the initial period, which is attributed to the environment-induced interatomic interaction dominated by the trigonometric term $\sin^2(4D\tau)$. The oscillation frequency decreases as the centripetal acceleration increases. In addition, this oscillation form decays to that of the Minkowski spacetime case after a long time, which is determined by the exponential term $e^{-8A_1\tau}$. Therefore, the differences in the entanglement dynamics between the $\kappa$-deformed spacetime case with an LDP and the Minkowski spacetime case are more obvious during the initial period when the environment-induced interatomic interaction is considered. This finding indicates that the environment-induced interatomic interaction between two circularly accelerated atoms is beneficial to discriminate between these two universes through entanglement generation dynamics.

\subsubsection{Two circularly accelerated atoms initially prepared in entangled state $|\psi\rangle=\sqrt{p}|A\rangle+\sqrt{1-p}|S\rangle$}
Here, we analyze the effects of the environment-induced interatomic interaction on the entanglement dynamics of two circularly accelerated atoms initially prepared in $|\psi\rangle=\sqrt{p}|A\rangle+\sqrt{1-p}|S\rangle$ $(0<p<1, p\neq 1/2)$.

\begin{figure}[H]
\centering
{\includegraphics[height=1.65in,width=2.65in]{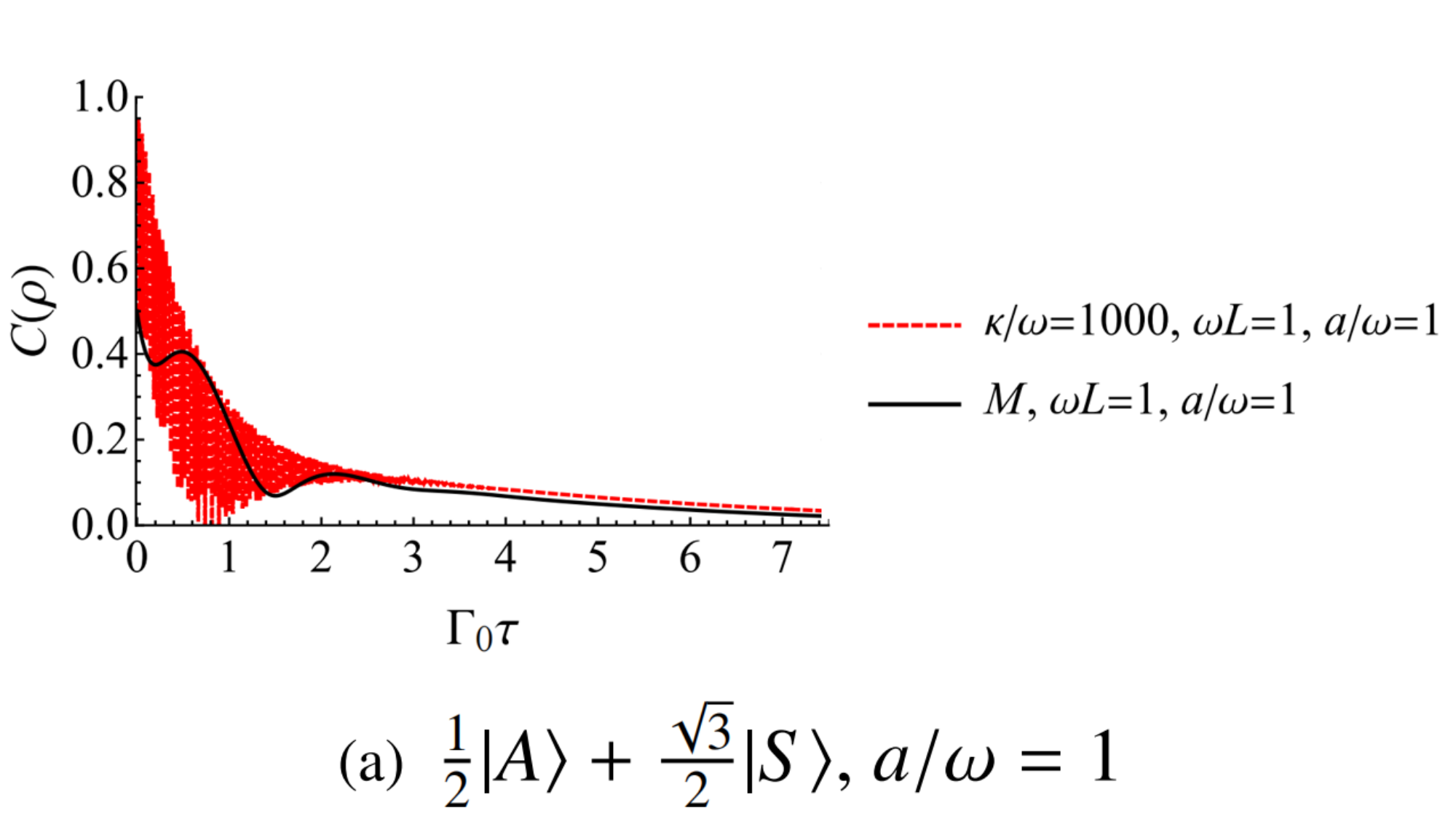}}
{\includegraphics[height=1.65in,width=2.65in]{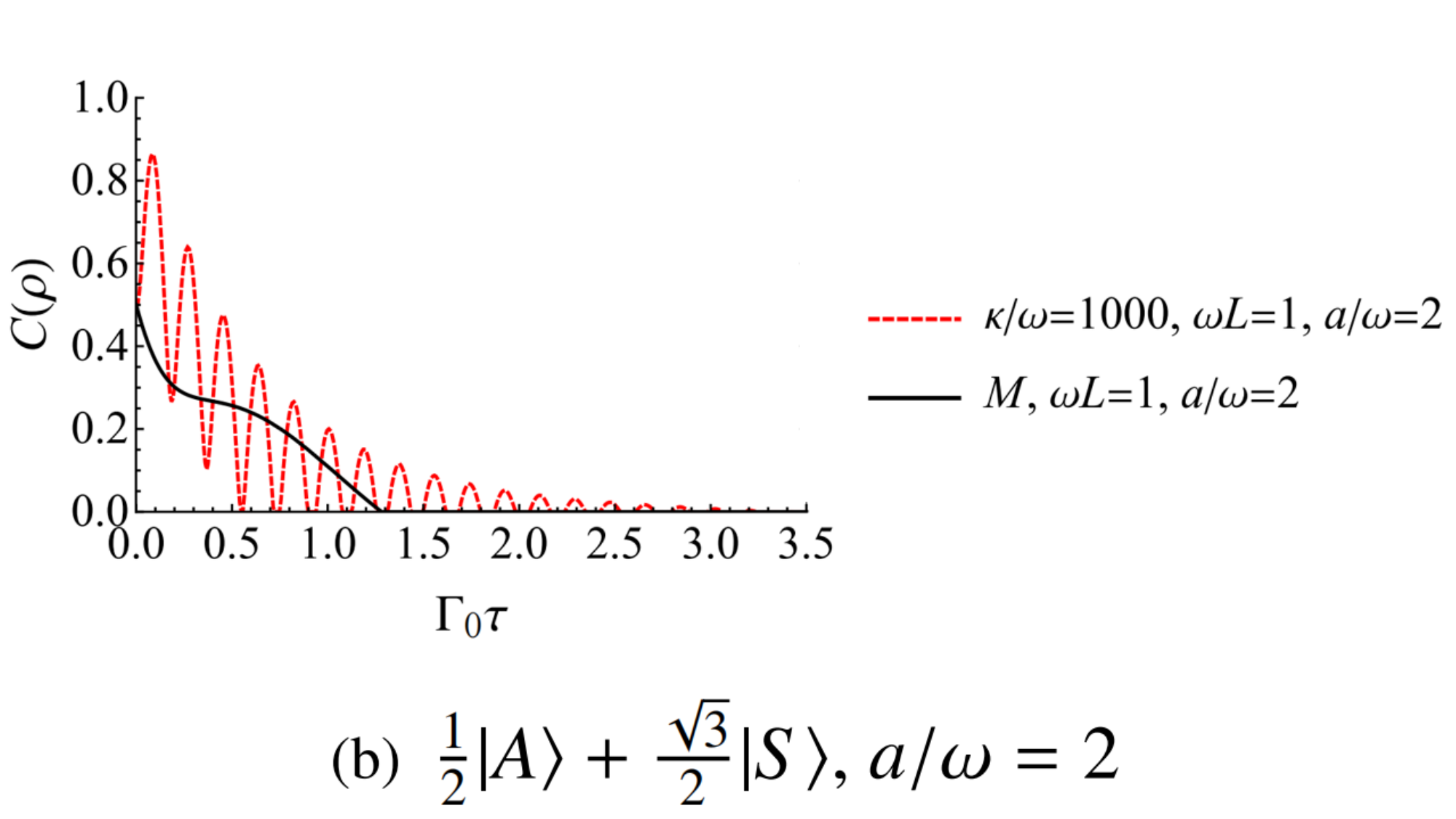}}
{\includegraphics[height=1.65in,width=2.65in]{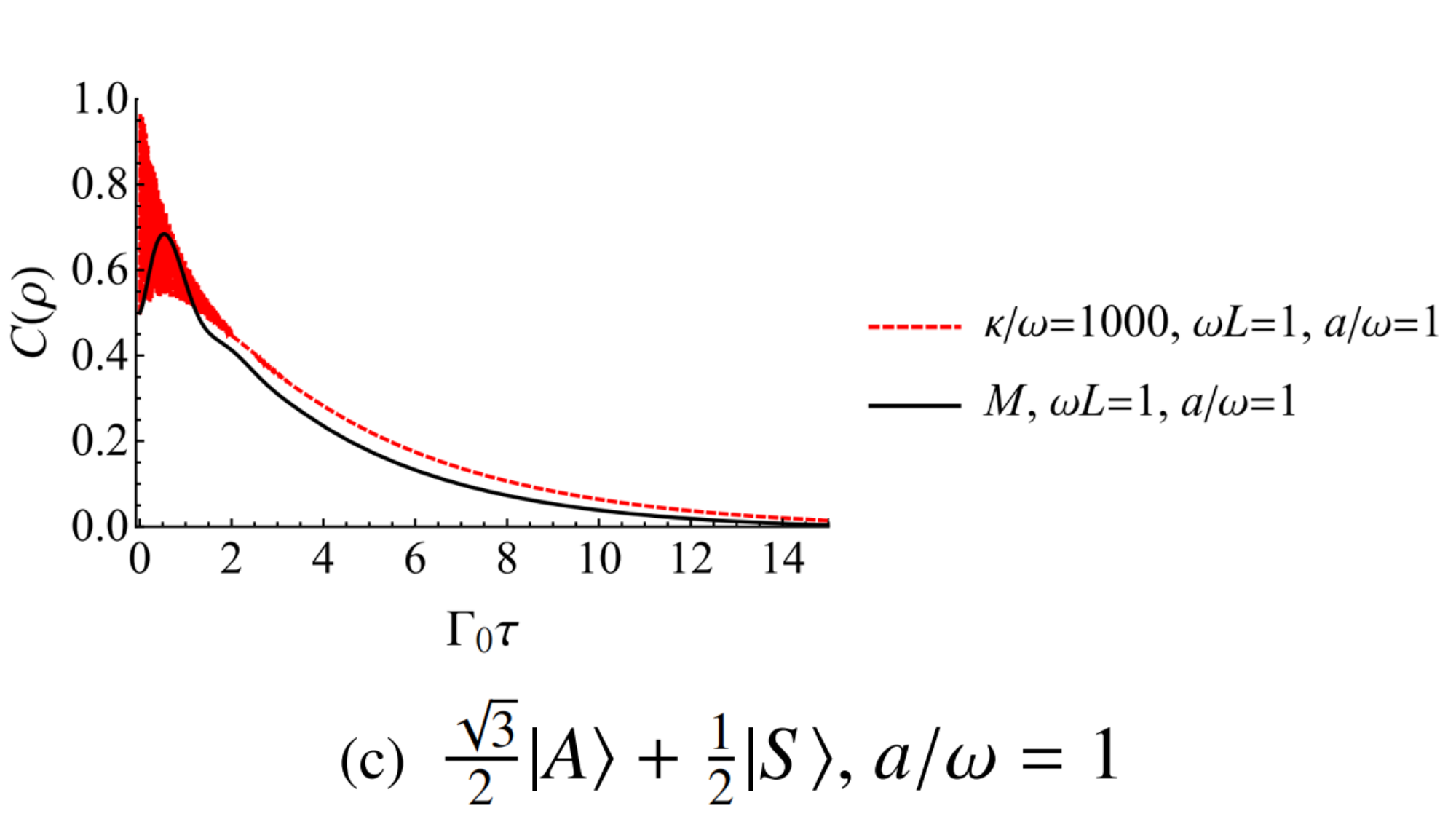}}
{\includegraphics[height=1.65in,width=2.65in]{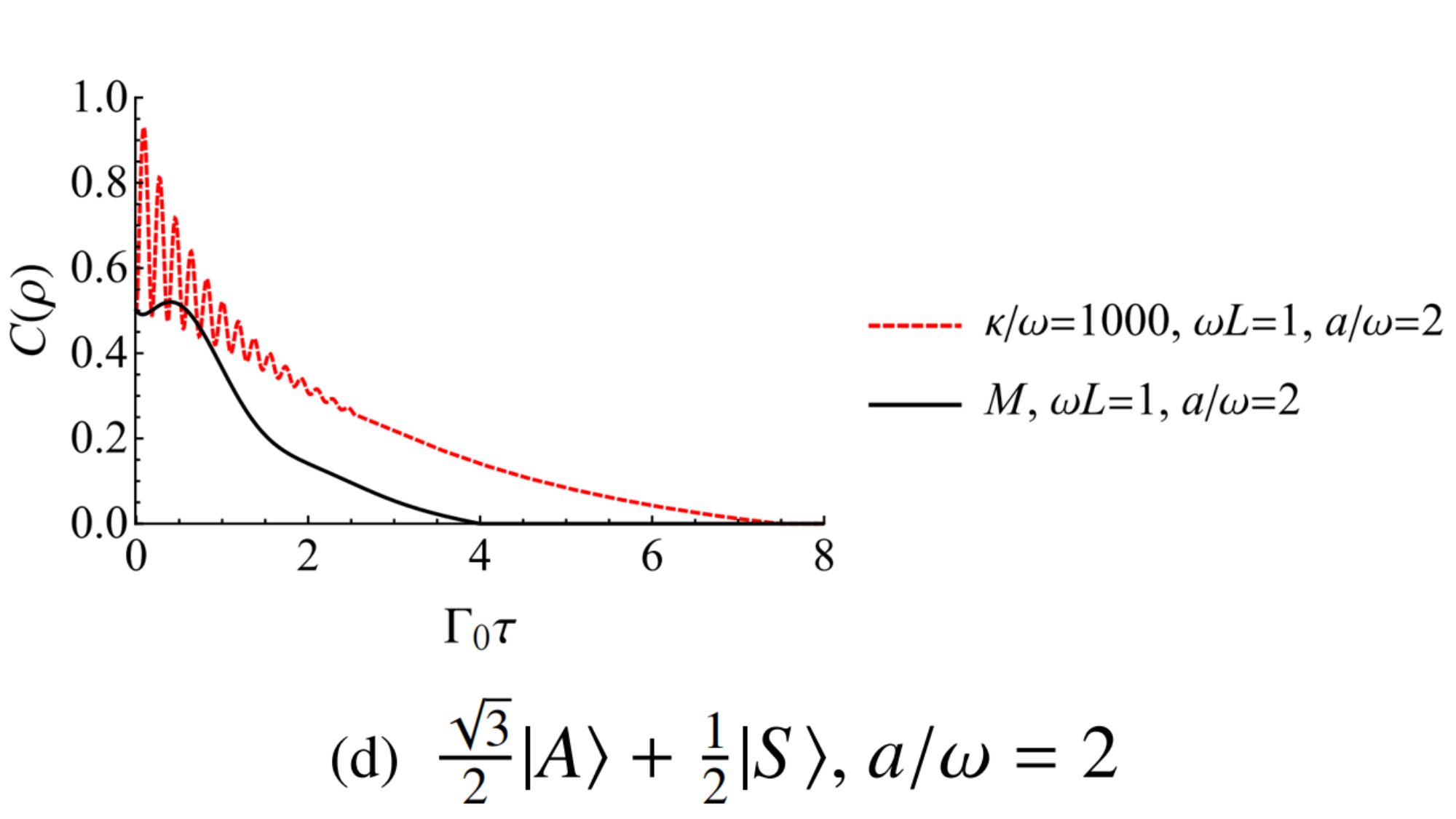}}
{\includegraphics[height=1.65in,width=2.65in]{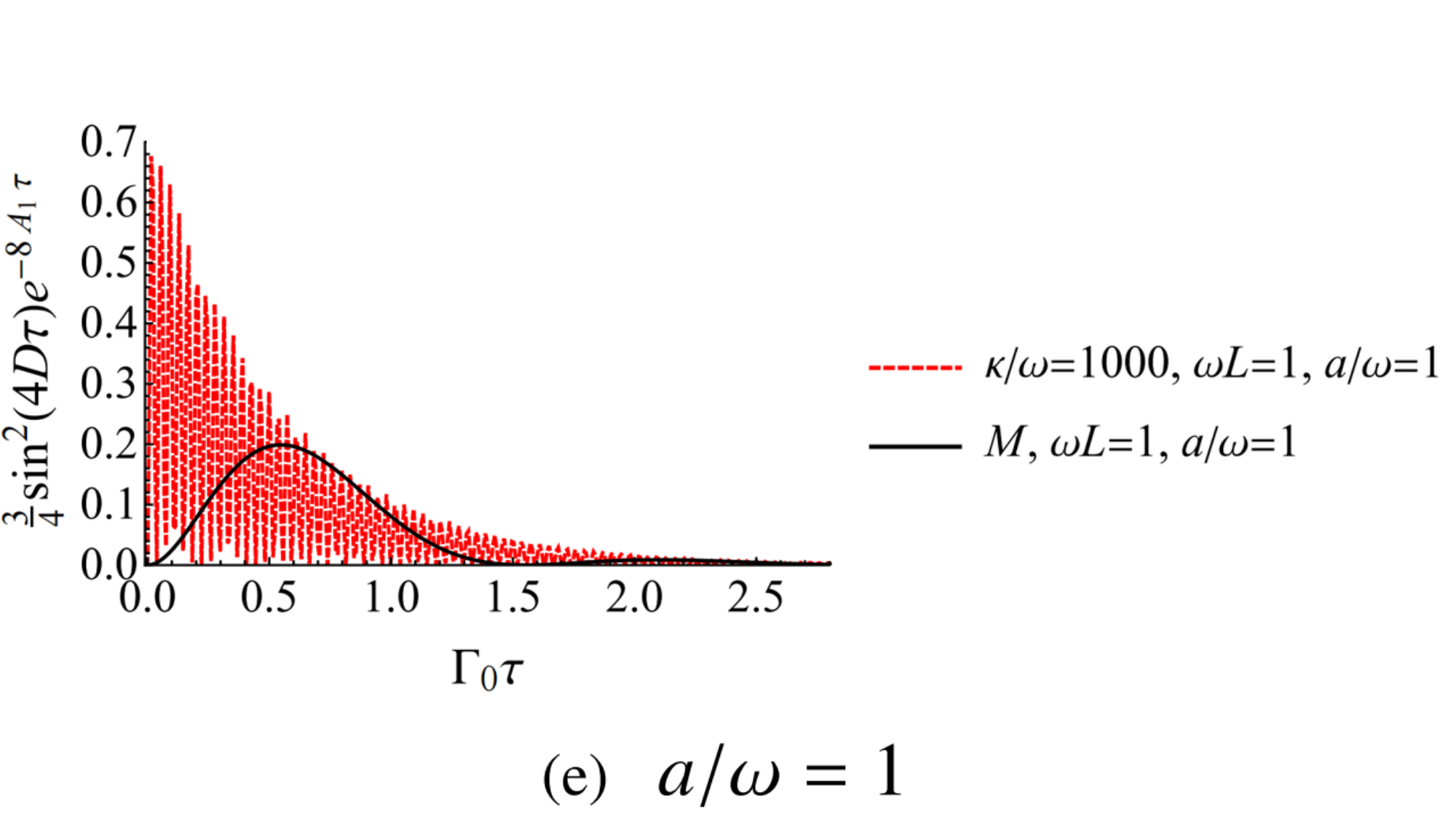}}
{\includegraphics[height=1.65in,width=2.65in]{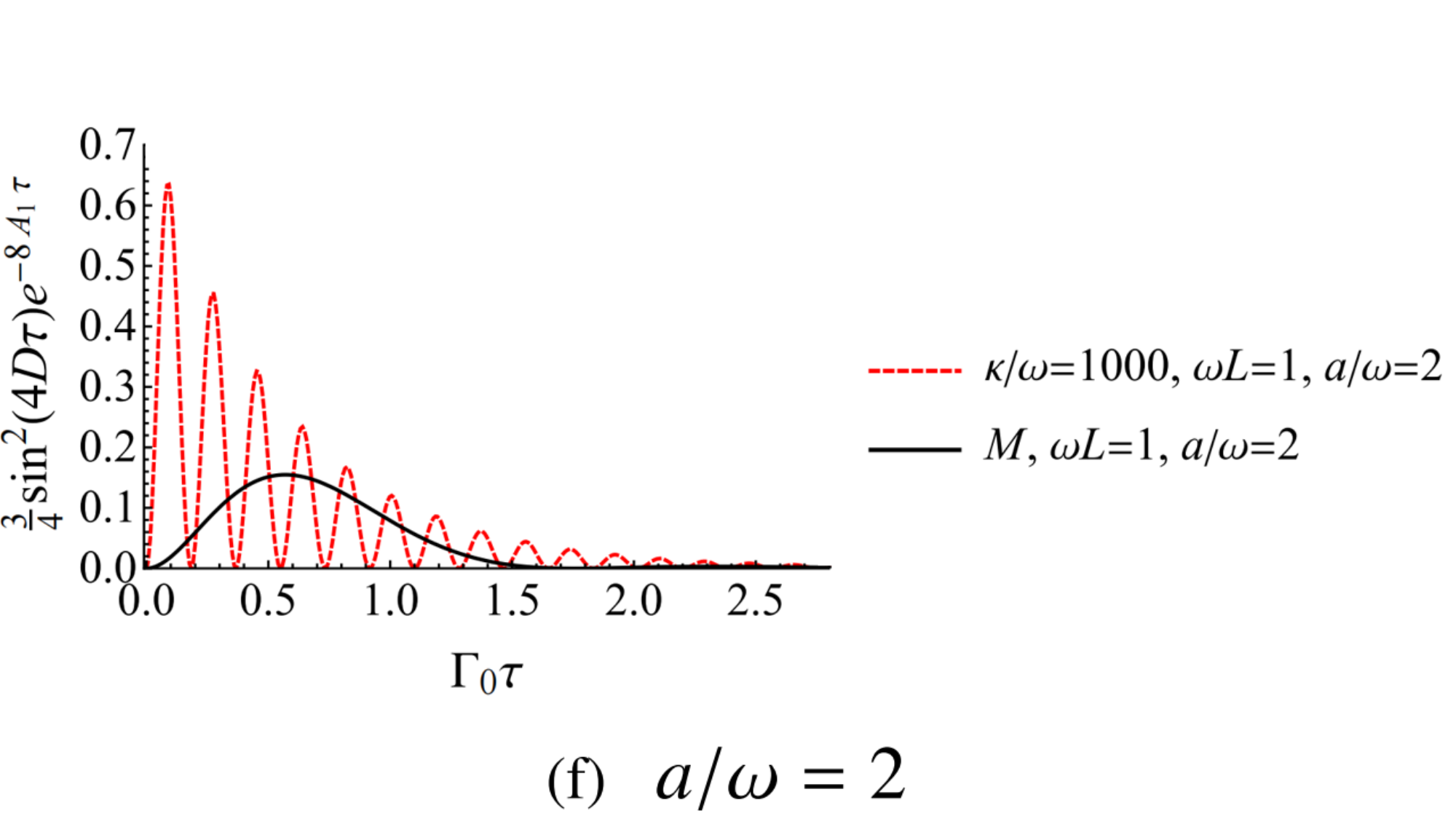}}
\caption{Time evolution of concurrence of two circularly accelerated atoms initially prepared in $\frac{1}{2}|A\rangle+\frac{\sqrt{3}}{2}|S\rangle$ for different values $a/\omega=1$ (a) and $a/\omega=2$ (b), and $\frac{\sqrt{3}}{2}|A\rangle+\frac{1}{2}|S\rangle$ for different values $a/\omega=1$ (c) and $a/\omega=2$ (d). The factor $\sin^2(4D\tau)e^{-8A_1\tau}$ is a function of time with the velocities $a/\omega=1$ (e) and $a/\omega=2$ (f).
}\label{circularly6-7}
\end{figure}

Fig.~\ref{circularly6-7} shows that the enhancement behavior depends on the state in which the atoms are initially prepared. When $p=1/4$ is considered, the symmetry entangled state mainly contributes to the initial state, and we infer from Figs.~\ref{circularly6-7} (a) and (b) that, under the effect of the environment-induced interatomic interaction, the two-atom entanglement exhibits an intriguing phenomenon: its decay and revival are different from those of static atoms, as shown in Fig.~\ref{R5} (a), and uniformly moving atoms, as shown in Figs.~\ref{f4} (a) and (b). In particular, when the centripetal acceleration is large, the entanglement will not be revived and will suffer ``sudden death." In addition, even in the late stage of evolution, in contrast to the aforementioned cases, the entanglement dynamics in $\kappa$-deformed spacetime does not coincide with that in Minkowski spacetime. In Figs. \ref{circularly6-7} (c) and (d), the entanglement dynamics of the $p=3/4$ initial state case is shown. We determine that the initial entanglement is enhanced in the initial phase and eventually decays to zero asymptotically. However, the entanglement dies off in an oscillatory manner in $\kappa$-deformed spacetime and will never coincide with the Minkowski spacetime case at any time. Remarkably, as the centripetal acceleration increases, the difference between $\kappa$-deformed and Minkowski spacetimes becomes larger. Moreover, the entanglement dynamics in both $\kappa$-deformed and Minkowski spacetimes could suffer entanglement ``sudden death," which means that the entanglement decays to zero at a finite time. An interesting phenomenon for the entanglement in the large acceleration situation is that, at a certain time interval, entanglement exists in the $\kappa$-deformed spacetime, whereas it vanishes in the Minkowski spacetime. This distinct characteristic of entanglement dynamics helps us distinguish these two universes.

We also note that the environment-induced interatomic interaction is embodied in the extra term $\frac{3}{4}\sin^2(4D\tau)e^{-8A_1\tau}$. In Figs. \ref{circularly6-7} (e) and (f), we show how this extra term performs under the effects of atomic acceleration in $\kappa$-deformed and Minkowski spacetimes. Because of acceleration, the extra term is always determined by the term $\sin^2(4D\tau)$ in the $\kappa$-deformed spacetime case, which is different from that in the Minkowski spacetime case. This characteristic is different from that of static and uniformly moving atoms that have been discussed previously. In addition, when the acceleration increases, the oscillation frequency and amplitude decrease in the $\kappa$-deformed spacetime case.

Therefore, we conclude that even when the spacetime deformation parameter $\kappa$ is large, by checking the atom entanglement dynamics, one may distinguish these two universes with the help of the environment-induced interatomic interaction between two circularly accelerated atoms in principle.

\section{Conclusions}\label{section5}
In this study, we have investigated the dynamic behavior of entanglement of a pair of static, uniformly moving, and circularly accelerated atoms with different initial states coupled to the massless scalar field in $\kappa$-deformed and Minkowski spacetimes. Through numerical evaluation, two different scenarios, i.e., with and without the environment-induced interatomic interaction, are considered. We have shown that the relativistic motion and environment-induced interatomic interaction have a significant effect on the entanglement dynamics of a two-atom system.

On the one hand, when two static atoms are initially prepared in the excited state, antisymmetric entangled state, and symmetric entangled state, we find that, without the environment-induced interatomic interaction, the differences in the entanglement dynamics between $\kappa$-deformed and Minkowski spacetimes are not obvious when the deformation parameter is large. However, when the inertial atoms move with a constant velocity, the entanglement dynamics are different between these two universes when the velocity is large. Furthermore, for two circularly accelerated atoms with a nonvanishing separation, the entanglement evolves differently with respect to the acceleration and interatomic distance or other parameters. More importantly, under certain conditions, the circularly accelerated atoms initially prepared in the excited state in $\kappa$-deformed spacetime (in Minkowski spacetime)  can get entangled, while they will not get entangled in Minkowski spacetime (in $\kappa$-deformed spacetime). Thus, the relativistic motion of the atoms significantly influences the differences in the entanglement dynamics between these two universes.

On the other hand, when the atoms are initially prepared in a separable state $|10\rangle$ and a superposition entangled state, the effect of environment-induced interatomic interaction cannot be ignored. First, for two static atoms, the behavior of entanglement dynamics in $\kappa$-deformed spacetime is different from that in Minkowski spacetime during the initial stage under the effect of environment-induced interatomic interaction, even when the deformation parameter is large. Moreover, for both cases of the uniform motion atoms and the uniform circular atoms, the relativistic motion enhances this difference in the entanglement dynamics between these two spacetimes. Thus, the numerical analysis results have shown that when the environment-induced interatomic interaction is considered, the $\kappa$-deformed spacetime can be easily distinguished from the Minkowski spacetime.

We have demonstrated how the relativistic motion and environment-induced interatomic interaction affect the entanglement dynamics of two atoms coupled to the massless scalar field. We anticipate that the relativistic motion can improve detection accuracy in the geometric phase or Lamb shift of $\kappa$-deformed spacetime, as investigated in Refs.~\cite{Arya2022,Arya2023}, making it easier for us to explore the structure and nature of spacetime. In addition, the many-body physics (or many detectors) may enhance the entanglement because of the large number of systems, thus improving the detection of $\kappa$ deformation of spacetime. How the entanglement estimation precision is improved has a direct impact on the detection of $\kappa$ deformation of spacetime in the possible experimental implementation, which is quite a challenge. We would rather leave such research to future work.

\begin{acknowledgments}
This work was supported by the Key Program of the National Natural Science Foundation of China (NSFC) (Grant No. 12035005). Xiaobao Liu
was supported by NSFC (Grant No. 12065016), and the Discipline-Team of
Liupanshui Normal University of China (Grant No. LPSSY2023XKTD11).
Zehua Tian was supported by NSFC (Grant No. 11905218) and the Scientific Research Start-Up Funds of Hangzhou Normal University (Grant No.
4245C50224204016). Xiaobao Liu thanks Shifeng Huang and Jiaozhen She
for advice and discussions.
\end{acknowledgments}

\end{document}